\documentclass[twocolumn,preprintnumbers,aps,pra,superscriptaddress,amsmath,amssymb]{revtex4-1}

\usepackage{graphicx}
\usepackage{dcolumn}
\usepackage{bm}
\usepackage{bbm}
\usepackage{hyperref}
\usepackage{color}
\usepackage{multirow}
\usepackage{physics}
\newcommand{\h}[1]{\color{black}{#1}}

\begin{document}

\title{\h Apparent delocalisation of the current flow in metallic wires observed with diamond nitrogen-vacancy magnetometry} 

\author{J.-P. Tetienne} 
\email{jtetienne@unimelb.edu.au}
\affiliation{School of Physics, The University of Melbourne, VIC 3010, Australia}	

\author{N. Dontschuk}
\affiliation{School of Physics, The University of Melbourne, VIC 3010, Australia}
\affiliation{Centre for Quantum Computation and Communication Technology, School of Physics, The University of Melbourne, VIC 3010, Australia}	
	
\author{D. A. Broadway}
\affiliation{School of Physics, The University of Melbourne, VIC 3010, Australia}
\affiliation{Centre for Quantum Computation and Communication Technology, School of Physics, The University of Melbourne, VIC 3010, Australia}

\author{S. E. Lillie}
\affiliation{School of Physics, The University of Melbourne, VIC 3010, Australia}	
\affiliation{Centre for Quantum Computation and Communication Technology, School of Physics, The University of Melbourne, VIC 3010, Australia}

\author{T. Teraji}
\affiliation{National Institute for Materials Science, Tsukuba, Ibaraki 305-0044, Japan}

\author{D. A. Simpson}
\affiliation{School of Physics, The University of Melbourne, VIC 3010, Australia}

\author{A. Stacey}
\affiliation{School of Physics, The University of Melbourne, VIC 3010, Australia}
\affiliation{Centre for Quantum Computation and Communication Technology, School of Physics, The University of Melbourne, VIC 3010, Australia}

\author{L. C. L. Hollenberg}
\affiliation{School of Physics, The University of Melbourne, VIC 3010, Australia}
\affiliation{Centre for Quantum Computation and Communication Technology, School of Physics, The University of Melbourne, VIC 3010, Australia}

\date{\today}
	
\begin{abstract}
\h
We report on a quantitative analysis of the magnetic field generated by a continuous current running in metallic micro-wires fabricated on an electrically insulating diamond substrate. A layer of nitrogen-vacancy (NV) centres engineered near the diamond surface is employed to obtain spatial maps of the vector magnetic field, by measuring Zeeman shifts through optically-detected magnetic resonance spectroscopy. The in-plane magnetic field (i.e. parallel to the diamond surface) is found to be significantly weaker than predicted, while the out-of-plane field also exhibits an unexpected modulation. We show that the measured magnetic field is incompatible with Amp{\`e}re's circuital law or Gauss's law for magnetism when we assume that the current is confined to the metal, independent of the details of the current density. This result was reproduced in several diamond samples, with a measured deviation from Amp{\`e}re's law by as much as 94(6)\% (i.e. a $15\sigma$ violation). To resolve this apparent magnetic anomaly, we introduce a generalised description whereby the current is allowed to flow both above the NV sensing layer (including in the metallic wire) and below the NV layer (i.e. in the diamond). Inversion of the Biot-Savart law within this two-channel description leads to a unique solution for the two current densities, which completely explains the data, is consistent with the laws of classical electrodynamics and indicates a total NV-measured current that closely matches the electrically-measured current. However, this description also leads to the surprising conclusion that in certain circumstances the majority of the current appears to flow in the diamond substrate rather than in the metallic wire, and to spread laterally in the diamond by several micrometres away from the wire. No electrical conduction was observed between nearby test wires, ruling out a conventional conductivity effect. Moreover, the apparent delocalisation of the current into the diamond persists when an insulating layer is inserted between the metallic wire and the diamond or when the metallic wire is replaced by a graphene ribbon. The possibilities of a measurement error, a problem in the data analysis or a current-induced magnetisation effect are discussed, but do not seem to offer a more plausible explanation for the effect. Understanding and mitigating this apparent anomaly will be crucial for future applications of NV magnetometry to charge transport studies.

\end{abstract}

\maketitle

\section{Introduction}

The nitrogen-vacancy (NV) defect centre in diamond is routinely used as an atomic-sized magnetometer through optical detection of its electron spin~\cite{Doherty2013,Rondin2014}. Thanks to its high sensitivity and small size, it is particularly well suited to applications in condensed matter physics~\cite{Casola2018}, where the quantitative measurements it provides can be precisely compared to theoretical models under diverse conditions, including from cryogenic temperatures up to 600 K~\cite{Acosta2010,Toyli2012}. Recent applications of NV sensing in this area include the study of nanoscale spin textures in ferromagnets~\cite{Rondin2013,Tetienne2014,Tetienne2015,Dussaux2016,Gross2016,Dovzhenko2018} and multiferroics~\cite{Gross2017}, vortices in superconductors~\cite{Waxman2014,Thiel2016,Pelliccione2016,Schlussel2018}, spin excitations in ferromagnets~\cite{VanderSar2015,Du2017,Page2018}, Johnson noise in metals~\cite{Kolkowitz2015,Agarwal2017,Ariyaratne2018} and current flow in conductors~\cite{Nowodzinski2015,Chang2017,Tetienne2017}. The latter is the focus of this work. By measuring the stray magnetic field produced by a stationary (DC) electric current, known as the Oersted field, it is possible to probe the properties of this current, and even in some situations to fully reconstruct its spatial distribution~\cite{Roth1989,Meltzer2018}. This capability offers potential applications to large-scale testing of integrated circuits~\cite{Nowodzinski2015}, as well as to real-space observation and investigation of exotic transport phenomena in condensed matter systems, such as electron refraction and viscous flow in van der Walls materials~\cite{Chen2016,Bandurin2016}. 

In this work, we use NV magnetic microscopy to image the stray field produced when injecting a DC current in metallic micro-wires fabricated on a NV-diamond sensing chip~\cite{Tetienne2017}. {\h Our vector magnetic field measurements are analysed in several ways of increasing generality: (i) by comparing to the predictions from the Biot-Savart law assuming a uniform current density in the metallic wire; (ii) by using Amp{\`e}re's circuital law in its integral form to derive an equality independent of the current density distribution in the wire; (iii) by using Gauss's law for magnetism and Amp{\`e}re's law in their differential forms ($\nabla\cdot{\bf B}=0$ and $\nabla\cross{\bf B}=0$, respectively) which give relationships between the magnetic field components independent of the nature and location of the sources of magnetic field (but all above the NV layer). These various analysis methods all point to an apparent magnetic anomaly, that is, the measured magnetic field seems to be incompatible with the laws of classical electrodynamics for a single current-carrying wire. To resolve this apparent violation of Amp{\`e}re's law and Gauss's law for magnetism, we propose to relax an assumption made in the application of these laws, namely we allow the sources of the measured magnetic field (including charge currents and magnetization) to be located anywhere in space including below the NV layer (i.e. in the diamond). In this case, our data become compatible with $\nabla\cdot{\bf B}=0$ and $\nabla\cross{\bf B}=0$. We then show that the Biot-Savart law can be inverted to obtain two current density distributions (projected in the NV plane), one for the sources located above the NV layer, one for the sources located below the NV plane. The solution is unique and, by construction, is an exact fit to the magnetic field data. However, it leads to the surprising conclusion that the majority of the current in some instances appears to flow in the diamond rather than in the metallic wire. The second part of the paper aims to gain an understanding of the reason for this apparent leakage of the current into an insulator, through further experimental tests and discussions of alternative explanations such as a measurement or analysis error. We conclude that the least implausible interpretation of our observations is that there is indeed an apparent long-range delocalisation of the current density (as seen via its associated magnetic field) which is not associated with a delocalisation of free charges since no conductivity between nearby contacts was observed.     

The manuscript is organised as follows. In Sec.~\ref{sec:methods}, we summarise our methods for sample fabrication, measurements and data analysis, which are described in more detail in Appendices~\ref{sec:fab}-\ref{sec:analysis}. In Sec.~\ref{sec:anomaly}, we present the magnetic field results for two representative samples and analyse them first under the natural assumption that the current is confined in the metallic wire (\ref{sec:conventional}), unveiling an apparent anomaly in the measured magnetic field which can be resolved by relaxing this assumption  (\ref{sec:resolving}); we then introduce a generalised description of the magnetic field in terms of a two-channel current density (\ref{sec:generalised}), indicating that the current flows in majority into the diamond, verify that this apparent delocalisation does not allow conventional electrical conduction between nearby contacts (\ref{sec:Nearby}), and discuss possible interpretations (\ref{sec:interpretations}). In Sec.~\ref{sec:tests}, we perform a number of experimental tests including varying the injected current (\ref{sec:CurrentDep}), the characteristics of the diamonds and fabricated devices (\ref{sec:DiamondDep}), the laser intensity (\ref{sec:LaserDep}), inserting an insulating layer (\ref{sec:Oxide}), and suspending the metal (\ref{sec:suspended}). Finally, we summarise the various possible interpretations and their respective plausibility (Sec.~\ref{sec:summary}) and conclude on the implications of the findings (Sec.~\ref{sec:conclusion}).}    

\begin{figure*}[t!]
	\begin{center}
		\includegraphics[width=0.99\textwidth]{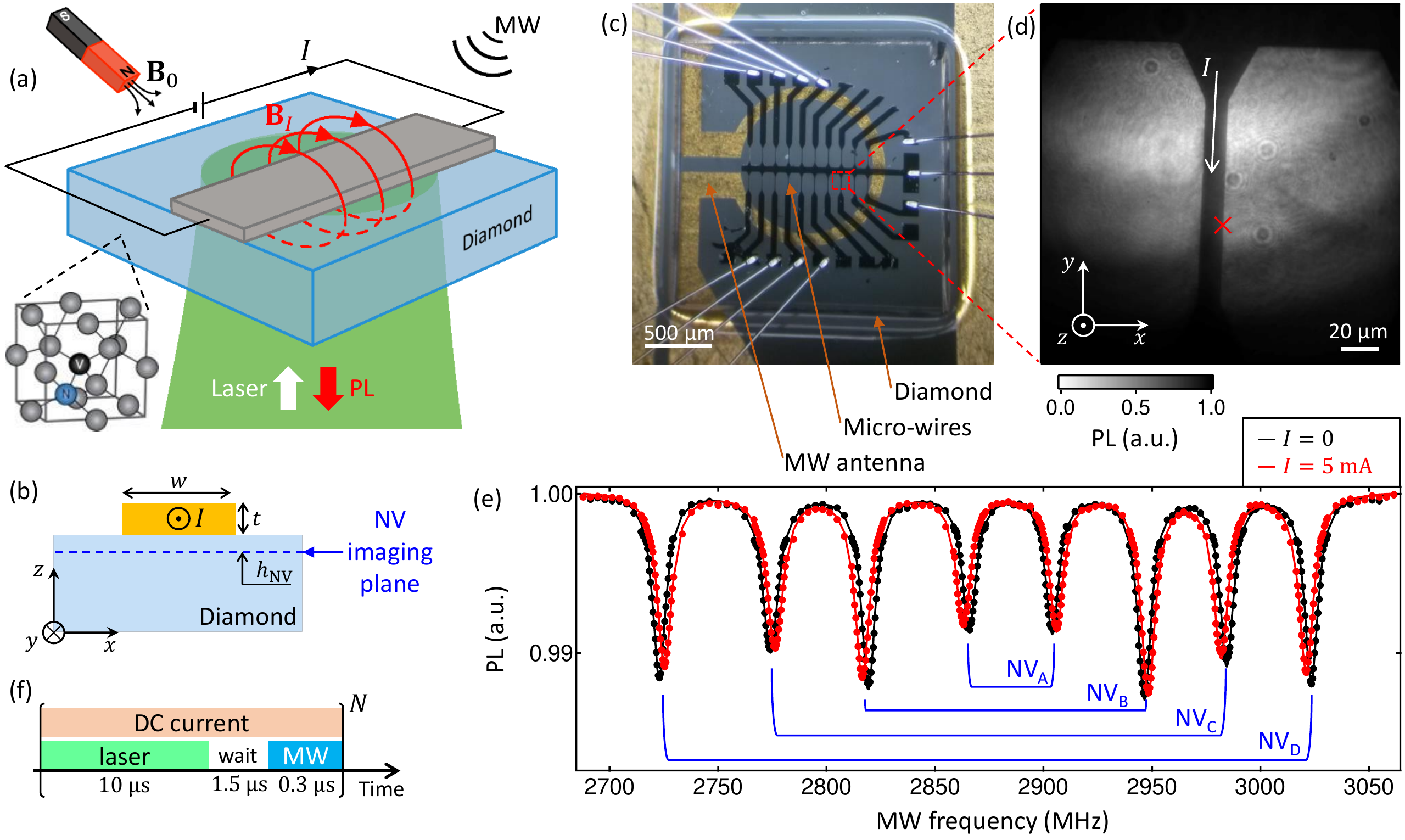}
		\caption{(a) Schematic of the experiment. A metallic strip carrying a DC current $I$ is fabricated on a diamond containing a layer of near-surface NV centres. The vector magnetic field is mapped by performing optically detected magnetic resonance (ODMR) spectroscopy on the NV centres, which requires laser and microwave (MW) excitations and collection of the NV photoluminescence (PL). (b) Cross-section of the device defining the geometrical parameters: the wire has a width $w$ and thickness $t$, the mean NV-surface distance is denoted as $h_{\rm NV}$. (c) Photograph showing the micro-wires fabricated on top of the diamond and the MW antenna placed underneath. (d) PL image of a typical device. (e) ODMR spectra from a single pixel near the edge of the wire indicated by the red cross in (d), with (red data) and without (black) an applied current $I=-5$~mA. Solid lines are multiple-Lorentzian fits. Dips from the different NV orientations are labelled NV$_{\rm A...D}$. (f) Pulse sequence used for the measurement, which is repeated typically $N\sim3000$ times for each MW frequency.}
		\label{FigIntro}
	\end{center}
\end{figure*} 

\section{Methods summary} \label{sec:methods}

The principle of the experiment is depicted in Fig.~\ref{FigIntro}a. A flat metallic wire (or strip) is fabricated on a diamond substrate comprising a layer of NV centres at a depth $h_{\rm NV}$ from the surface (Fig.~\ref{FigIntro}b). The goal of the experiment is to image the stray magnetic field ${\bf B}_I$ generated by a current $I$ running through the wire, using the NV layer as an array of vector magnetometers~\cite{Steinert2010,Maertz2010,Pham2011}. Precisely, we prepared several single-crystal diamond plates implanted with nitrogen ions at various energies and fluences to form the NV centres (see details in Appendix~\ref{sec:fab}). The mean depth in a given sample, $h_{\rm NV}$, ranged from $h_{\rm NV}\sim8$~nm to $h_{\rm NV}\sim28$~nm, set by the implantation energy. On each diamond plate, we fabricated Ti/Au or Cr/Au wires by photolithography and electron-beam evaporation. The wires are between 9 and $23~\mu$m in width, at least $100~\mu$m in length, and the Au layer is 50-100 nm thick on top of a 10-nm adhesion layer made of either Ti or Cr. A photograph of a typical mounted device is shown in Fig.~\ref{FigIntro}c. 

The time-averaged magnetic field was imaged using pulsed optically detected magnetic resonance (ODMR) spectroscopy on the layer of NV centres, using a custom-built wide-field fluorescence microscope \cite{Simpson2016,Tetienne2017}. The set-up comprises a green laser (wavelength $\lambda=532$~nm) to excite the NV centres over a wide field of view ($\sim100~\mu$m diameter spot), a camera to image the red photoluminescence (PL), a microwave (MW) antenna to drive the NV spin resonances, and a DC current source connected to the device under study (see further details in Appendix~\ref{sec:meas}). A PL image of a typical device (from underneath) is shown in Fig.~\ref{FigIntro}d, where the wire appears darker because of some non-radiative decay induced by the metal (see Appendix~\ref{sec:optical}). ODMR spectra from a single pixel (containing several hundreds of NVs typically) close to the wire are shown in Fig.~\ref{FigIntro}e with a current $I=-5$~mA (red data) and no current (black), with the typical pulse sequence shown in Fig.~\ref{FigIntro}f. The negative sign of $I$ denotes that the current flows in the $-y$ direction. The spectra comprise eight lines due to two electron spin resonances for each of the four possible NV orientations (labelled NV$_{\rm A...D}$). These eight lines would be degenerate in the absence of a magnetic field, but can be resolved via the application of a purposefully oriented bias magnetic field ${\bf B}_0$~\cite{Steinert2010,Chipaux2015,Glenn2017,Tetienne2017} produced by a permanent magnet. The amplitude of this bias field satisfies $|{\bf B}_0|\gg|{\bf B}_I|$, hence the current-induced field ${\bf B}_I$ manifests as small shifts in the ODMR frequencies, as illustrated in Fig.~\ref{FigIntro}e. 

To analyse the ODMR data, we fit the spectrum at each pixel with a sum of eight Lorentzian functions (solid lines in Fig.~\ref{FigIntro}e). The eight resulting frequencies are then used to infer the total magnetic field ${\bf B}_{\rm tot}$ by numerical fitting of the calculated frequencies obtained from the spin Hamiltonian for each NV orientation (see Appendix~\ref{sec:analysis} for details). For each sample studied, we first measure the field without applying any current ($I=0$) yielding the background field ${\bf B}_0$, before measuring the field with a given current $I$, corresponding to a total field ${\bf B}_{\rm tot}={\bf B}_0+{\bf B}_I$. Subtraction of the two maps then gives the current-induced field alone, ${\bf B}_I$, which is the field we will show and discuss in the next section.   
 
\section{From a magnetic anomaly to a conduction anomaly} \label{sec:anomaly}

\subsection{\h A conventional analysis of the magnetic field} \label{sec:conventional} 

\begin{figure*}[t!]
	\begin{center}
		\includegraphics[width=0.95\textwidth]{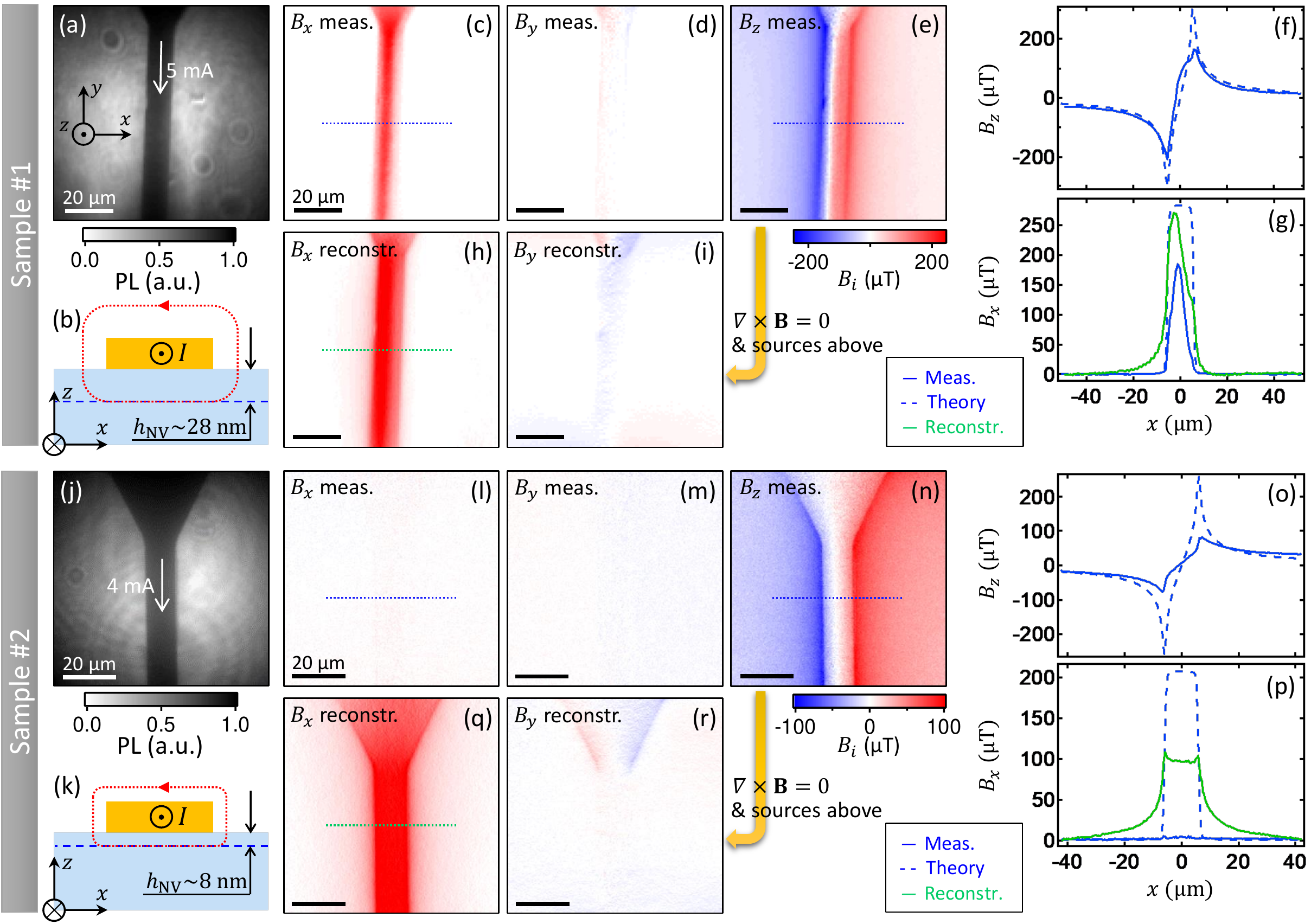}
		\caption{(a) PL image of a device in sample \#1, which has a mean NV depth $h_{\rm NV}\sim28$~nm. (b) Schematic cross-section of the device, with a typical magnetic field line depicted by a red dotted line. (c-e) Maps of the magnetic field components $B_x$ (c), $B_y$ (d) and $B_z$ (e) measured for the device shown in (a) with a DC current $I=-5$~mA, where the bias field ${\bf B}_0$ measured with $I=0$ was subtracted to show only the current-induced field ${\bf B}_I$. (f,g) Solid blue lines: line cuts of $B_z$ (f) and $B_x$ (g) taken along the horizontal dotted line shown in (c,e). Dashed blue lines: prediction from the Biot-Savart law assuming a uniform current density confined within the metallic wire. A convolution with a Gaussian function (full width at half maximum of 1~$\mu$m) was applied to account for the finite optical resolution of the microscope~\cite{Tetienne2018b}. In (g), the green line is the reconstructed $B_x$ profile as defined in (h,i). (h,i) Maps of the reconstructed $B_x$ and $B_y$ components based on the measured $B_z$ and Eqs.~(\ref{eq:bx},\ref{eq:by}). Since these equations are not valid for $k=0$, a constant offset was added to cancel the field at the boundaries of the images (away from the wire). (j-r) Same as (a-i) for a device in sample \#2, which has a mean NV depth $h_{\rm NV}\sim8$~nm. Here the injected current is $I=-4$~mA.}
		\label{FigAnomaly}
	\end{center}
\end{figure*} 

We first consider two different samples with NV centres at mean depths $h_{\rm NV}\sim28$~nm (sample \#1) and $h_{\rm NV}\sim8$~nm (sample \#2). The data for sample \#1 are shown in Fig.~\ref{FigAnomaly}a-i. Figure~\ref{FigAnomaly}a-e shows the PL image of the device under study (a), a schematic cross-section of the device (b), and the measured magnetic field components $B_x$ (c), $B_y$ (d) and $B_z$ (e) under a DC current $I=-5$~mA. The $B_y$ component is found to be mostly null (as expected from the device symmetry), whereas $B_x\approx200~\mu$T near the centre of the wire and $B_z\approx\pm150~\mu$T near the edges of the wire, consistent with the $\Delta f\approx3$~MHz Zeeman shifts observed in the ODMR spectra at this location (Fig.~\ref{FigIntro}e). In these images, the pixel-to-pixel noise is about $1~\mu$T (standard deviation over an ensemble of pixels) and systematic errors are estimated to be less than $2~\mu$T (see Appendix~\ref{sec:uncertainties}). 

To compare with theoretical expectations, we use the Biot-Savart law which expresses the magnetic field generated by a current density ${\bf J}({\bf r})$,
\begin{eqnarray} \label{eq:BS}
{\bf B}_I({\bf r}) = \frac{\mu_0}{4\pi}\int d^3{\bf r}'\frac{{\bf J}({\bf r}')\times({\bf r}-{\bf r}')}{|{\bf r}-{\bf r}'|^3}~,
\end{eqnarray}
where $\mu_0$ is the vacuum permeability and the integration is over all space. Assuming a uniform current density that is perfectly contained inside the wire, we can compute the magnetic field in the NV plane by integration of Eq.~(\ref{eq:BS}). The results are shown in Fig.~\ref{FigAnomaly}f,g for the $B_z$ and $B_x$ components, respectively (dashed lines, $B_y$ is null in this scenario), along with line cuts extracted from the measurements (solid blue lines). There is a large discrepancy especially in the $B_x$ component, where the measured field is significantly lower overall than the predicted field. To quantify the deviation from theory, we apply Amp{\`e}re's circuital law in its integral form to an appropriately chosen closed curve $C$ (see Appendix~\ref{sec:Ampere}), which allows us to write  
\begin{eqnarray} \label{eq:Ampere0}
\mu_0 I=\oint_C{\bf B}_I\cdot d{\bf l} \approx 2\int_{-x_b}^{+x_b}B_x(x)dx~,
\end{eqnarray}  
where $x=\pm x_b$ are the bounds of the measurements (i.e. $2x_b$ is the width of the images) satisfying $x_b\gg w$ ($w$ is the width of the wire, see Fig.~\ref{FigIntro}b). To a very good approximation (see Appendix~\ref{sec:Ampere}), the equality in Eq.~(\ref{eq:Ampere0}) should hold for any current density distribution ${\bf J}({\bf r})$ as long as it is confined inside the wire, and tells us that the area under the $B_x$ profile (as plotted in Fig.~\ref{FigAnomaly}g) should be independent of the details of ${\bf J}({\bf r})$ and equal to $\mu_0 I/2$. We quantify the deviation from Amp{\`e}re's law as $\chi \doteq 1-\frac{2\int_{-x_b}^{+x_b}B_x(x)dx}{\mu_0 I}$ and find $\chi\approx61(6)\%$ in this case, where the quoted uncertainty is based on the possibility of systematic errors in the measured $B_x$ (see Appendix~\ref{sec:uncertainties}). 

The data for sample \#2 are shown in Fig.~\ref{FigAnomaly}j-r, for a DC current $I=-4$~mA. Here we find that the $B_x$ component is close to the noise floor, with a value of $B_x=3(2)~\mu$T under the wire and $B_x=2(2)~\mu$T elsewhere. This result is consistent with the lack of observable Zeeman shifts in the ODMR data at the centre of the wire (see Fig.~\ref{FigODMR} in Appendix~\ref{sec:analysis}), and is in clear disagreement with the Biot-Savart law which predicts a value of $B_x\approx200~\mu$T under the wire. Consequently, the deviation from Amp{\`e}re's law is extremely high, $\chi\approx94(6)\%$, corresponding to a $15\sigma$ violation (where $\sigma$ is the standard error). The $B_z$ component also deviates significantly from theory (Fig.~\ref{FigAnomaly}o), with the measured $B_z$ peaking a factor $2-3$ smaller than the predicted field. 

{\h 
To further analyse this magnetic anomaly, we recall another fundamental law of classical electrodynamics, namely Gauss's law for magnetism, or $\nabla\cdot{\bf B}=0$ in its differential form. In Cartesian coordinates, this law writes $\frac{\partial B_x}{\partial x}+\frac{\partial B_y}{\partial y}+\frac{\partial B_z}{\partial z}=0$, hence the components of the magnetic field are not completely independent~\cite{Lima2009}. For sample \#2 where $B_x$ and $B_y$ are essentially null everywhere in the NV plane, this implies that $\frac{\partial B_z}{\partial z}=0$. However, since the magnetic field originates from a conduction current localised outside the diamond (and so above the NV layer), $B_z$ should decay monotonically with distance from the current-carying wire and hence should never satisfy $\frac{\partial B_z}{\partial z}=0$ unless $B_z=0$. To verify this quantitatively, we move to the two-dimensional (2D) Fourier space where real-space coordinates $x$ and $y$ become $k$-space coordinates $k_x$ and $k_y$. Gauss's law for magnetism then writes (see derivation in Ref.~\cite{Lima2009} and Appendix~\ref{sec:relationships})   
\begin{eqnarray} \label{eq:GaussLaw}
ik_xb_x^\pm(k_x,k_y,z)+ik_yb_y^\pm(k_x,k_y,z)=\pm kb_z^\pm(k_x,k_y,z)
\end{eqnarray}
where $b_p^\pm(k_x,k_y,z)$ is the 2D Fourier transform of $B_p^\pm(x,y,z)$, ${\bf k}=(k_x,k_y)$ is the spatial frequency vector, and $k=\sqrt{k_x^2+k_y^2}$. The $\pm$ sign refers to the magnetic field produced by sources located above ($+$ sign) and below ($-$ sign) the $z$ plane. In our experiments, assuming all the sources are located above the measurement plane (we neglect the weak diamagnetic response of diamond, which has a magnetic susceptibility of $-2.1\times10^{-5}$), we should have $ik_xb_x+ik_yb_y=kb_z$. In Appendix~\ref{sec:gauss}, we plot the expected $B_z$ map as reconstructed from the measured $B_x$ and $B_y$ components for samples \#1 and \#2 using this equation, in clear disagreement with the measured $B_z$ with a difference up to an order of magnitude larger than the measurement uncertainty.

Likewise, we can apply the differential form of Amp{\`e}re's law in a source-free region ($\nabla\cross{\bf B}=0$) to write relationships between the magnetic field components (see Appendix~\ref{sec:relationships}), namely
\begin{eqnarray} \label{eq:bx}
b_x^\pm(k_x,k_y,z)=\mp \frac{ik_x}{k}b_z^\pm(k_x,k_y,z) \\
b_y^\pm(k_x,k_y,z)=\mp\frac{ik_y}{k}b_z^\pm(k_x,k_y,z).  \label{eq:by}
\end{eqnarray}
For the field generated by a current-carrying wire located above the diamond, we should have $b_x=-\frac{ik_x}{k}b_z$ and $b_y=-\frac{ik_y}{k}b_z$. In other words, the components of the magnetic field in a given plane are completely inter-related~\cite{Lima2009,Casola2018,Dovzhenko2018}. Applying these relations to the measured out-of-plane component $B_z$, we can obtain the reconstructed in-plane components $B_x$ and $B_y$ as shown in Fig.~\ref{FigAnomaly}h,i for sample \#1 and in Fig.~\ref{FigAnomaly}q,r for sample \#2, revealing large discrepancies with the measured field. In particular, the reconstructed $B_x$ is much larger and closer to the Biot-Savart prediction than the measured $B_x$ (see green lines in Fig.~\ref{FigAnomaly}g,p). Interestingly, the equality in Eq.~(\ref{eq:Ampere0}) is satisfied (within error) when using the reconstructed $B_x$, that is, the integral form of Amp{\`e}re's law is satisfied when using the measured $B_z$ component but not when using the measured $B_x$ component (which appears abnormally suppressed). In other words, the measured $B_z$ profile is quantitatively consistent with the injected current $I$. However, there is still an apparent anomaly in the measured $B_z$, because the reconstructed $B_x$ spreads beyond the width of the wire in the $x$ direction (especially for sample \#2), which is not expected from a current confined to the wire in our geometry. 

}

\subsection{\h Resolving the magnetic field anomaly} \label{sec:resolving}
 
{\h Let us briefly summarise our findings so far. We measured the vector components of the current-induced magnetic field in the NV plane, but found that these components are apparently not inter-related as they should be according to Gauss's law for magnetism ($\nabla\cdot{\bf B}=0$) or Amp{\`e}re's law ($\nabla\cross{\bf B}=0$), which are independent of the detail of the current density in the metallic wire. In other words, our measurements appear to be incompatible with the laws of classical electrodynamics. This implies that either the magnetic field measurements are erroneous, or that these laws have not been applied correctly.      

Our measurements rely on the conversion of precisely determined spin resonance frequencies into a magnetic field through a well-characterised Hamiltonian~\cite{Doherty2013,Rondin2014}. Examining ODMR spectra at different wire currents $I$ (Fig.~\ref{FigIntro}e) or at different locations with respect to the wire (Fig.~\ref{FigODMR}c in Appendix~\ref{sec:analysis}) does not reveal any significant modification of the NV charge state or spin resonance character. In all cases, the set of resonance frequencies is well fit by the standard NV spin Hamiltonian, with a fit error comparable to the measurement uncertainty (see details in Appendix~\ref{sec:analysis}). In other words, there is no evidence that this Hamiltonian may be incorrect or incomplete for our purpose. Moreover, the anomaly concerns a small differential magnetic field (induced by the current) on top of a much larger background magnetic field (produced by a permanent magnet), which as expected is seen to be uniform and unaffected by the presence of the metallic wire (Fig.~\ref{FigB0} in Appendix~\ref{sec:analysis}). This rules out a modification of the purely magnetic response of the NVs, as this would affect the total magnetic field, not just the small current-induced magnetic field. The possibility of a problem in the analysis will be re-analysed in detail in Sec.~\ref{sec:summary}, and representative raw ODMR data are available at the link~\cite{data} to allow independent verifications to be carried out.  

Beside the possibility of a measurement error in this differential magnetic field, the other way to reconcile experiment and theory is to question the assumptions that led to the apparent violation of the laws of classical electrodynamics. To apply $\nabla\cdot{\bf B}=0$ and $\nabla\cross{\bf B}=0$ to the data, one assumption was made: it was assumed that the sources of magnetic field are located only on one side of the NV layer, namely above the NV layer where the metallic wire is located. This led to Eq.~(\ref{eq:GaussLaw}) with the plus sign for $\nabla\cdot{\bf B}=0$, and Eqs.~(\ref{eq:bx},\ref{eq:by}) with the minus sign for $\nabla\cross{\bf B}=0$. Such an assumption is needed as measurements in the $xy$ plane do not have direct access to the $\partial/\partial z$ terms of the differential equations (see Appendix~\ref{sec:relationships}). Although this assumption seems very reasonable a priori, we will see that removing this assumption not only resolves the magnetic anomaly problem, i.e. there is no longer a violation of Gauss's law for magnetism and Amp{\`e}re's law, but also leads to an excellent match between the total current deduced from the magnetic field measurements and the electrically measured current. However, this will also lead to the surprising conclusion that the majority of the current (or more generally, the dominant source of magnetic field) is located in the diamond rather than in the metallic wire. 

}

\subsection{\h A generalised analysis of the magnetic field} \label{sec:generalised}

{\h Instead of making an assumption on the location of the magnetic sources, here we generalise our description} to the situation where the measured (total) magnetic field ${\bf B}$ has contributions from sources that are both above (current density ${\bf J}^+$ producing a field ${\bf B}^+$) and below (${\bf J}^-$ producing ${\bf B}^-$) the NV plane. We emphasise that at this stage the sources are not specified and could be in the form, for instance, of a magnetised object (permanent or induced) equivalent to a current density ${\bf J}=\nabla\times{\bf M}$ where ${\bf M}$ is the magnetisation density. In this generic scenario, Eqs.~(\ref{eq:bx},\ref{eq:by}) become $b_x=-\frac{ik_x}{k}(b_z^+-b_z^-)$ and $b_y=-\frac{ik_y}{k}(b_z^+-b_z^-)$, implying that the total out-of-plane component ($b_z=b_z^++b_z^-$) is completely decoupled from the total in-plane components, which are themselves still related to each other via $k_yb_x=k_xb_y$. Likewise, Gauss's law for magnetism no longer imposes any relationship on the magnetic field components. As a result, the experimental data becomes compatible with the laws of classical electrodynamics. Moreover, it is then easy to find a source that can explain (qualitatively) the data of sample \#2: if the current $I$ is allowed to flow partly above and partly below the NV plane in a symmetric fashion, the in-plane field components will be identically null (because the in-plane field from the two sides interfere destructively, i.e. $B_x^+=-B_x^-$) while the out-of-plane field will be essentially unchanged (because of constructive interference, $B_z^+=B_z^-$). Likewise, the reduction in $B_x$ observed for sample \#1 is consistent with a current that flows partly below the NV plane (although still mostly above). {\h This thus resolves the apparent discrepancy between the measured and predicted in-plane field}. As for the discrepancy in the out-of-plane component $B_z$ (related to the anomalous lateral spread in the reconstructed $B_x$), it can be explained by a lateral spread in the current density beyond the width of the wire. 

We now quantify these effects by inferring the current densities ${\bf J}^+$ and ${\bf J}^-$ from the measured (total) magnetic field ${\bf B}={\bf B}^++{\bf B}^-$. To do so, we make the assumption that ${\bf J}^+$ and ${\bf J}^-$ are confined within a distance $h_{\rm max}$ to the NV plane such that $h_{\rm max}\ll\Delta x_{\rm min}$ where $\Delta x_{\rm min}\approx500$~nm is the lateral spatial resolution of our measurements (which limits the maximum spatial frequency accessible, $k_{\rm max}=1/\Delta x_{\rm min}$), close to the diffraction limit~\cite{Simpson2016} and roughly matched to the pixel size. Under this assumption, valid here since $h_{\rm max}\approx88$~nm and $h_{\rm max}\approx118$~nm in samples \#1 and \#2, respectively (based on the current flowing in the Au layer), the magnetic field depends only on the projected current density $\tilde{\bf J}^\pm=\int{\bf J}^\pm dz$ ($\tilde{\bf J}^\pm$ is a lineal current density, in units of A/m), with no experimental parameter. Namely, we have in the Fourier plane (see derivation in Appendix~\ref{sec:relationships2})
\begin{eqnarray} \label{eq:jy}
b_x &=& -\frac{\mu_0}{2}\left(\tilde{j}_y^+-\tilde{j}_y^-\right) \\
b_y &=& \frac{\mu_0}{2}\left(\tilde{j}_x^+-\tilde{j}_x^-\right) \\ 
b_z &=& -\frac{\mu_0}{2}\frac{ik}{k_x}\left(\tilde{j}_y^++\tilde{j}_y^-\right) \label{eq:jy3} \\
b_z &=& \frac{\mu_0}{2}\frac{ik}{k_y}\left(\tilde{j}_x^++\tilde{j}_x^-\right)~, \label{eq:jy2} 
\end{eqnarray}
showing that $B_z$ is related to the total projected current density, $\tilde{\bf J}=\tilde{\bf J}^++\tilde{\bf J}^-$, whereas $B_x$ and $B_y$ are related to the difference $\Delta\tilde{\bf J}=\tilde{\bf J}^+-\tilde{\bf J}^-$. Eqs.~(\ref{eq:jy}-\ref{eq:jy2}) are very general and apply to any situation where the magnetic sources (charge currents, magnetic moments etc.) are located within a distance $h_{\rm max}\ll\Delta x_{\rm min}$ of the magnetic field measurement plane. In reality, the NV centres exhibit a spread in $z$ due to the implantation process, with a typical standard deviation of $h_{\rm NV}/2$ where $h_{\rm NV}$ is the mean implantation depth~\cite{Lehtinen2016}. Therefore, the distinction `above' and `below' is to be understood as `mostly above' and `mostly below', respectively, with an appropriate weighting for sources located within the NV layer.  

\begin{figure*}[t!]
	\begin{center}
		\includegraphics[width=0.95\textwidth]{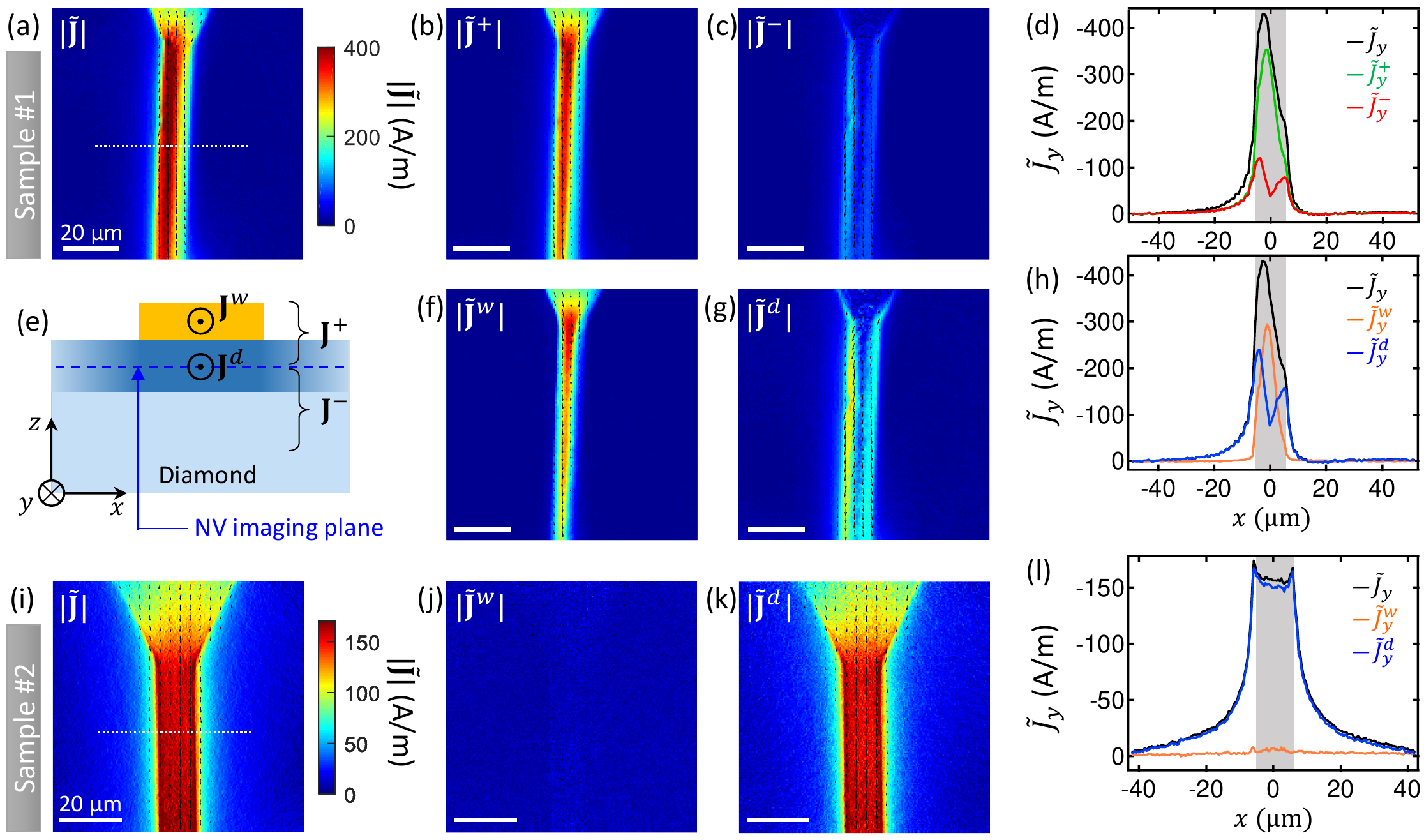}
		\caption{(a) Map of the total current density projected in the NV plane (denoted $\tilde{\bf J}$) deduced from the measured magnetic field component $B_z$ for sample \#1 (device imaged in Fig.~\ref{FigAnomaly}c-e) via Eqs.~(\ref{eq:jy3},\ref{eq:jy2}). Since these equations are not valid for $k=0$, a constant offset was added to cancel the current density at the boundaries of the images (away from the wire). The colour codes for the norm $\tilde{|{\bf J}|}$ whereas the direction of $\tilde{\bf J}$ is indicated by black arrows overlaid on the image. The arrows have a length proportional to $\tilde{|{\bf J}|}$ and are shown only if $|\tilde{{\bf J}}|>100$~A/m. (b,c) Maps of the current density above ($\tilde{\bf J}^+$, panel b) and below ($\tilde{\bf J}^-$, panel c) the NV plane, deduced from Eqs.~(\ref{eq:jy}-\ref{eq:jy2}). (d) Line cuts of $\tilde{J}_y$, $\tilde{J}_y^+$ and $\tilde{J}_y^-$ taken along the horizontal dotted line shown in (a). The grey shading indicates the location of the wire as extracted from the PL image. (e) Model of the current flow: the injected current $I$ is split into two separate paths, a current $I_w$ confined to the metallic wire (current density ${\bf J}^w$) and a current $I_d$ flowing in the diamond symmetrically with respect to the NV plane and unbounded laterally (current density ${\bf J}^d$). (f,g) Maps of ${\bf J}^w$ and ${\bf J}^d$ deduced from (a-c). (h) Line cuts of $\tilde{J}_y$, $\tilde{J}_y^w$ and $\tilde{J}_y^d$ taken along the horizontal dotted line shown in (a). (i-k) Maps of $\tilde{\bf J}$, $\tilde{\bf J}^w$ and $\tilde{\bf J}^d$ obtained for sample \#2 (device imaged in Fig.~\ref{FigAnomaly}l-n). The threshold for the arrows is $|\tilde{{\bf J}}|>40$~A/m. (l) Line cuts of $\tilde{J}_y$, $\tilde{J}_y^w$ and $\tilde{J}_y^d$ taken along the horizontal dotted line shown in (i).}
		\label{FigModel}
	\end{center}
\end{figure*} 

{\h Equations~(\ref{eq:jy}-\ref{eq:jy2}) form a system of four equations with four unknowns ($\tilde{j}_x^\pm$, $\tilde{j}_y^\pm$) and has a unique solution. Traditionally, one of the conduction channels is neglected (e.g., $\tilde{\bf J}^-$) and the system is then overdetermined, i.e. the vector components of the magnetic field are used as redundant information to improve the reconstruction of the single-channel current density~\cite{Nowodzinski2015,Tetienne2017}. Alternatively, when only a single field component is available, the current continuity condition ($\nabla\cdot{\bf J}=0$) must be imposed to provide a unique solution for the single-channel current density~\cite{Roth1989}. Here, we make full use of the vector information available to reconstruct the two-channel current density, without any unnecessary assumption.} The results of the reconstruction for sample \#1 are shown in Fig.~\ref{FigModel}a-c where we plotted $\tilde{\bf J}$, $\tilde{\bf J}^+$ and $\tilde{\bf J}^-$, respectively. In these maps, the colour codes for the norm of the current density vector ($|\tilde{\bf J}|$) whereas the direction of the vector is indicated by overlaid arrows. Line cuts of $\tilde{J}_y$, $\tilde{J}_y^+$ and $\tilde{J}_y^-$ ($y$ is the main direction of current flow) across the wire are shown in Fig.~\ref{FigModel}d, revealing that $\tilde{J}_y^+$ is maximum near the centre of the wire while $\tilde{J}_y^-$ is peaked near the edges. Importantly, integrating $\tilde{J}_y$ over the transverse direction $x$ gives a total NV-measured current $I_{\rm tot}=\int_{-x_b}^{+x_b}\tilde{J}_y(x)dx=-4.9(3)~$mA, in agreement with the electrically measured current of $I=-5.000(5)$~mA (elsewhere quoted as -5~mA for brevity). This agreement indicates that the injected current is completely accounted for by our measurements, and validates our reconstruction method. The uncertainty in $I_{\rm tot}$ is dominated by truncation artefacts due to the finite size of the measured $B_z$ map (see Appendix~\ref{sec:truncation}). 

Similarly to the total current, we can integrate $\tilde{J}_y^+$ and $\tilde{J}_y^-$ over $x$ to obtain the total current flowing above the NV plane, $I_+\approx-3.4(4)$~mA, and below the NV plane, $I_-\approx-1.5(4)$~mA. This implies that a significant portion of the current, namely $I_-/I_{\rm tot}\approx30\%$ flows below the NV plane, setting a lower bound for the portion of the current flowing in the diamond, $|I_d|\geq|I_-|$. Furthermore, about 20\% of $I_+$ is localised at positions $x$ laterally offset from the wire ($|x|>w/2$) and hence must also flow inside the diamond, raising the lower bound to $|I_d|\geq2.1$~mA (i.e. 43\% of the total current). The remaining 80\% of $I_+$ is localised either in or under the wire, which the measurements cannot distinguish, therefore the upper bound for $I_d$ is simply $I_d\leq I$, corresponding to the case where the current flows entirely in the diamond. 

As a metric to characterise the leakage current $I_d$, we will use $I_d\doteq2I_-$, which amounts to assuming that the current flowing in the diamond is distributed equally above and below the NV plane. Moreover, the lateral distribution of this current is likely to be similar above and below the NV plane given the expected vertical confinement, therefore we define the projected current density in the diamond as $\tilde{\bf J}^d\doteq2\tilde{\bf J}^-$. This scenario is illustrated in Fig.~\ref{FigModel}e, where the current density flowing in the metallic wire is denoted as $\tilde{\bf J}^w$ such that the total current density is simply the sum $\tilde{\bf J}=\tilde{\bf J}^w+\tilde{\bf J}^d$. Within this model, we have that
\begin{eqnarray} \label{eq:Jwd}
\tilde{\bf J}^w & \doteq & \tilde{\bf J}^+-\tilde{\bf J}^- \\
\tilde{\bf J}^d & \doteq & 2\tilde{\bf J}^- ~. \label{eq:Jwd2} 
\end{eqnarray}
That is, there is a simple one-to-one correspondence between the model-independent quantities $(\tilde{\bf J}^\pm)$ and the model-specific $(\tilde{\bf J}^w,\tilde{\bf J}^d)$. To facilitate the discussions, in the following we will analyse the data using the quantities $\tilde{\bf J}^w$ and $\tilde{\bf J}^d$ and describe them as the current flowing in the wire and in the diamond, respectively, knowing that the total current flowing in the diamond may in fact be lower by a factor up to 2 (bounded by $\tilde{\bf J}^-$) or larger (bounded by the total current $\tilde{\bf J}$).

Applying this conversion to the data of sample \#1, we obtain the maps shown in Fig.~\ref{FigModel}f,g, with line cuts plotted in Fig.~\ref{FigModel}h. The integrated current flowing in the diamond is then $I_d=2I_-\approx-2.9(4)$~mA while the current in the wire is only $I_w\approx-2.0(3)$~mA, i.e. a leakage of $I_d/I_{\rm tot}\approx60\%$. It can be seen that $\tilde{\bf J}^w$ is laterally confined to the region delimited by the width of the wire (grey shading in Fig.~\ref{FigModel}h) while $\tilde{\bf J}^d$ spreads several micrometres beyond in the $x$ direction, suggesting that $\tilde{\bf J}^w$ and $\tilde{\bf J}^d$ have been sensibly separated. Applying the same analysis to the data of sample \#2 (current maps are shown in Fig.~\ref{FigModel}i-k), we find that the current flows mostly in the diamond, with a ratio $I_d/I_{\rm tot}\approx94\%$. Like for sample \#1, the current spreads laterally beyond the width of the wire, here by as much as $\sim20~\mu$m. In fact, only a portion $\int_{-w/2}^{+w/2}\tilde{J}_y^d dx/I_d\approx57\%$ of the current flows right under the wire, with the remaining 43\% of $I_d$ flowing a distance of $1~\mu$m or more (laterally) from the wire, and 18\% of $I_d$ flowing at a distance larger than $10~\mu$m. This significant lateral spread explains the apparent discrepancy between measured and calculated $B_z$ in Fig.~\ref{FigAnomaly}o, which cannot be explained by a current purely confined to the width of the wire (whether above or below the NV plane). Again, the total current obtained by integrating $\tilde{J}_y$ is $I_{\rm tot}\approx-3.9(3)~$mA, in agreement with the injected current of $I=-4.000(4)$~mA. This shows that the magnetic field data are completely consistent (within error) with the injected current flowing near the metallic wire (i.e. within our field of view) but simply delocalised into the diamond both vertically and laterally. 

While our measurements provide direct access to the lateral distribution of the projected current density, the estimation of the vertical extent of $\tilde{\bf J}^d$ necessitates further discussion. Let us first consider the case of sample \#2, for which $\tilde{\bf J}^w\doteq\tilde{\bf J}^+-\tilde{\bf J}^-\approx0$ -- a consequence of $B_x$ and $B_y$ being null. Assuming for simplicity that the current flows entirely in the diamond (which is formally true in the regions not under the metal), this implies that for any lateral position $(x,y)$, the following equality must hold: $\int_{-h_{\rm NV}}^{0}{\bf J}dz=\int_{-\infty}^{-h_{\rm NV}}{\bf J}dz$, where $z=0$ is the diamond surface and $z=-h_{\rm NV}$ the NV plane. 
Therefore, any well-behaved function describing the $z$-dependence of ${\bf J}$ must decay over a length scale of the order of $h_{\rm NV}$ under the NV plane. Thus, in sample \#2 the current is likely confined within a distance of the order of $h_{\rm NV}\sim8$~nm from the surface or from the NV plane, whereas it spreads laterally over several micrometres (17\% of the total current flows at a distance larger than $10~\mu$m from the edges of the wire). The same reasoning applies to sample \#1 for the regions outside the wire ($|x|>w/2$) where the current is necessarily confined to the diamond $z<0$, implying again that the current must be vertically confined to an extent of the order of $h_{\rm NV}$. Thus, it is likely that the current be confined within $h_{\rm NV}\sim28$~nm of the surface everywhere in sample \#1 as well.

\subsection{\h The case of nearby wires} \label{sec:Nearby}

{\h

\begin{figure*}[t!]
	\begin{center}
		\includegraphics[width=1.02\textwidth]{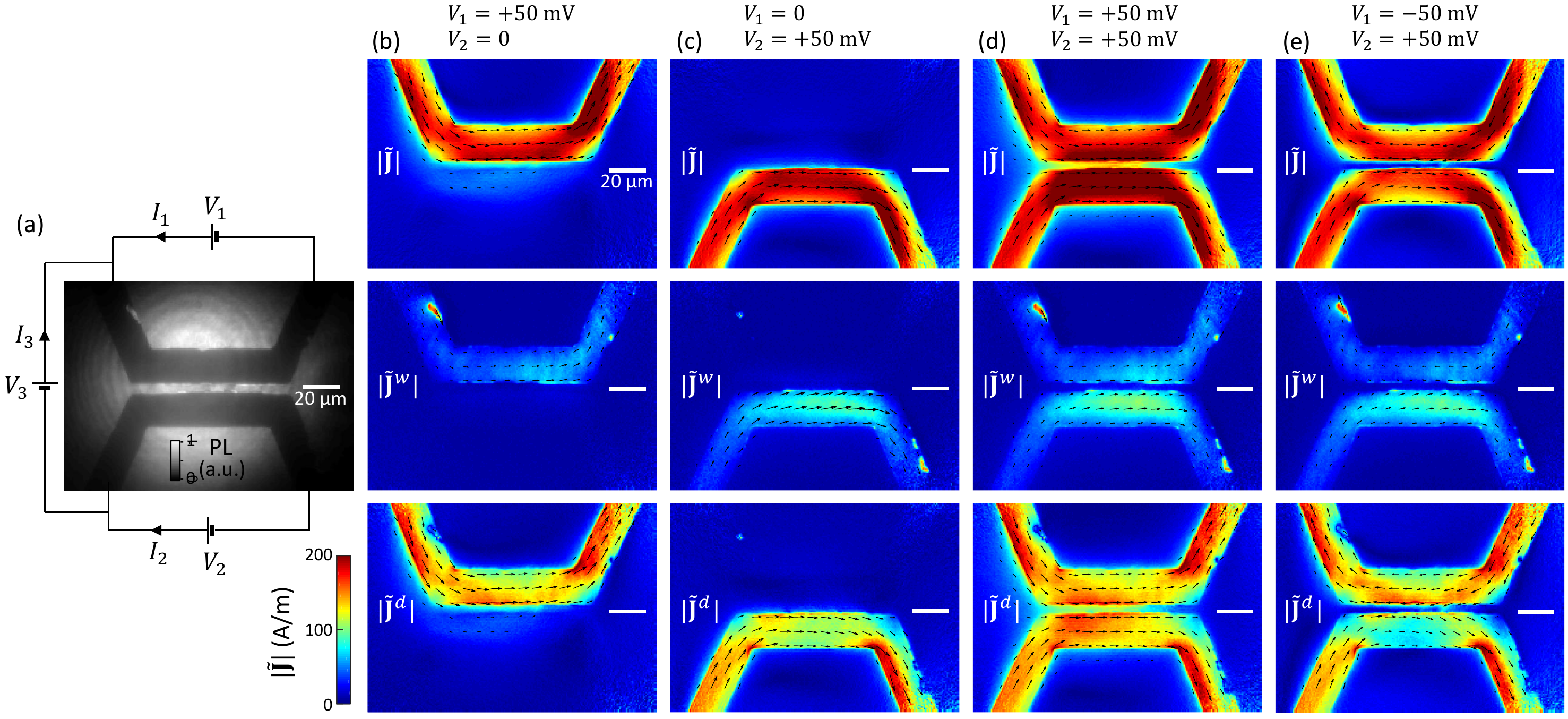}
		\caption{\h (a) PL image of a set of nearby wires on sample \#2c and schematic of the electrical setup. (b-e) Maps of $\tilde{\bf J}$ (top), $\tilde{\bf J}^w$ (middle) and $\tilde{\bf J}^d$ (bottom) under a voltage applied only to the top wire in (b), only to the bottom wire in (c), to both wires simultaneously in (d,e) with identical (d) or opposite (e) current directions. In all cases the voltage between the two wires ($V_3$) is set to zero, but was verified to have no effect on the results. All the current density maps share the same colour bar capped to a maximum value of 200~A/m. The threshold for the arrows is $|\tilde{{\bf J}}|>40$~A/m.}
		\label{FigCoupling}
	\end{center}
\end{figure*} 

One of the most surprising conclusions of the above analysis is the fact that the current appears to leak several micrometers away from the metallic wire laterally, just underneath the diamond surface (about the NV layer). If this apparent leakage was associated with a conventional conduction current, one would expect a nearby metallic contact on the diamond to be able to collect some of this current. To test this, we fabricated a set of wires with a minimum lateral separation of $4~\mu$m on sample \#2c (same diamond substrate as in sample \#2, but different fabrication parameters, see Table~\ref{T:diamonds}). A PL image is shown in Fig.~\ref{FigCoupling}a, also indicating the connections to three different power supplies.

The two wires have a similar resistance of about $10~\Omega$ each, measured by applying a DC voltage of 50~mV ($V_1$ or $V_2$) and reading the corresponding current ($I_1$ or $I_2$) i.e. about 5~mA here. However, applying a DC voltage between the two wires, e.g. $V_3=10$~V, does not produce any measurable current, namely $I_3<1$~pA limited by the noise floor of our instrumentation, i.e. a resistance between the two wires of $R>10^{13}~\Omega$. This result is independent of whether a current is injected in the wires (including in both simultaneously) and whether the laser is illuminating the wires during the measurement. We observed a small increase in the resistance of each wire (by about 1\%) upon turning the laser on, however this change exhibited little dependence on the exact position of the laser spot on the diamond and hence is attributed to laser-induced heating and the expected temperature-dependence of the resistance of the metal.    

Thus, we conclude that there is no actual electrical conduction between nearby wires on our diamond, as expected. To verify that the distance between the two wires was sufficiently small to allow the apparent leakage from one wire to reach the other, we used the NV centres to map the total current density as well as the contributions $\tilde{\bf J}^w$ and $\tilde{\bf J}^d$ (Fig.~\ref{FigCoupling}b-e). Here a voltage source was connected to the wires instead of a current source as in previous sections, and the voltage between the two wires was set to $V_3=0$ (no difference in the current density maps was observed with $V_3=10$~V). Figure~\ref{FigCoupling}b shows the case where a current is injected into the top wire only ($V_1=50$~mV, $I_1\approx5$~mA). The portion of the total current that flows in the diamond ($I_d/I_{\rm tot}$) varies between 60\% (near the centre of the image) and 80\% (near the bottom boundary). This spatial variation and the fact that the leakage is smaller than measured previously on the same diamond (Fig.~\ref{FigModel}) can be explained by the dependence of the leakage effect on the laser intensity, as discussed in Sec.~\ref{sec:LaserDep}. Nevertheless, the apparent leakage is significant (at least 3 mA appears to flow in the diamond) and a sizeable portion of this leakage current (more than $100~\mu$A) spatially overlaps with the footprint of the second wire. Likewise, when a current is injected into the bottom wire only ($V_2=50$~mV, $I_2\approx5$~mA), at least 60\% of the total current flows in the diamond, although here the leakage current remains mostly laterally confined to under the wire (Fig.~\ref{FigCoupling}c). Thus, there is a clear spatial overlap between the current densities $\tilde{\bf J}^d$ associated with the two wires. Yet, there is no actual conduction between the two wires. 

In other words, the conduction electrons in the metal are not able to tunnel through a $4~\mu$m insulating gap (an obvious result) but the magnetic field they generate suggests that the current density associated with these conduction electrons is delocalised over such distances. Again, the reconstructed current densities are completely satisfying from the classical electrodynamics point of view, in that the total current deduced from the current density maps are within error of the electrically measured current. Moreover, running a current in both wires simultaneously gives a net current density that is consistent with the addition of the current densities obtained previously (Fig.~\ref{FigCoupling}d,e), with a constructive (destructive) interference effect visible in $\tilde{\bf J}^d$ when the current flows in identical (opposite) directions. Again, the deduced net current flowing along the $x$-axis is in excellent agreement with the electrically measured current, namely we find 10.4(4) mA with identical current directions and 0.1(4) mA with opposite current directions (against 10.1(1) mA and 0.0(1) mA expected). This adds to the evidence that the analysis of the magnetic field is sound.   

}

\subsection{\h Examination of a few possible interpretations} \label{sec:interpretations}

{\h Before proceeding to further experimental tests, we discuss here a few possible interpretations of this apparent anomaly. To summarise the situation, we analysed the magnetic field data with a minimal set of assumptions, yielding a unique solution for the current density in the system that perfectly fits the magnetic field data and completely accounts for the total injected current. However, this analysis suggests that the current density extends far out of the metallic wire, into the diamond and along the diamond surface, even though no actual electrical conduction was measured between two nearby contacts on the diamond. 

Although the absence of electrical conduction through the diamond rules out a conventional conduction effect, it is useful to discuss this possibility quantitatively. Consider the case of sample \#1, where a sheet resistance of $R_s\approx2.2~\Omega/$sq was measured using four-terminal sensing (allowed by the network of wires visible in Fig.~\ref{FigIntro}c). Given the $t=50$~nm thickness of the wire, we deduce a wire resistivity of $\rho_w=R_st\approx 10^{-5}~\Omega$~cm or a conductivity $\sigma_w\approx10^5$~S/cm, consistent with typical values for evaporated gold~\cite{Ma2010}. On the other hand, a 50\% current leakage through the diamond would indicate that the resistance of the diamond channel is comparable to that of the metal, and so with a comparable conductivity $\sigma_d\sim\sigma_w\sim5\times10^4$~S/cm, given that the vertical extent of the diamond conductive channel is also of the order of $t\sim50$~nm. Such conductivity is two orders of magnitude larger than the record values reported so far at room temperature ($\sim10^2$~S/cm), obtained for boron-doped metallic diamond~\cite{Barjon2009}. This implies that an unprecedently efficient doping mechanism would have to take place. One could imagine an induced conductivity effect at the metal/diamond interface, but the conductive region would likely be localised within a few nanometres from the interface, not the tens of nanometers indicated by our NV measurements, and would not extend laterally over several micrometres. Another possible mechanism could involve photo-induced doping caused by the laser illumination present in our experiments, however there is no evidence of such an effect (explored in Sec.~\ref{sec:LaserDep}). Moreover, the carrier density required to explain a conductivity of $\sigma_d=5\times10^4$~S/cm is unrealistically large: assuming an optimistic mobility of $\mu=3000$~cm$^2$/Vs as achieved in high quality CVD diamond for both electrons and holes~\cite{Isberg2002,Nesladek2008}, the required carrier density must be $n=\sigma_d/q\mu\sim10^{20}$~cm$^{-3}$ where $q$ is the electron charge. 

An alternative explanation could be that the current density $\tilde{\bf J}^d$ does not correspond to a conduction current. Indeed, our analysis does not distinguish between the different types of magnetic field sources provided they are induced by the electrical current, i.e. $\tilde{\bf J}^d$ could include effective currents associated with bound charges or magnetization. Let us first discuss the case of bound charges. Changes in the electric polarisation density ${\bf P}$ of the diamond would produce a polarisation current ${\bf J}_{\rm P}=\frac{\partial{\bf P}}{\partial t}$. A current $I=1$~mA in the metallic wire corresponds to a drift velocity $v_d\sim0.1$~m/s. Assuming a similar density of charge carriers as in the metal ($\sim60$~nm$^{-3}$, just bound instead of free) and a similar cross-section area for the effective diamond channel, this would require a net displacement of the charges by 30 nm over the 300-ns duration of a single measurement run (the $\pi$-pulse duration in pulsed ODMR), which is not compatible with bound charges. 

Magnetisation of the diamond induced by the charge current is another possible candidate to explain the effective current flowing in the diamond, ${\bf J}^d$. For instance, the spin Hall effect in the metallic wire may induce a spin accumulation in the diamond, characterised by a current-induced magnetisation density ${\bf M}$ and corresponding to an effective current density ${\bf J}_{\rm M}=\nabla\times{\bf M}$. For such a source to be responsible for the measured ${\bf J}^d$, the magnetisation density must verify ${\bf J}^d={\bf J}_{\rm M}=\nabla\times{\bf M}$, where ${\bf J}^d$ is parallel to the main charge current flowing in the metallic wire, and relatively uniform under the wire. This requires that ${\bf M}$ lie in the $xz$ plane, be confined roughly in the region under the wire, and exhibit a curling distribution, i.e. ${\bf M}$ would point towards $+x$ at some depth below the NV centres, but towards $-x$ at some deeper depth. This would be quite a peculiar distribution, inconsistent with spin injection from the metal. Furthermore, this would require magnetisations up to $\sim I/t_d\sim10^5$~A/m locally in the case of sample \#2 for instance ($t_d\sim20$~nm is the maximum thickness of the effective diamond channel carrying $I\sim4$~mA). Such large magnetisations are typically found in strong ferromagnets, and would correspond to $\sim1~\mu_B$ (Bohr magneton) per carbon atom of the diamond. 

It is important to note that theories involving bound currents would raise another problem, which is that of charge conservation. Indeed, if the current $I_d$ is to be explained by e.g. a current-induced magnetisation, then the conduction current in the wire as seen by our NV measurements is the remaining part $I_w=I_{\rm tot}-I_d$, which is far below the electrically measured current. For sample \#2, this means that $\approx94\%$ of the current injected between the two contacts on the diamond would be unaccounted for.

Summarising, there seems to be no plausible explanation for the apparent leakage of the current in the diamond as identified by the NV measurements. It is therefore natural to question the measurements and the analysis. However, we will argue in Sec.~\ref{sec:summary} that it is even less plausible that a measurement error or a problem in the analysis may provide a complete explanation of all of our observations, and that an effective delocalisation of the conduction current into the diamond seems to be an overall more satisfying interpretation. In the following, we will therefore focus on this interpretation and perform further experiments aiming to gain some insight into the underlying phenomenon by varying parameters such as the total current, the NV density, the material composing the wire, the wire-diamond distance, and the intensity of the laser used in the experiments. 
}

\section{Further experimental tests} \label{sec:tests}

\subsection{Dependence on the total current} \label{sec:CurrentDep}

\begin{figure}[b!]
	\begin{center}
		\includegraphics[width=0.48\textwidth]{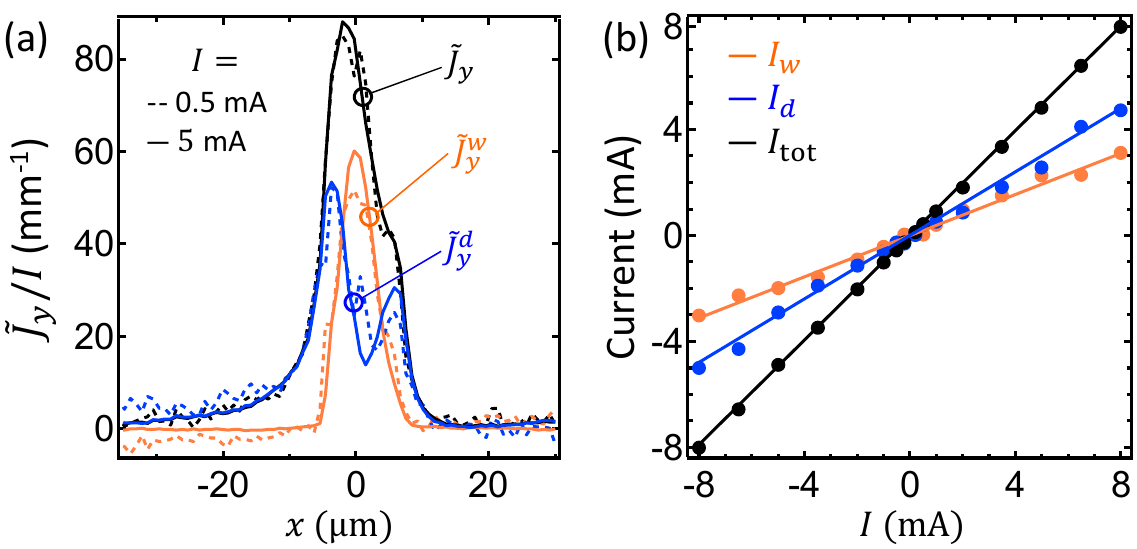}
		\caption{(a) Normalised line cuts of $\tilde{J}_y$, $\tilde{J}_y^w$, $\tilde{J}_y^d$ in sample \#1 (taken along the dotted line shown in Fig.~\ref{FigModel}a) measured for an injected current $I=+5$~mA (solid lines) and $I=+0.5$~mA (dashed lines). (b) Integrated current $I_w$, $I_d$ and $I_{\rm tot}$ as a function of $I$. The solid lines are linear fits to the data, yielding slopes of 0.39(2), 0.60(2) and 0.99(1), respectively.}
		\label{FigCurrentDep}
	\end{center}
\end{figure} 

We used sample \#1 to study the dependence of the effect on the total injected current, $I$. Namely, we recorded the magnetic field for various values of $I$ between 0.2 mA and 8 mA (both positive and negative) and for each $I$ we reconstructed the current density as explained previously. Line cuts of $\tilde{J}_y$, $\tilde{J}_y^w$ and $\tilde{J}_y^d$ obtained for $I=+5$~mA and $I=+0.5$~mA (normalised by the value of $I$) are compared in Fig.~\ref{FigCurrentDep}a, showing very similar profiles (within noise). In Fig.~\ref{FigCurrentDep}b, we plot the integrated currents $I_w$, $I_d$ and the sum $I_{\rm tot}=I_w+I_d$ as a function of $I$, showing a good linearity across the range studied. Linear fit to the data gives average ratios $I_d/I_{\rm tot}=60(2)\%$ and $I_{\rm tot}/I=99(1)\%$. We conclude that the current leakage effect does not depend on the injected current within the range of currents applied, which was limited on one end by the sensitivity of the measurements (due to systematic errors in excess of 0.1~mA, see Appendix~\ref{sec:uncertainties}), and on the other end by the maximum current density that can be handled by the devices (currents above 8~mA typically irreversibly damaged the device, presumably due to electromigration-induced failure).   

\subsection{Dependence on the device/diamond characteristics} \label{sec:DiamondDep}

\begin{figure*}[t!]
	\begin{center}
		\includegraphics[width=0.99\textwidth]{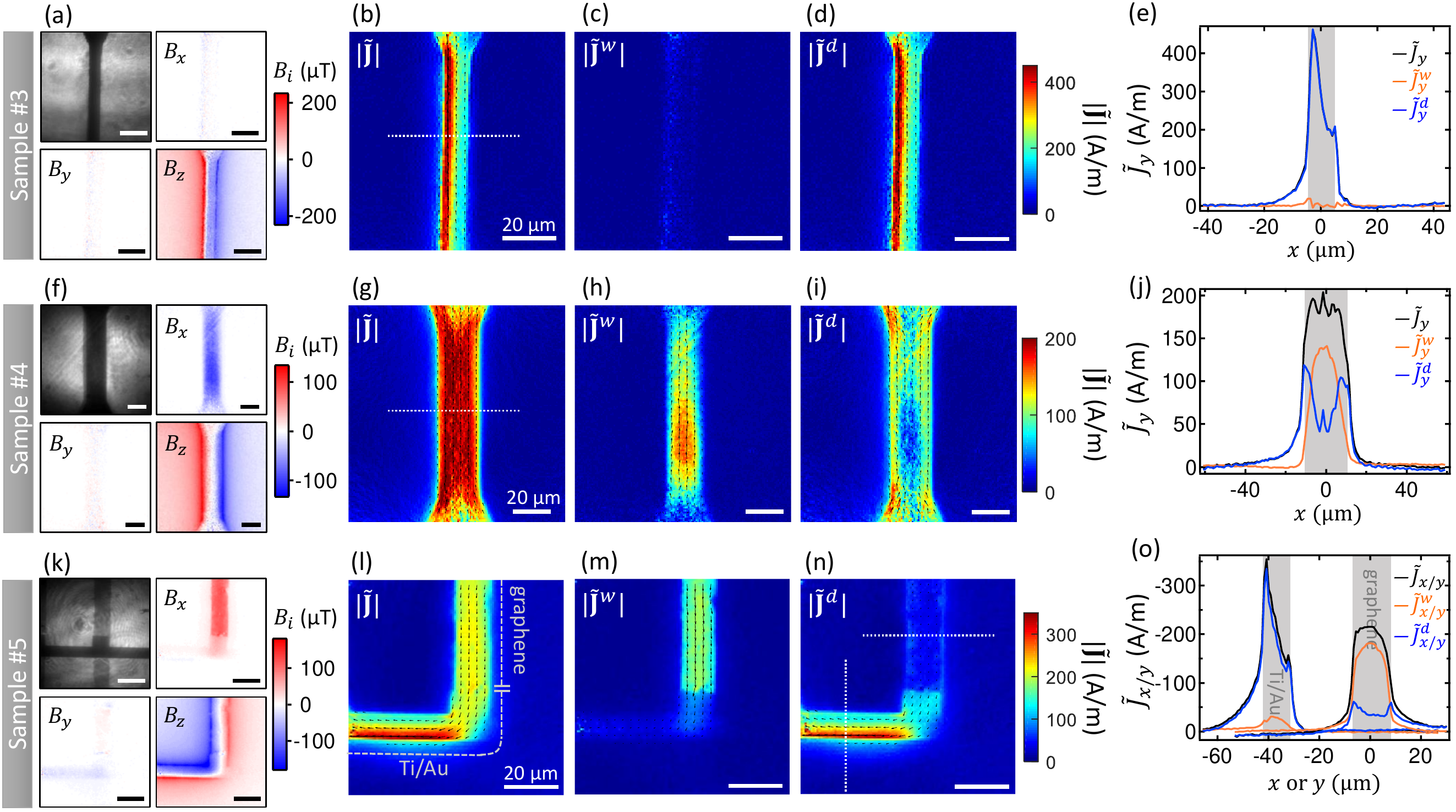}
		\caption{(a) PL image and measured magnetic field for a device in sample \#3 with an injected current $I=+5$~mA. (b-d) Maps of $\tilde{\bf J}$ (b), $\tilde{\bf J}^w$ (c) and $\tilde{\bf J}^d$ (d) deduced from (a). (e) Line cuts of $\tilde{J}_y$, $\tilde{J}_y^w$ and $\tilde{J}_y^d$ taken along the horizontal dotted line shown in (b). The grey shading indicates the location of the wire as extracted from the PL image. (f-j) Same as (a-e) for a device in sample \#4 with $I=+5$~mA. (k-o) Same as (a-e) for a device in sample \#5. Here the device comprises a graphene ribbon along the $y$ direction connected to a metallic (Ti/Au) wire along the $x$ direction, with a current $I=-4$~mA flowing from the graphene into the Ti/Au wire. The line cuts in (o) show the $\tilde{J}_y$ components across the graphene ribbon (horizontal dotted line shown in (n)) and the $\tilde{J}_x$ components across the Ti/Au wire (vertical dotted line in (n)). In the current density maps, the threshold for the arrows is $|\tilde{{\bf J}}|>100$~A/m for samples \#3 and \#5, and $|\tilde{{\bf J}}|>40$~A/m for sample \#4.}
		\label{FigMoreSamples}
	\end{center}
\end{figure*} 

To test the reproducibility of the effect, we varied a number of parameters in the fabricated devices and used a set of different diamonds, the main parameters being listed in Table~\ref{T:diamonds} of Appendix~\ref{sec:fab}. First, we note that samples \#1 and \#2 had a number of differences besides the nitrogen implantation depth mentioned before. Namely, the fabricated devices differed in material composition (Ti/Au vs Cr/Au) and thickness (10/50 nm vs 10/100 nm), and the diamonds were prepared differently prior to implantation: in sample \#2 the diamond surface was as polished whereas in sample \#1 the polished surface was overgrown with $2~\mu$m of CVD diamond. The fact that the two samples showed a strong current leakage through the diamond suggests that these differences did not play a major role, and that the effect is relatively robust with respect to the quality of the diamond surface and the nature of the metal in contact. Instead, it is likely that the difference between the results of samples \#1 and \#2 is mostly related to the difference in implantation depth ($h_{\rm NV}\sim28$~nm vs 8~nm).

In Fig.~\ref{FigMoreSamples}, we show the results obtained for three other samples, labelled \#3 to \#5. For each sample, we show the PL image and measured magnetic field maps, the reconstructed current densities separated in terms of $\tilde{\bf J}^w$ and $\tilde{\bf J}^d$, and line cuts across the wire. Sample \#3 was implanted at the same energy as sample \#2 (hence same depth $h_{\rm NV}\sim8$~nm) but with a fluence 20 times lower ($5\times10^{11}$ against $10^{13}$~nitrogen/cm$^2$), thus creating about 20 times fewer NV centres and related implantation defects. Yet, the results are broadly similar to sample \#2, with a large suppression of the $B_x$ field component under the wire (Fig.~\ref{FigMoreSamples}a) indicating that the current flows mostly in the diamond. From the reconstructed current densities (Fig.~\ref{FigMoreSamples}b-e), we obtain a ratio $I_d/I_{\rm tot}\approx97\%$. We therefore conclude that the density of NV centres and associated defects (such as substitutional nitrogen and vacancy clusters~\cite{Tetienne2018}) in the implanted layer does not play a key role in the effect, or that the smallest density in our samples already exceeds a threshold required to activate the effect.

Sample \#4 was implanted deeper ($h_{\rm NV}\sim20$~nm) with a fluence of $10^{12}$~ions/cm$^2$. Similar to sample \#1, there is a partial recovery of the $B_x$ component (Fig.~\ref{FigMoreSamples}f) leading to current densities that are relatively balanced between wire and diamond paths (Fig.~\ref{FigMoreSamples}g-j) although the ratio $I_d/I_{\rm tot}$ varies along the wire from 47\% (near the top of the image) to 64\% (towards the bottom). This confirms that the implantation depth $h_{\rm NV}$ is a key parameter whereas the fluence appears not to be. 

In all the samples measured so far, the quantity $\tilde{\bf J}^w\doteq\tilde{\bf J}^+-\tilde{\bf J}^-$ was found to be approximately null outside the wire ($|x|>w/2$) even when the total current density $\tilde{\bf J}$ is not, regardless of the NV depth. This indicates that $\tilde{\bf J}^d\doteq2\tilde{\bf J}^-$ is a good measure of the current density in the diamond, at least away from the wire. Furthermore, $\tilde{\bf J}^w=0$ implies $\tilde{\bf J}^+=\tilde{\bf J}^-$, which requires that the NV layer be at the centre of the current density in the diamond regardless of the NV depth. Thus, this observation suggests that the NV layer plays a role in the leakage effect by dictating the $z$-dependence of $\tilde{\bf J}^d$. In summary, a larger NV depth results in a smaller overall leakage $I_d/I_{\rm tot}$ but $\tilde{\bf J}^d$ remains always centred with respect to the NV layer.  

Finally, sample \#5 was implanted at $h_{\rm NV}\sim12$~nm and comprises not only metallic wires (Ti/Au) but also graphene ribbons (see Ref.~\cite{Tetienne2017} for fabrication details). Figure~\ref{FigMoreSamples}k-n show the data for a junction between a graphene ribbon (along the $y$ direction) and a Ti/Au wire (along $x$), with an injected current $I=-4$~mA. Interestingly, the ratio $I_d/I_{\rm tot}$ changes across the junction, as clearly seen from the line cuts in Fig.~\ref{FigMoreSamples}o. Namely, we have $I_d/I_{\rm tot}\sim90\%$ near the Ti/Au wire, consistent with samples \#2 and \#3 (which had a comparable implantation depth $h_{\rm NV}$), but this ratio drops to $I_d/I_{\rm tot}\sim27\%$ near the graphene ribbon. The fact that there is still a significant leakage through the diamond under the graphene ribbon suggests that the effect does not rely on a specific interfacial mechanism and may possibly be present with any conductive material in close proximity to the diamond.


\subsection{Dependence on the laser intensity} \label{sec:LaserDep}

\begin{figure}[t!]
	\begin{center}
		\includegraphics[width=0.48\textwidth]{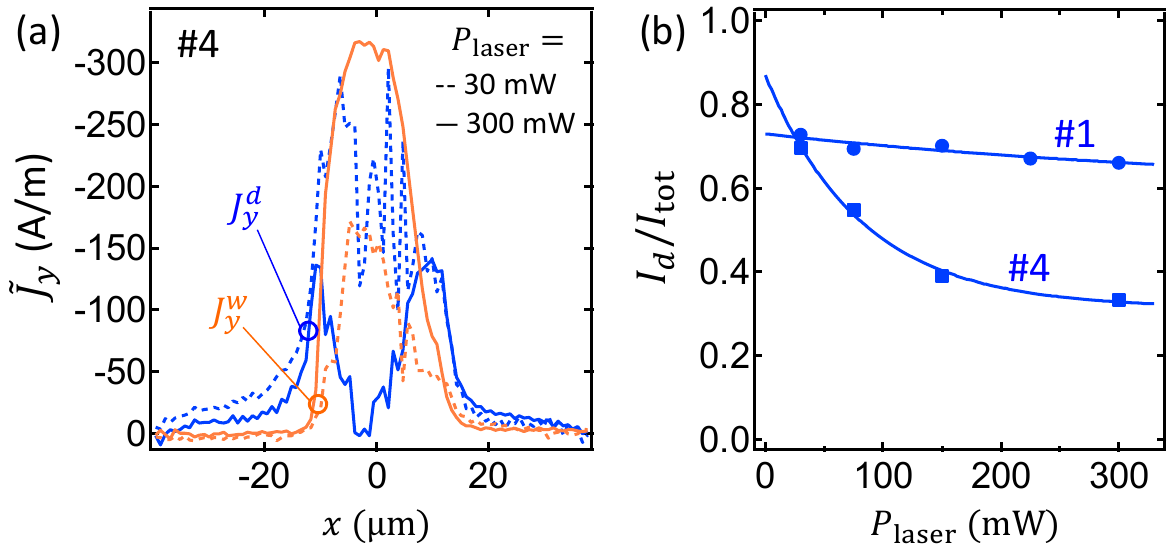}
		\caption{(a) Line cuts of $\tilde{J}_y^w$ and $\tilde{J}_y^d$ across a wire in sample \#4 with $I=-8$~mA, for two different CW laser powers, $P_{\rm laser}=300$~mW (solid lines) and $P_{\rm laser}=30$~mW (dashed lines). (b) Ratio $I_d/I_{\rm tot}$ as a function of $P_{\rm laser}$ from a given line cut across a wire in sample \#4 (squares) and sample \#1 (circles). The total measured current $I_{\rm tot}$ was approximately constant across the range of $P_{\rm laser}$, with $I_{\rm tot}=-8.0(1)$~mA for sample \#4 (for an injected current $I=-8$~mA) and $I_{\rm tot}=-4.9(1)$~mA for sample \#1 ($I=-5$~mA), the quoted uncertainty being the standard deviation. Solid lines are an exponential fit to the data.}
		\label{FigLaserDep}
	\end{center}
\end{figure} 

In some samples, we noticed a correlation between the PL intensity and the amplitude of the $B_x$ magnetic field component, indicating a change in the ratio $I_d/I_{\rm tot}$. This can be seen for instance in sample \#4 (see Fig.~\ref{FigMoreSamples}f-i) and in sample \#2c (Fig.~\ref{FigCoupling}) where the leakage through the diamond seems smallest where the PL under the wire is brightest, i.e. near the centre of the laser illumination spot. This observation prompted us to study the effect of the laser intensity impinging on the sample. Namely, we kept the size of the laser spot constant ($\approx120~\mu$m~$1/e^2$ diameter) and varied the total continuous-wave (CW) laser power $P_{\rm laser}$ from 300 mW (the value used so far) to 30 mW, corresponding to a maximum power density at the centre of the spot (ignoring interference effects from the sample) varied from about 5~kW/cm$^2$ to 0.5~kW/cm$^2$. Note that the pulse sequence used for the measurements (see Fig.~\ref{FigIntro}f) gives a laser duty cycle of $\alpha\approx0.85$, hence the average laser power is $\alpha P_{\rm laser}$. In Fig.~\ref{FigLaserDep}a, we plotted line cuts of the reconstructed current density $\tilde{J}_y^w$ and $\tilde{J}_y^d$ for sample \#4, obtained with two different laser powers $P_{\rm laser}=300$~mW and $30$~mW. While the total measured current is unchanged, i.e. $I_{\rm tot}=-8.0(3)$~mA (for an injected current $I=-8$~mA), the spatial distribution is clearly affected with more current flowing in the diamond at lower laser power. This is quantified in Fig.~\ref{FigLaserDep}b which plots the ratio $I_d/I_{\rm tot}$ against $P_{\rm laser}$, showing a roughly exponential decrease as $P_{\rm laser}$ is increased, with a value of $I_d/I_{\rm tot}\approx70\%$ at $P_{\rm laser}=30$~mW and $I_d/I_{\rm tot}\approx33\%$ at $P_{\rm laser}=300$~mW. In other words, increasing the laser intensity seems to decrease the leakage effect, suggesting that the presence of the laser acts against the mechanism leading to the leakage. Other samples showed a milder effect, for instance in sample \#1 the ratio $I_d/I_{\rm tot}$ decreases from $\approx73\%$ at $P_{\rm laser}=30$~mW to $\approx66\%$ at $P_{\rm laser}=300$~mW (Fig.~\ref{FigLaserDep}b). Moreover, for samples that showed a nearly complete leakage through the diamond at the maximum available power ($P_{\rm laser}=300$~mW), such as samples \#2 and \#3, decreasing the laser power did not noticeably changed the ratio $I_d/I_{\rm tot}$.

This laser dependence calls for caution when comparing samples with different NV depths or different wire materials. Indeed, although the laser power entering the objective lens was kept constant in Figs.~\ref{FigAnomaly}-\ref{FigMoreSamples}, namely $P_{\rm laser}=300$~mW, the laser intensity is locally modulated by the presence of the devices. In particular, the metallic wires largely reflect the laser beam resulting in a laser intensity that is about twice as large in the NV layer at $h_{\rm NV}=28$~nm as at $h_{\rm NV}=8$~nm due to an interference effect (see Appendix~\ref{sec:optical}). Likewise, in sample \#5 the laser intensity at the NVs is expected to be almost twice as large under the (unreflective) graphene as under the metal. 
Thus, the variations in the ratio $I_d/I_{\rm tot}$ observed across samples could be potentially partly due to differences in the local laser intensity.
 
\subsection{Effect of an insulating layer} \label{sec:Oxide} 
 
 \begin{figure*}[t!]
	\begin{center}
		\includegraphics[width=0.59\textwidth]{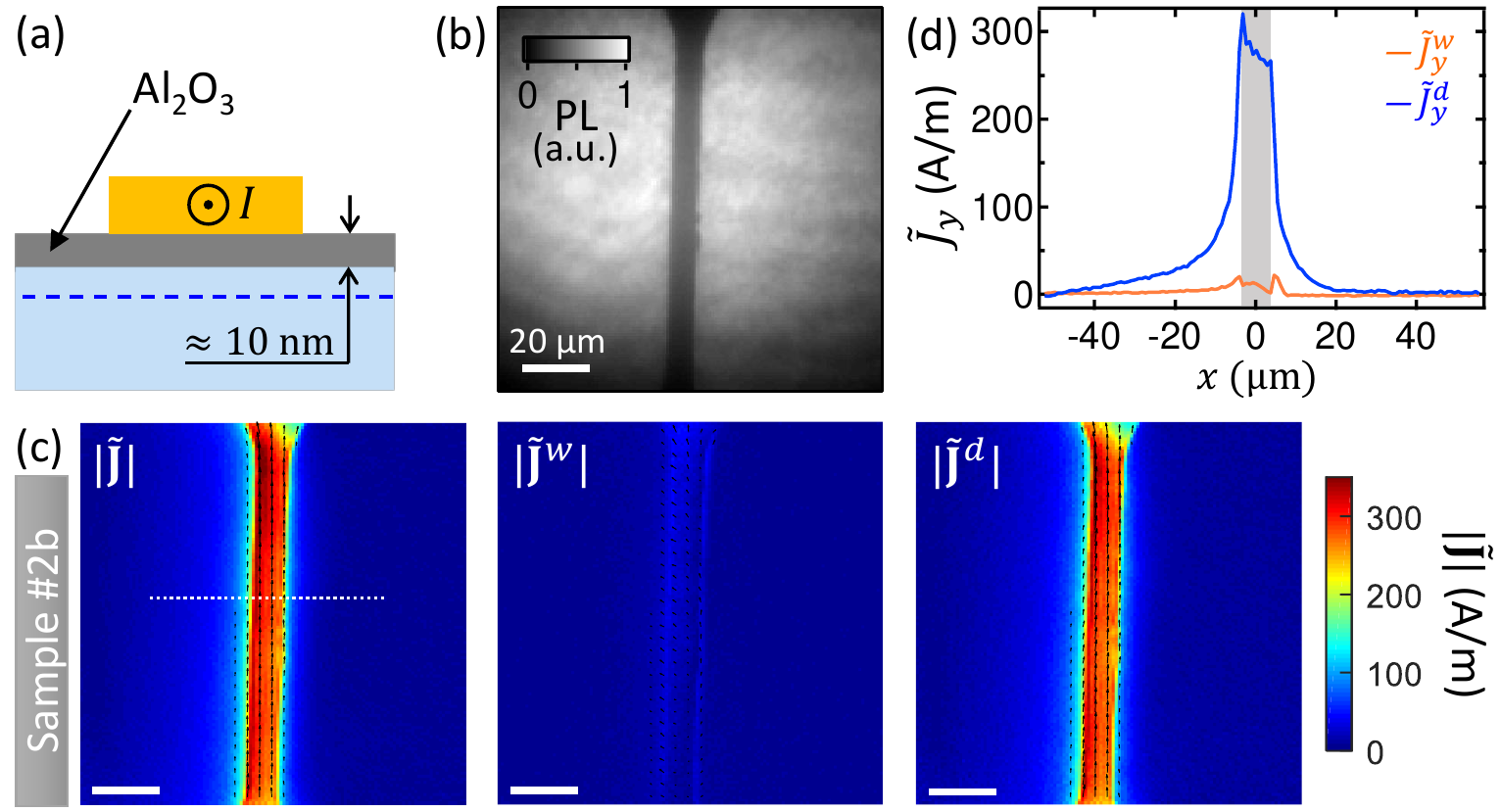}
		\caption{(a) Cross-sectional schematic of sample \#2b, where a 10-nm thick layer of Al$_2$O$_3$ was deposited on the diamond before fabricating Cr/Au wires. (b) PL image of a device. (c)   Maps of $\tilde{\bf J}$ (left image), $\tilde{\bf J}^w$ (middle) and $\tilde{\bf J}^d$ (right) with an injected current $I=+5$~mA and a laser power $P_{\rm laser}=300$~mW. The threshold for the arrows is $|\tilde{{\bf J}}|>80$~A/m. (d) Line cuts of $\tilde{J}_y^w$ and $\tilde{J}_y^d$ taken along the dotted line shown in (c). The total integrated current is $I_{\rm tot}=+4.8(3)$~mA. The grey shading indicates the location of the wire as extracted from the PL image.}
		\label{FigOxide}
	\end{center}
\end{figure*} 

\begin{figure*}[t!]
	\begin{center}
		\includegraphics[width=0.99\textwidth]{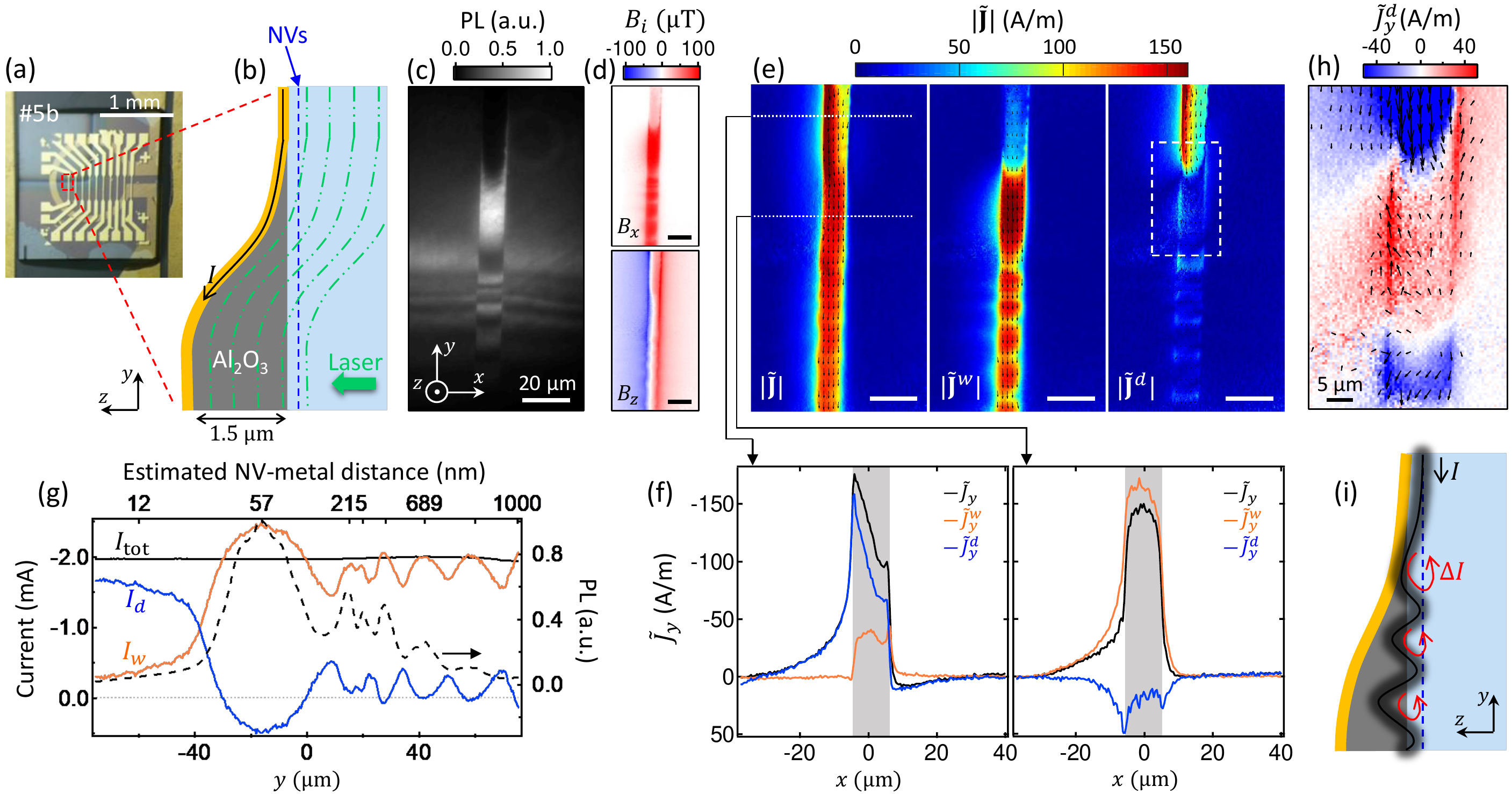}
		\caption{(a) Photograph of sample \#5b, which has a Al$_2$O$_3$ ramp made by evaporation through a shadow mask. (b) Schematic of the cross-section of the device (not to scale). The green dash-dotted lines represent the anti-nodes of the standing wave formed by reflection of the laser on the metal. (c) PL image of a metallic wire. (d) Maps of the magnetic field components $B_x$ and $B_z$ with an injected current $I=-2$~mA and a laser power $P_{\rm laser}=300$~mW. (e) Maps of the current density $\tilde{\bf J}$ (left image), $\tilde{\bf J}^w$ (middle) and $\tilde{\bf J}^d$ (right). The threshold for the arrows is $|\tilde{{\bf J}}|>100$~A/m. (f) Line cuts across the wire taken at two different locations as indicated by the dotted lines shown in (e). The grey shading indicates the location of the wire as extracted from the PL image. (g) Integrated current $I_w$ (orange line), $I_d$ (blue) and $I_{\rm tot}$ (black) as a function of the position $y$ along the wire. The dashed line is the PL intensity measured at the centre of the wire. The top axis gives the approximate NV-metal distance as estimated from the fringe pattern seen in the PL (see Appendix~\ref{sec:optical}). (h) Zoom-in of the current density $\tilde{J}_y^d$ in the region delimited by the dashed rectangle shown in (e). (i) Schematic representation (cross-sectional view) of a possible current flow pattern qualitatively consistent with the data.}
		\label{FigRamp}
	\end{center}
\end{figure*} 

Next, we fabricated a sample with an electrically insulating layer between the metallic wires and the diamond. Precisely, we removed the metallic wires from sample \#2 and deposited a 10-nm layer of Al$_2$O$_3$ on the whole diamond by atomic layer deposition, before fabricating a new set of metallic wires (Cr/Au), labelled sample \#2b (Fig.~\ref{FigOxide}a). Such films are commonly used as a gate oxide in field effect transistors based on the conductive hydrogen-terminated diamond surface~\cite{Pakes2014}, and were confirmed to be highly insulating on similarly prepared diamonds~\cite{Broadway2018}. The measured current densities are shown in Fig.~\ref{FigOxide}c, with the corresponding PL image shown in Fig.~\ref{FigOxide}b. Similar to the no-oxide case, the current flows mostly in the diamond, with a ratio $I_d/I_{\rm tot}\approx95\%$ according to the line cuts shown in Fig.~\ref{FigOxide}d. Looking more closely at $\tilde{\bf J}^w$, we find that the remaining 5\% of the total current is in fact localised just outside the wire (laterally), as clearly seen by comparing the $\tilde{\bf J}^w$ map to the PL image, suggesting that the portion of current flowing in the metallic wire may be even less than 5\%. This result is consistent with the picture (possibly non-physical) of an apparent long-range delocalisation of the current density through insulating materials (whether diamond or Al$_2$O$_3$), even though there is no possibility for the free charges to actually escape the metal.

To study the role of the distance between the metallic wire and the diamond, we fabricated a sample (labelled \#5b, same diamond substrate as in sample \#5) where a $1.5~\mu$m layer of Al$_2$O$_3$ was evaporated through a shadow mask resulting in a ramp with a thickness increasing from 0 to $1.5~\mu$m over a lateral distance of $\sim150~\mu$m (i.e. an average slope of 1\%), before fabricating Cr/Au wires (Fig.~\ref{FigRamp}a,b). A PL image of a typical device is shown in Fig.~\ref{FigRamp}c, revealing interference fringes due to reflection of the laser light at the oxide/metal interface (under the metallic wire) or at the oxide/air interface (elsewhere). These fringes can be used as a ruler to estimate the oxide thickness (see Appendix~\ref{sec:optical}). At the top of the image, the metallic wire sits on the bare diamond surface, causing a strong reduction in PL intensity due to near-field coupling.

The current-induced magnetic field for $I=-2$~mA is shown in Fig.~\ref{FigRamp}d and reveals a correlation with the PL intensity. This is particularly clear in the $B_x$ component, where the largest fields correspond to the bright fringes seen in PL, but a correlation can also be seen in the $B_z$ component. The reconstructed current densities (Fig.~\ref{FigRamp}e) reveal that the current oscillates between $\tilde{\bf J}^w$ and $\tilde{\bf J}^d$ in correlation with the PL intensity. Precisely, while the current flows mostly in the diamond where the wire sits on the bare diamond surface (left graph in Fig.~\ref{FigRamp}f) with a ratio $I_d/I_{\rm tot}\approx81\%$, the ratio $I_d/I_{\rm tot}$ decreases under the first bright fringe to zero and even turns negative (right graph in Fig.~\ref{FigRamp}f) before increasing again ($I_d/I_{\rm tot}\approx26\%$ for the first dark fringe) and so on. This oscillatory behaviour is clearly seen in Fig.~\ref{FigRamp}g, which plots the integrated currents as a function of the position along the wire and confirm that $I_w$ and $I_d$ are correlated with the PL intensity. The total current $I_{\rm tot}=I_w+I_d$ is relatively constant along the wire ranging between $-1.9(3)$~mA and $-2.0(3)$~mA, in agreement with the injected current $I$. This confirms that the reconstruction is sound even near the bottom of the image where the assumption $h_{\rm max}\ll\Delta x_{\rm min}$ breaks down due to the large oxide thickness; the main effect of this assumption is to over-smooth the reconstructed current density, but this does not affect our conclusions.  

The negative sign of $I_d$ for some of the bright fringes is particularly intriguing, and is highlighted by the zoomed-in $\tilde{J}_y^d$ map plotted in Fig.~\ref{FigRamp}h. As can be seen in the line cuts (Fig.~\ref{FigRamp}f, right graph), $\tilde{J}_y^d$ is negative especially near the edges of the wire, while $\tilde{J}_y^w$ becomes larger than $\tilde{J}_y$ (thus conserving the net current). Moreover, unlike all previous measurements, here $\tilde{\bf J}^w$ spreads beyond the region delimited by the wire, indicating that $\tilde{\bf J}^w$ comprises a contribution that is not confined to the wire and may be localised in the diamond above the NV plane or in the oxide layer. These observations are broadly consistent with a current flow pattern as depicted in Fig.~\ref{FigRamp}i, where the main current $I=-2$~mA oscillates in the $z$ direction between the wire and the diamond, accompanied by current loops that cross the NV layer giving rise to the negative current in $\tilde{J}_y^w$ (intensity $\Delta I\sim0.4$~mA for the main bright fringe in the experiment). 

It is important to note that the bright PL fringes correspond to an increased laser intensity in the NV layer, while the laser intensity penetrating into the metal is essentially unchanged (see Appendix~\ref{sec:optical}). Therefore, the correlation between PL and current leakage observed in Fig.~\ref{FigRamp} shows that $I_d$ is governed by the laser intensity at the NVs, where an increase in laser intensity appears to disturb the leakage mechanism and reduce $I_d$. This is consistent with the conclusion drawn in Sec.~\ref{sec:DiamondDep} that the current density in the diamond appears to be centred with respect to the NV layer. 

\subsection{\h The case of suspended metal} \label{sec:suspended}

\begin{figure}[b!]
	\begin{center}
		\includegraphics[width=1\columnwidth]{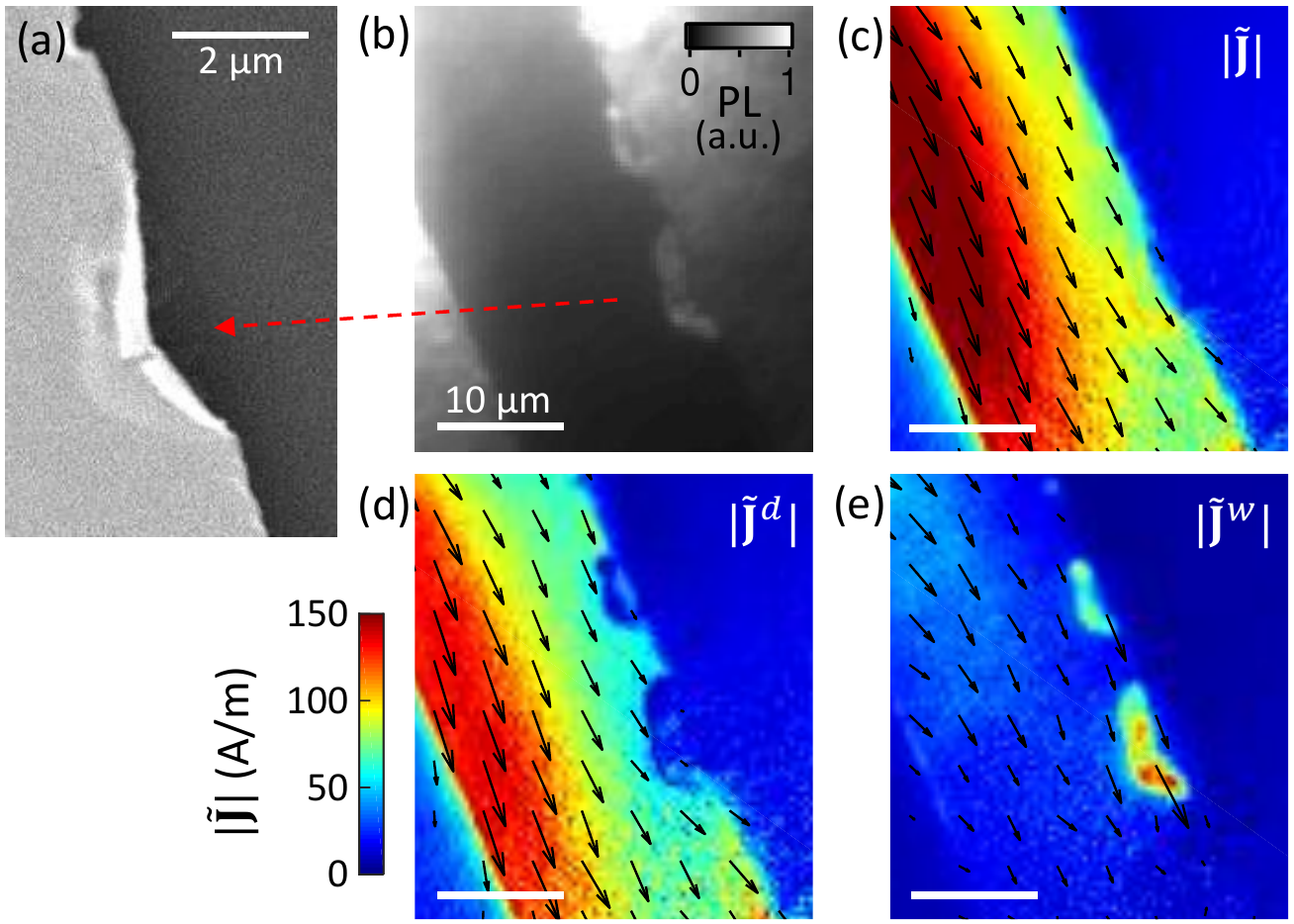}
		\caption{\h (a) Scanning electron microscopy (SEM) image of the edge of a wire in sample \#2c, taken with a $41^\circ$ tilt. (b) PL image of the same region. (c-e) Corresponding maps of $\tilde{\bf J}$ (c), $\tilde{\bf J}^d$ (d) and $\tilde{\bf J}^w$ (e) with an injected current $I=5$~mA and a laser power $P_{\rm laser}=300$~mW. The threshold for the arrows is $|\tilde{{\bf J}}|>40$~A/m.}
		\label{FigSuspended}
	\end{center}
\end{figure} 

{\h 
An interesting question is whether the leakage effect would still occur through an air gap, i.e. without physical contact between the metallic wire and the diamond. This is a situation that is naturally present in some of our samples because of fabrication imperfections at the edges of the wire, where the metal sometimes raises up during the lift-off process, leaving a gap between metal and substrate. An example of this is shown in Fig.~\ref{FigSuspended}a, for one of the wires of sample \#2c imaged in Fig.~\ref{FigCoupling}. Figure~\ref{FigSuspended}b shows a PL image, where the bright regions near the edge confirm that the metal is not in contact with the diamond, giving a PL enhancement instead of a PL quenching. The current density maps (Fig.~\ref{FigSuspended}c-e) reveal that the current appears to flow exclusively in the metal wherever the metal is suspended, while it flows mostly in the diamond where the metal is in contact with the diamond. This may suggest that a physical contact is a necessary condition for the apparent leakage to occur, however the fact that the metal-diamond interface is changed as well as the laser intensity seen by the NVs under the suspended metal prevents a definitive conclusion.     
}

\subsection{Effect of the pulse sequence} \label{sec:sequence}

\begin{figure*}[t!]
	\begin{center}
		\includegraphics[width=0.95\textwidth]{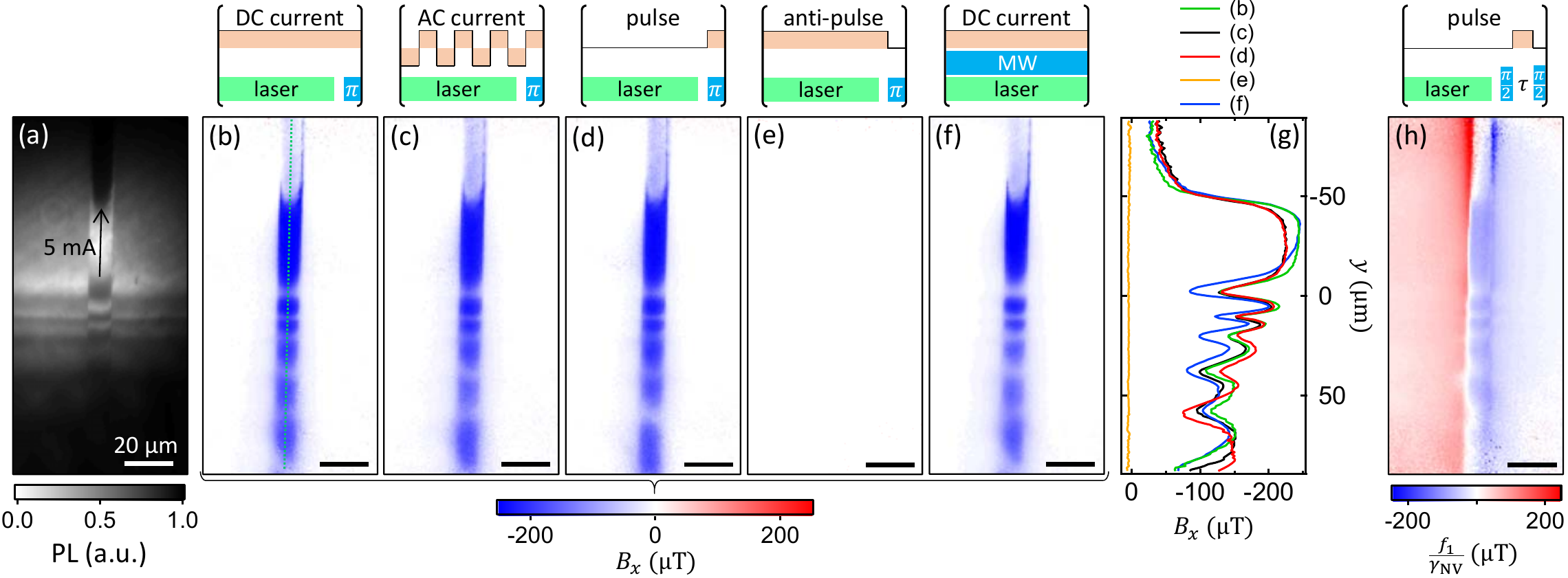}
		\caption{(a) PL image of a metallic wire from sample \#5b. (b-f) $B_x$ maps under various pulse sequences for the measurement and current injection, as depicted above each image and described in the text. In all the images, the CW laser power is $P_{\rm laser}=300$~mW and the maximum current $I=5$~mA. (g) Line cuts of $B_x$ extracted from (b-f) taken along the vertical dotted line shown in (b). (h) Map of the frequency shift $f_1/\gamma_{\rm NV}$ (normalised by the zero-current case) measured by Ramsey interferometry.}
		\label{FigSequences}
	\end{center}
\end{figure*} 

Finally, we investigated the effect of the pulse sequence used for the magnetic field measurements. So far, we used pulsed ODMR while injecting a DC current. Using sample \#5b with the Al$_2$O$_3$ ramp as a test sample, we compared a number of other measurement schemes, varying the laser/MW sequence and/or the way the current $I$ is injected with respect to this sequence (i.e. DC, AC or pulsed). The results are shown in Fig.~\ref{FigSequences}, where the PL of the wire under study is shown in (a), the magnetic field measurements for the different sequences in (b-f), and line cuts along the wire in (g). Figure~\ref{FigSequences}b shows the reference $B_x$ map obtained with the standard pulsed ODMR sequence ($P_{\rm laser}=300$~mW) with a DC current $I=5$~mA.  In Fig.~\ref{FigSequences}c-e, we kept the same pulsed ODMR sequence but changed the current injection. In Fig.~\ref{FigSequences}c, we applied a square AC modulation at 1 MHz, i.e. the sign of $I$ is alternated every 500 ns, and synchronised such that $I=+5$~mA during the 500-ns segment overlapping the 300-ns MW pulse. The resulting $B_x$ shows little change compared to the reference measurement of Fig.~\ref{FigSequences}b. In Fig.~\ref{FigSequences}d, the current is on only during the MW pulse, again with little difference in $B_x$. In Fig.~\ref{FigSequences}e, the current in on except during the MW pulse when it is turned off, giving no field at all. 

In pulsed ODMR, the measurement of the field occurs during the MW pulse, when the Zeeman shifts are encoded into a change of spin population subsequently readout via a laser pulse. The tests performed in Fig.~\ref{FigSequences}c-e therefore show that the history of the current injection makes no substantial difference, i.e. the stray field depends on the instantaneous value of the current at the time of the measurement. This means that the leakage current through the diamond settles in a time much faster than the 500-ns pulse duration used e.g. in Fig.~\ref{FigSequences}e, and its steady state value is independent of whether the current is on or off or alternating the rest of the time. We also varied parameters of the pulsed ODMR sequence: (i) the laser pulse duration was decreased to $2~\mu$s or increased to $20~\mu$s instead of the nominal $10~\mu$s, while keeping the CW laser power constant $P_{\rm laser}=300$~mW; (ii) the wait time of $1.5~\mu$s was increased to $100~\mu$s, also keeping $P_{\rm laser}=300$~mW constant; (iii) the MW pulse duration was shortened to 75 ns while keeping the CW MW power constant (such that 300 ns corresponds to a $\pi$-flip of the NV spins); none of these alterations resulted in a significant change in the measured $B_x$ and, hence, in the leakage current.

Since the measured field was previously observed to depend on the laser intensity, even though the laser is not applied during the actual field measurement (i.e. during the MW $\pi$-pulse), it is useful to look at the effect of the measurement sequence itself. In Fig.~\ref{FigSequences}f, we applied a DC current but employed CW ODMR for the measurement, i.e. the laser and MW were applied continuously throughout the measurement with the same CW laser and MW powers as in Fig.~\ref{FigSequences}b. The resulting field is essentially unchanged, although $B_x$ appears slightly reduced near the centre of the image (see line cuts in Fig.~\ref{FigSequences}g). We also compared ODMR spectroscopy with Ramsey interferometry. In the latter, the Zeeman shift of a given ODMR line is estimated from the phase accumulated during the free interval $\tau$ between two $\frac{\pi}{2}$ MW pulses~\cite{Dreau2011}. In Fig.~\ref{FigSequences}h, we tuned the MW frequency to be near-on resonance with the lowest-frequency ODMR line (labelled $f_1$ in Fig.~\ref{FigODMR}c) and varied the time $\tau$ while applying a current pulse to the wire. The resulting Ramsey oscillations are fit to extract the value of $f_1$, which is shown in Fig.~\ref{FigSequences}g after subtracting the background field (i.e. $f_1$ measured using the same protocol but under $I=0$). The frequency $f_1$ is a function of not just $B_x$ but also $B_z$ hence cannot be directly compared with Fig.~\ref{FigSequences}b, however $f_1$ shows a similar modulation to the $B_x$ measured via ODMR, with in particular a sharp change where the wire sits on the bare diamond, therefore we can conclude that the leakage effect is still present in this measurement. This suggests that the MW field, which is off during the field measurement in the Ramsey sequence unlike in ODMR, does not play an essential role in the effect.

From the experiments presented in Fig.~\ref{FigSequences}, we conclude that the apparent leakage current though the diamond quickly settles after the current is switched on (in a few tens of nanoseconds at most), and does not primarily depend on whether the laser and/or the MW are applied during the measurement. However, the fact that the leakage current does depend on the laser intensity prior to the measurement (even after a $100~\mu$s wait) indicates that the laser illumination has a long lasting effect ($>100~\mu$s) that does affect the amplitude of this leakage current when the current is switched on.

\section{Summary of the findings and possible explanations} \label{sec:summary}

{\h
The main finding of this work is the observation of an anomaly in the magnetic field generated by a DC current in a metallic wire in physical contact with the diamond surface. Precisely, the vector components of the magnetic field measured in the NV layer do not satisfy Gauss's law for magnetism ($\nabla\cdot{\bf B}=0$) or Amp{\`e}re's law ($\nabla\cross{\bf B}=0$). In short, the in-plane magnetic field is strongly attenuated compared to theoretical expectations, whereas the out-of-plane field appears distorted although it still exhibits values that are of the expected order of magnitude. The strong attenuation (nearly total in sample \#2) of the in-plane field is not permitted by Gauss's law for magnetism and Amp{\`e}re's law, which impose strict relationships between the different components. The only assumption made to apply these laws to the data is that the current is confined to the metallic wire (as opposed to having magnetic field sources on both sides of the NV layer). We therefore explored the possibility that this assumption may be incorrect, and by allowing the sources to be located anywhere in space a unique solution that fully explains the measured magnetic field is found. This solution leads to the surprising result that a significant portion of the current density is located below the NV plane within the diamond. This is only an apparent delocalisation of the current (and its characteristic magnetic field), however, as we verified that no actual electrical conduction can take place between two nearby wires via the diamond. 

We discussed the possibility that this anomalous current density, i.e. the part that is delocalised into the diamond, may be associated with current-induced magnetisation, however this would require a very peculiar magnetisation distribution with an extremely large magnetic moment density. Furthermore, it would raise another problem, which is that there would then be a large portion of the electrically measured current unaccounted for by the magnetic field measurements. 

Another possibility to consider is that of a major measurement error or a problem in the analysis of the raw data. We remind the reader that the raw data consists of a set of ODMR spectra (a full representative data set is available at the link~\cite{data} to allow an independent analysis to be undertaken), one for each pixel, exhibiting eight resonances split through the application of a background magnetic field (of amplitude 4 mT) generated by a permanent magnet. It is only the small current-induced component, and not the total magnetic field, that is anomalous. The anomaly observed on this current-induced field takes two different forms. On the one hand, the in-plane field appears to be strongly suppressed, for instance in the case of sample \#2 it is nearly null under the wire ($B_x=3(2)~\mu$T) when it should be about $B_x\approx200~\mu$T. This can be directly seen in the raw ODMR data (Fig.~\ref{FigODMR}c in Appendix~\ref{sec:analysis}), showing that the resonances do not shift upon turning on the current, their positions remaining set by the background field. On the other hand, the out-of-plane field $B_z$ appears modulated in such a way that the field is less intense than predicted at the edges of the wire but the tails extend over larger distances (which is interpreted as a lateral spread of the current outside the wire in our generalised analysis). Importantly, the integral $I_{\rm tot}=\int_{-x_b}^{+x_b}\tilde{J}_y(x)dx$, where $\tilde{J}_y$ is related to $B_z$ via Eq.~(\ref{eq:jy3}) in the Fourier space, always remains in agreement with the electrically measured current $I$, for all the different samples and measurement conditions (or experimental parameters) we tested. 

These two different and very specific observations make an explanation based on a measurement or analysis error extremely unlikely, including an error based on some unknown physical mechanism affecting the NV response. Indeed, the underlying mechanism would have to meet a number of peculiar requirements. First, it would have to be able to distinguish between the background magnetic field and the current-induced field. That is, it cannot be magnetically activated otherwise it would respond to the total magnetic field, instead it must be activated by the charge current. Second, it must be able to distinguish between the in-plane and out-of-plane components of the current-induced field, since the response to each is very different (suppression vs modulation). This is problematic since the positions of the ODMR resonances are not dictated by the Cartesian components of the magnetic field. Instead, each pair of resonances splits and shifts according to the direction of the local magnetic field with respect to the symmetry axis of the corresponding NV family, which does not coincide with any of the Cartesian directions. So the mechanism underlying the error would have to correlate the information gained from multiple NV centres separated by 30~nm on average to retrieve the direction of the local magnetic field. Third, since the in-plane field appears suppressed uniformly across the image, the mechanism for the error would have to know the value of the  current-induced in-plane field at each pixel of the image in order to exactly cancel its effect pixel by pixel, or NV by NV. Fourth, in order to keep the integral $I_{\rm tot}=\int_{-x_b}^{+x_b}\tilde{J}_y(x)dx$ constant, it would have to know the value of the out-of-plane current-induced field across the whole image and then apply a non-local correction to this field. 

The combination of these four requirements clearly rules out a simple measurement error. As for an analysis error, the only way to satisfy all four requirements is for the underlying mechanism to have a complete knowledge of the current density in the metallic wire so that it can deduce the true current-induced magnetic field from the Biot-Savart law and then apply both a local correction and a non-local correction to change the response of each NV centre to this magnetic field, based on the knowledge of the crystallographic orientation of this NV centre. We argue that such a scenario is far less plausible than the solution proposed in this paper, namely that the current density is partly delocalised into the diamond. This simple solution suffices to explain all the above observations, and therefore there must be a physical explanation for the apparent long-range delocalisation of the current density despite the absence of conductivity through the diamond.    

}

\section{Conclusion} \label{sec:conclusion}

{\h 
In this work, we identified an anomaly in magnetic field measurements of the current-induced field from various metallic wires fabricated on different diamonds. Regardless of the explanation for this anomaly, whether it is due to a measurement error or the signature of an actual physical phenomenon, it has immediate consequences for experiments that use NV-based magnetic sensing to study charge transport in DC~\cite{Nowodzinski2015,Chang2017,Tetienne2017} but also possibly for fluctuating signals ~\cite{Kolkowitz2015,Agarwal2017,Ariyaratne2018}. Indeed, since we used very standard methods to measure and analyse the ODMR data and found the effect to be very robust against many technical details, it is likely that the effect was and will be present in other related works. In our own previous work where the current in graphene ribbons was imaged~\cite{Tetienne2017}, the current flow patterns were dominated by structural defects in the graphene layer and therefore clearly visible in the current density maps despite a possible leakage through the diamond. However, the presence of the effect may be problematic in the investigation of more subtle transport phenomena in graphene and other two-dimensional electronic systems. In this context, the methodology introduced in this paper to identify the anomaly and reconstruct the two-channel current density will be a valuable tool. It could be employed, for instance, to find empirically a way to prevent the anomaly from occurring. Here we found that adding a solid insulating layer between the conductor and the diamond is not sufficient, however increasing the laser intensity as well as an air gap were seen to partially mitigate the effect.   

On the other hand, understanding why this anomaly occurs may unveil some interesting physics, either about the measurement system (the NV-diamond physics) or about the magnetic field generated by a conduction current in the near-field regime, or about the current density near conductor-insulator interfaces. We made several observations that may guide future theoretical work in these directions. First, the layer of NV centres seems to play a central role because the current density in the diamond appears to be roughly centred about the NV layer, and because the effect is modulated by the laser intensity seen by the NV centres rather than by the metal. Second, the long-lived effect of the laser (which reduces the apparent current leakage even several microseconds after the laser was turned off) suggests that the underlying mechanism depends on long-lived states in the diamond, possibly defects states that are photo-ionized. Third, the effect exists for different conductive materials in contact with the diamond, even for graphene, and persists through an oxide spacing layer. In fact, it is possible that the effect is completely independent of the conducting wire materials, as the differences between different samples may be possibly explained by the laser dependence only. 

}

\section*{Acknowledgements}

We acknowledge interesting discussions with M. Doherty, M. Usman, A. Wood, S. Rachel, B. Johnson, J. McCallum, R. Scholten, A. Martin, M. Barson, J. McCoey, L. Hall and D. McCloskey. 
This work was supported by the Australian Research Council (ARC) through grants DE170100129, CE170100012 and FL130100119. D.A.B. and S.E.L. are supported by an Australian Government Research Training Program Scholarship. T.T. acknowledges the support of Grants-in-Aid for Scientific Research (Grant Nos. 15H03980, 26220903, and 16H06326), the ``Nanotechnology Platform Project'' of MEXT, Japan, and CREST (Grant No. JPMJCR1773) of JST, Japan.

\appendix

\section{Sample fabrication} \label{sec:fab}

The NV-diamond samples used in these experiments were made from 4 mm $\times$ 4 mm $\times$ 50 $\mu$m electronic-grade ([N]~$<1$~ppb) single-crystal diamond plates with \{110\} edges and a (100) top facet, purchased from Delaware Diamond Knives. The plates were used as received (i.e. polished with a best surface roughness $<5$~nm Ra) or overgrown with $2~\mu$m of CVD diamond ([N]~$<1$~ppb) using $^{12}$C-enriched (99.95\%) methane, leaving an as-grown surface with roughness below 1~nm~\cite{Teraji2015,Lillie2018}. All the plates were laser cut into smaller 2 mm $\times$ 2 mm $\times$~50~$\mu$m plates and acid cleaned (15 minutes in a boiling mixture of sulphuric acid and sodium nitrate). Each plate was then implanted with $^{15}$N$^+$ ions (InnovIon) at various energies and fluences (see Table~\ref{T:diamonds}), with a tilt angle of 7$^\circ$. Following implantation, the diamonds were annealed in a vacuum of $\sim10^{-5}$~Torr to form the NV centres, using the following sequence \cite{Tetienne2018}: 6h at 400$^\circ$C, 2h ramp to 800$^\circ$C, 6h at 800$^\circ$C, 2h ramp to 1100$^\circ$C, 2h at 1100$^\circ$C, 2h ramp to room temperature. The depth profile of the resulting NV centres is mostly governed by the implantation energy, $E_{\rm imp}$; as a rule of thumb approximately valid in this regime of shallow implants~\cite{Lehtinen2016}, in the discussions we assume a mean NV depth $h_{\rm NV}=2E_{\rm imp}$, although the value of $h_{\rm NV}$ is not actually used in the current density reconstruction. 

\begin{table}[t!]
	\begin{tabular}{c | c | c | c | c | c | c}
		Sample & Surface & Energy & Fluence & Material & $t$ & $w$ \\
		& & (keV) & (ions/cm$^2$) & & (nm) & ($\mu$m) \\
		\hline
		\#1 & O & 14 & $5\times10^{12}$ & Cr/Au & 10/50 & 11  \\
		\#2 & P & 4 & $1\times10^{13}$ & Ti/Au & 10/100 & 12  \\
		\#2b & P & 4 & $1\times10^{13}$ & Cr/Au & 10/50 & 9  \\
		\#2c & P & 4 & $1\times10^{13}$ & Cr/Au & 10/70 & 20  \\
		\#3 & O & 4 & $5\times10^{11}$ & Cr/Au & 10/50 & 9  \\
		\#4 & O & 10 & $1\times10^{12}$ & Cr/Au & 5/100 & 23  \\
		\#5 & P & 6 & $1\times10^{13}$ & Ti/Au & 20/40 & 10  \\
		\#5b & P & 6 & $1\times10^{13}$ & Cr/Au & 10/80 & 11  \\
	\end{tabular}
	\caption{Details of the samples used in this work. Samples \#2, \#2b and \#2c (\#5 and \#5b) correspond to the same diamond substrate used in two different fabrications. Column~2: indicates if the diamond was used as received (polished, `P') or overgrown by CVD prior to implantation (`O'). Columns~3,4: energy and fluence of the $^{15}$N$^+$ ion implantation used to create the NV layer. Columns~5-7: materials and dimensions of the metallic wires (thickness $t$, width $w$) fabricated on each diamond.}\label{T:diamonds}
\end{table}

To remove the graphitic layer formed during the annealing at the elevated temperatures, the samples were acid cleaned (as before). The metallic wires were fabricated by photolithography (except for sample \#2c where electron-beam lithography was used), electron-beam evaporation of the metallic stack, and lift-off. The metallic stack used for each sample is indicated in Table~\ref{T:diamonds} and is typically composed of 5-10 nm of an adhesion layer (Cr or Ti) and 50-100 nm of Au. The electrical conductivity of Cr and Ti is about an order of magnitude lower than that of Au, hence the current should dominantly flow in the Au. After fabrication, the diamond was glued face-up onto a glass coverslip patterned with metallic strips for microwave excitation and electrical control of the devices, which were wire-bonded to the coverslip. Finally, the coverslip was glued onto a printed circuit board (PCB) mounted on the microscope, with the electrical connection between coverslip and PCB achieved using silver epoxy. Photographs of a typical mounted device are shown in Fig.~\ref{FigSetup}.

\section{Measurements} \label{sec:meas}

The magnetic field was imaged using pulsed optically detected magnetic resonance (ODMR) spectroscopy on the layer of NV centres (except in Sec.~\ref{sec:sequence} where other protocols were tested), using a custom-built wide-field fluorescence microscope \cite{Simpson2016,Tetienne2017}. Optical excitation from a $\lambda=532$~nm continuous-wave (CW) laser (Laser Quantum Opus) was gated using an acousto-optic modulator (AA Opto-Electronic MQ180-A0,25-VIS), beam expanded (5x) and focused using a wide-field lens ($f=200$~mm) to the back aperture of an oil immersion objective lens (Nikon CFI S Fluor 40x, NA = 1.3). The photoluminescence (PL) from the NV centres is separated from the excitation light with a dichroic mirror and filtered using a bandpass filter before being imaged using a tube lens ($f=300$~mm) onto a sCMOS camera (Andor Neo). Microwave (MW) excitation was provided by a signal generator (Rohde \& Schwarz SMBV100A) gated using the built-in IQ modulation and amplified (Amplifier Research 60S1G4A) before being sent to the PCB. A pulse pattern generator (SpinCore PulseBlasterESR-PRO 500 MHz) was used to gate the excitation laser and MW and to synchronise the image acquisition. 

\begin{figure}[t!]
	\begin{center}
		\includegraphics[width=0.48\textwidth]{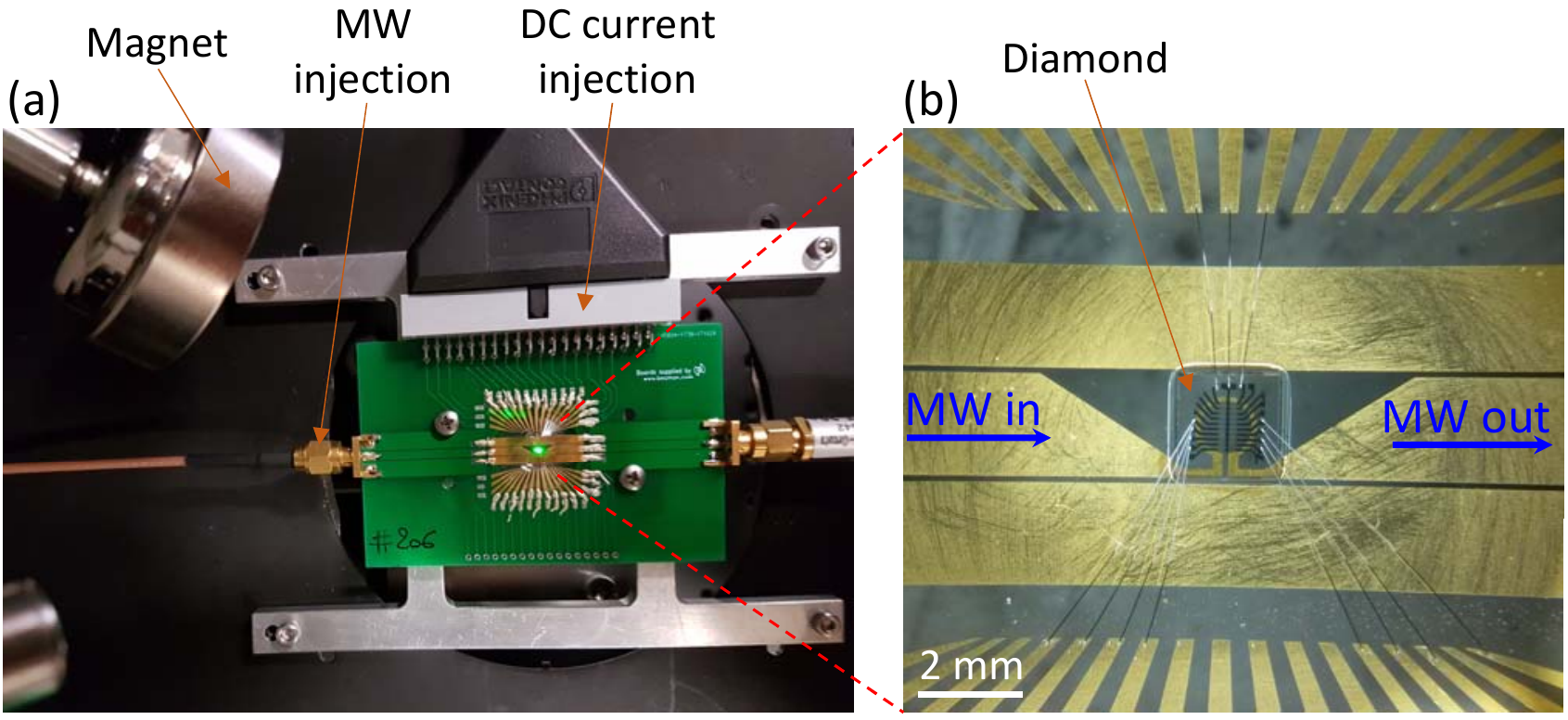}
		\caption{(a,b) Photographs of the diamond as mounted for NV measurements, showing the permanent magnet used to apply the bias field ${\bf B}_0$ as well as the electrical interfacing for MW and DC control.}
		\label{FigSetup}
	\end{center}
\end{figure} 

\begin{figure*}[t!]
\begin{center}
\includegraphics[width=0.6\textwidth]{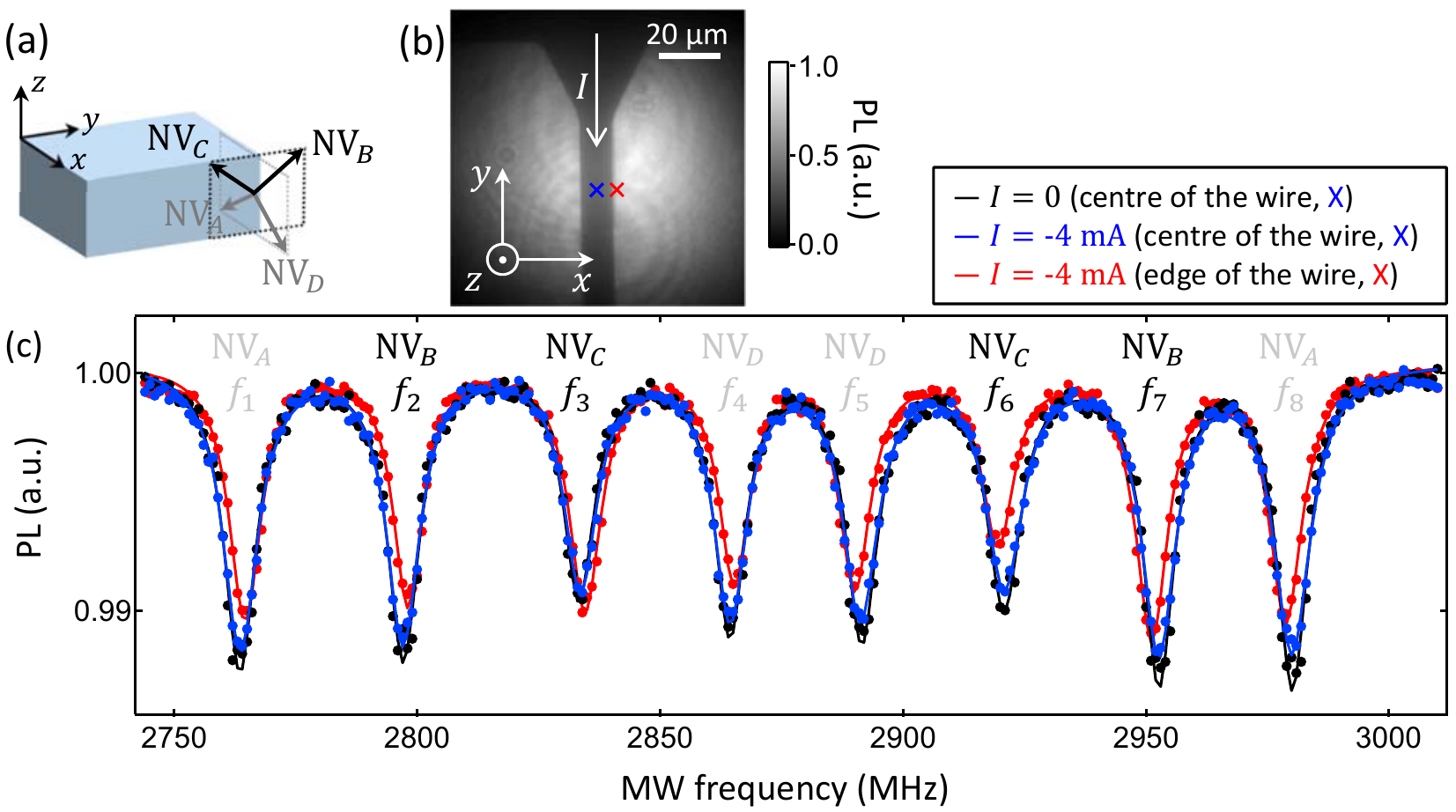}
\caption{(a) Schematic showing the four possible tetrahedral orientations of the NV bond with respect to the sample reference frame $xyz$. (b) PL image of a device in sample \#2. (c) ODMR spectra from a single pixel at the centre of the wire as indicated by the blue cross in (b), with (blue data) and without (black) an applied current $I=-4$~mA. Also shown for comparison is the spectrum from a pixel near the edge of the wire (red). Solid lines are multiple-Lorentzian fits.}
\label{FigODMR}
\end{center}
\end{figure*} 

The typical pulse sequence is shown in Fig.~\ref{FigIntro}f, and comprises a 10-$\mu$s laser pulse, a 1.5-$\mu$s wait time and a 300-ns MW pulse (corresponding approximately to a $\pi$-flip of the NV spins when on resonance). This sequence is repeated $N\sim3000$ times for each MW frequency (hence $\sim30$~ms per frequency, matching the exposure time of the sCMOS camera), and the MW frequency is swept while alternating MW on/off to allow removal of common mode fluctuations in the PL signal. A single frequency sweep takes typically 20 seconds and is repeated 50-500 times, hence total acquisition times of tens of minutes to hours. The total CW laser power at the sample was $P_{\rm laser}=300$~mW unless otherwise stated, which corresponds to a maximum power density of about 5~kW/cm$^2$ given the $\approx120~\mu$m~$1/e^2$ beam diameter. This power density is about two orders of magnitude below the saturation power of the NV optical cycling. The average laser power impinging on the sample during a pulsed ODMR measurement is $\alpha P_{\rm laser}$, where $\alpha\approx0.85$ is the laser duty cycle of the pulsed ODMR sequence. 

The DC current through the device under study was applied using a source-meter unit (Keithley SMU 2450) operated in constant current mode, and applied continuously during the whole acquisition. This source has an accuracy of about 0.1\% in the range of currents considered and a noise an order of magnitude smaller, hence a current $I=5$~mA actually means $I=5.000\pm0.005$~mA. All measurements were performed in an ambient environment at room temperature, under a bias magnetic field ${\bf B}_0$ applied using a permanent magnet (visible in Fig. \ref{FigSetup}a). To allow subtraction of ${\bf B}_0$ to the field measured with the current on, a separate measurement was performed with the current set to zero, with otherwise the exact same conditions and a similar total acquisition time.

\section{Data analysis}   \label{sec:analysis}

In our samples, the NV centres are randomly oriented along the four tetrahedral directions of the diamond crystal (Fig.~\ref{FigODMR}a). To lift the degeneracy of the different orientations in the ODMR spectrum, we apply a bias field ${\bf B}_0$ allowing all eight electron spin resonances (two for each NV orientation) to be resolved~\cite{Steinert2010,Chipaux2015,Glenn2017,Tetienne2017}. Example ODMR spectra from sample \#2 are shown in Fig.~\ref{FigODMR}c, with the pixel locations indicated on the PL image in Fig.~\ref{FigODMR}b. Upon turning on the current $I$, the total field becomes ${\bf B}_{\rm tot}={\bf B}_0+{\bf B}_I$ where $|{\bf B}_I|\ll|{\bf B}_0|$ for the currents considered in this work, so that there is no overlap or swapping of ODMR lines induced by the current~\cite{Tetienne2018b}. 
 
To analyse the ODMR data, we first fit the spectrum at each pixel with a sum of eight Lorentzian functions with free frequencies, amplitudes and widths (solid lines in Fig.~\ref{FigODMR}c). The eight resulting frequencies $\{f_i\}_{i=1\dots8}$ are then used to infer the total magnetic field ${\bf B}_{\rm tot}$ by minimising the root-mean-square error function
\begin{equation} \label{Eq:error}
\varepsilon(D,{\bf B}_{\rm tot}) = \sqrt{\frac{1}{8}\sum_{i=1}^8 \left[f_i - f_i^{\rm calc}(D,{\bf B}_{\rm tot}) \right]^2}
\end{equation}
where $\{f_i^{\rm calc}(D,{\bf B}_{\rm tot})\}_{i=1\dots8}$ are the calculated frequencies obtained by numerically computing the eigenvalues of the spin Hamiltonian for each NV orientation,
\begin{eqnarray} \label{eq:Ham}
{\cal H} &= DS_Z^2+\gamma_{\rm NV}{\bf S}\cdot{\bf B}~,
\end{eqnarray}
and deducing the electron spin transition frequencies. Here ${\bf S}=(S_X,S_Y,S_Z)$ are the spin-1 operators, $D$ is the temperature-dependent zero-field splitting, $\gamma_{\rm NV}=28.035(3)$~GHz/T is the isotropic gyromagnetic ratio, and $XYZ$ is the reference frame specific to each NV orientation, $Z$ being the symmetry axis of the defect~\cite{Doherty2012,Doherty2013}. For the ODMR spectrum shown in Fig.~\ref{FigODMR}c (with $I=0$), we find $D=2870.27(3)$~MHz, $B_x=2.075(2)$~mT, $B_y=-0.745(2)$~mT and $B_z=-3.806(2)$~mT, where the quoted uncertainty is the standard deviation obtained by interrogating adjacent pixels (i.e., the pixel-to-pixel noise). The residual error $\varepsilon\approx100$~kHz is relatively uniform across the image, is independent of whether the current is on or off, and is of the order of the uncertainty for the individual frequencies $\{f_i\}$ (as estimated from the pixel-to-pixel noise), indicating that the spin Hamiltonian considered in Eq.~(\ref{eq:Ham}) captures well the ODMR data. We note that the presence of residual electric field or strain in the sample could lead to a systematic bias on the magnetic field of up to $\sim\varepsilon/\gamma_{\rm NV}\approx40~\mu$T~\cite{Broadway2018,Broadway2018c}, however it should be efficiently rejected by background subtraction (current on/off) and was therefore neglected. 

{\h We stress that the results are extremely robust against the details of the analysis. For instance, instead of fitting the ODMR frequencies using the full NV Hamiltonian, one can use the approximation employed in many works~\cite{Chipaux2015,Tetienne2017,Tetienne2018b} that relates the splitting of each pair of resonances to the projection of the magnetic field along the corresponding NV axis, ignoring the effect of the transverse field. The same magnetic anomaly is observed when using this method. We also tested an alternative method to obtain the vector magnetic field, which involves aligning the background field along each NV axis sequentially, and measure the ODMR splitting of the aligned NV family. Combining the four measurements (or at least three) allows the Cartesian components to be reconstructed, again with the same outcome. The raw ODMR data corresponding to Fig.~\ref{FigAnomaly}j-n (sample \#2) is available at the public link~\cite{data} to allow independent analysis to be carried out, with the data for the other samples and situations discussed in the paper being available upon request.} 

\begin{figure}[b!]
\begin{center}
		\includegraphics[width=0.4\textwidth]{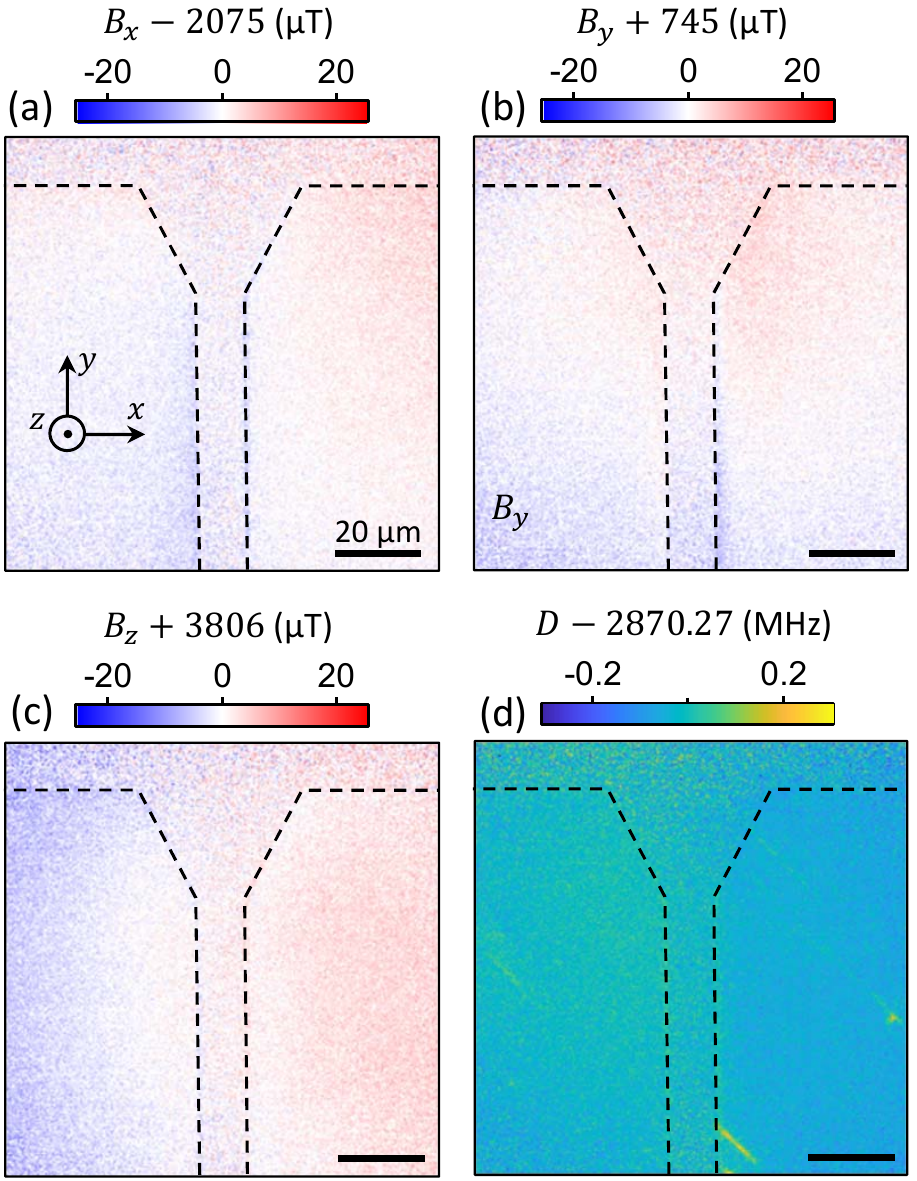}
		\caption{(a-c) Maps of the magnetic field components, $B_x$ (a), $B_y$ (b) and $B_z$ (c) obtained when no current is applied, i.e. showing the externally applied bias field ${\bf B}_0$. The scale bars indicate the central value of each component, e.g. $B_x$ is centred around $2075~\mu$T. (d) Zero-field splitting parameter, $D$, obtained from the fit. The yellow features are attributed to strain induced by polishing marks. The area imaged in (a-d) is the same as that imaged in Fig.~\ref{FigODMR}b, and corresponds to the wire (from sample \#2) imaged in Fig.~\ref{FigAnomaly}j-n. The contour of the metallic wire is indicated by dashed black lines.}
\label{FigB0}
\end{center}
\end{figure} 

The results of the fit for the whole image in sample \#2 are shown in Fig.~\ref{FigB0}, in the case where no current is applied ($I=0$). The magnetic field images (Fig.~\ref{FigB0}a-c) reveal small gradients caused by the non-uniform magnetic field ${\bf B}_0$ produced by the permanent magnet, resulting in peak-to-peak variations of up to $20~\mu$T across the images. The wire (shown as dashed lines) is barely visible, indicating minimal artefacts ($\lesssim5~\mu$T) despite the strong change in PL under the wire compared to the bare diamond. The zero-field splitting parameter, $D$, is also fairly uniform across the field of view (Fig.~\ref{FigB0}d), with only a few isolated features attributed to strain due to polishing damage. Overall, these results confirm that the fitting method is sound, and that the metallic wire is not magnetic nor perturbs the externally applied magnetic field ${\bf B}_0$. Injecting a current $I$ into the wire and performing the same magnetic field reconstruction, we obtain the total field ${\bf B}_{\rm tot}={\bf B}_0+{\bf B}_I$, from which we deduce the field induced by the current alone, ${\bf B}_I={\bf B}_{\rm tot}-{\bf B}_0$.

In Sec.~\ref{sec:anomaly} we found that in sample \#2 the current-induced field was nearly null under the wire, inconsistent with the prediction from the Biot-Savart law. The spectra taken near the centre of the wire (blue and black data in Fig.~\ref{FigODMR}c) confirm that there is indeed no apparent shift of the ODMR lines upon turning on the current, in contrast with the edges of the wire (red data in Fig.~\ref{FigODMR}c) which showed visible shifts of all the lines by $\approx2$~MHz corresponding to an out-of-plane field $B_z\approx100~\mu$T (the variation in ${\bf B}_0$ between the two pixels is negligible here). Importantly, there is no significant change in the shape of the ODMR lines (i.e. the contrast and width are essentially unchanged) upon turning on the current, indicating that the current does not add any significant magnetic noise that may measurably perturb the ODMR measurement. 

Comparing ODMR spectra at different locations, small differences in contrast can be seen, for instance the contrast is larger overall under the wire compared to the edges of the wire (compare blue and red spectra in Fig.~\ref{FigODMR}c), and the contrast is further reduced under the bare diamond surface. These are attributed to differences in the optical transition rates of the NVs due either to differences in the local laser intensity or to non-radiative decay processes (see Appendix~\ref{sec:optical}). In our pulsed ODMR measurements, the laser pulse duration is fixed (chosen as a trade-off between readout contrast and fidelity of the spin re-initialisation) and therefore variations in laser intensity are expected to result in variations in ODMR contrast (namely, the larger the laser intensity the smaller the contrast). However, these variations do not change the position of the ODMR lines, as confirmed by the absence of noticeable change under the wire in the zero-current magnetic field maps (see Fig.~\ref{FigB0}).

\section{Optical effects} \label{sec:optical}

\begin{figure}[b!]
	\begin{center}
		\includegraphics[width=0.48\textwidth]{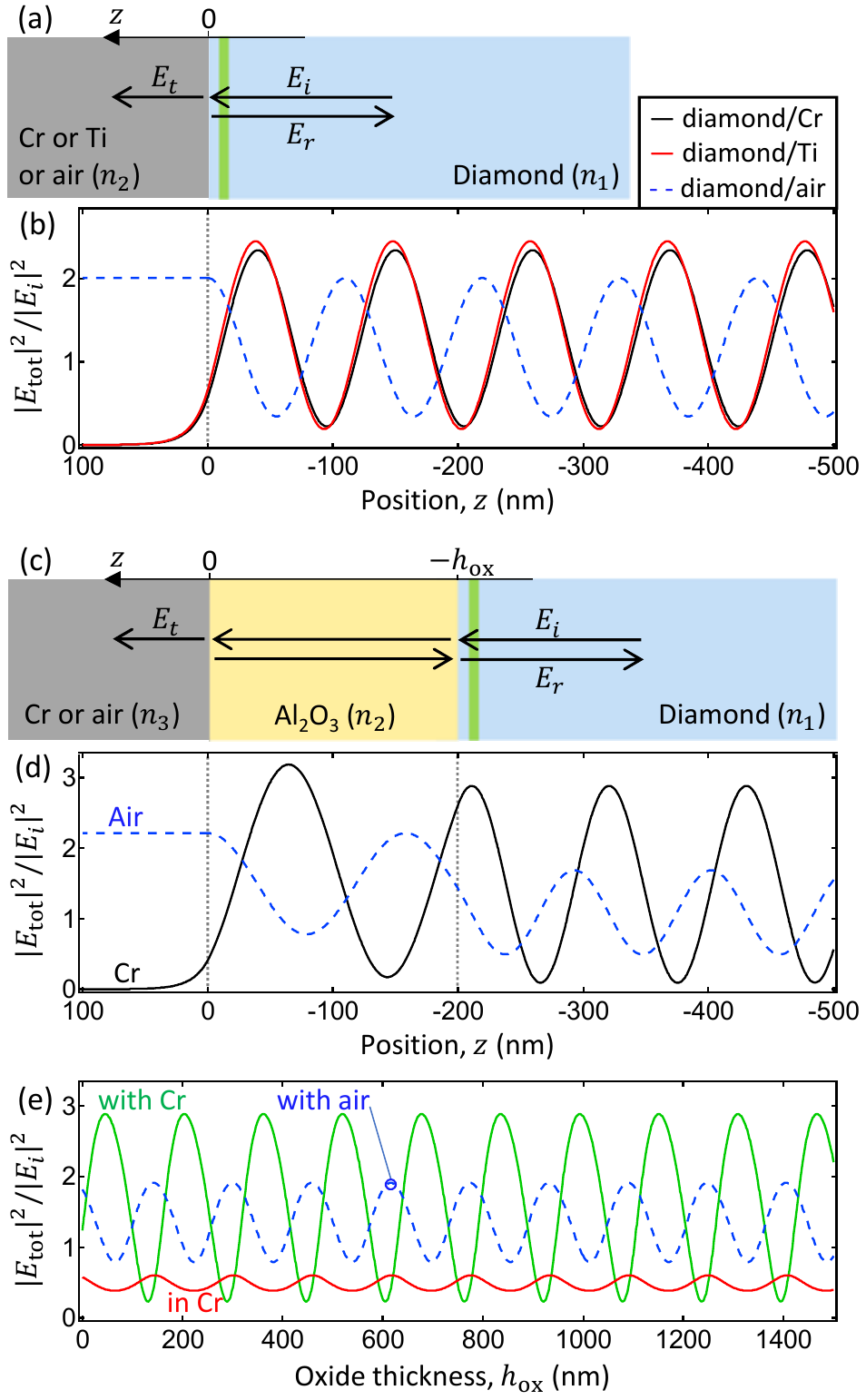}
		\caption{(a) Schematic illustrating the reflection of the laser light ($\lambda=532$~nm) at the interface between the diamond and the metal (Cr or Ti) or air. The green shading represents the NV plane. (b) Normalised electric field intensity as a function of position $z$ for the case of Cr, Ti and air as the second medium. (c) Schematic illustrating the multiple reflections of the laser light at the diamond-oxide and oxide-metal interfaces. (d) Normalised electric field intensity as a function of position $z$ for the case of a diamond-Al$_2$O$_3$-Cr structure with an oxide thickness of $h_{\rm ox}=200$~nm. (e) Green curve: normalised electric field intensity in the NV plane (precisely at $z=-h_{\rm ox}-h_{\rm NV}$ with $h_{\rm NV}=12$~nm) as a function of the oxide thickness of $h_{\rm ox}$. Red curve: normalised electric field intensity at $z=0$, indicative of the intensity transmitted to the metal. In (d,e), the blue dashed line shows the case of the diamond-Al$_2$O$_3$-air structure for comparison.}
		\label{FigInterferences}
	\end{center}
\end{figure} 

In this Appendix, we discuss the various optical effects occurring in the experiments. For a single optical emitter (two-level system) excited by a CW laser, the total photon emission rate in the steady state is $\frac{k_L k_r}{k_L+k_r+k_{nr}}$ where $k_L$ is the excitation rate (proportional to the laser intensity and to the polarisability of the emitter), and $k_r$ and $k_{nr}$ are the radiative and non-radiative decay rates. In our experiments, the laser intensity is well below saturation of the NV optical cycling, i.e. $k_L\ll k_r$. Furthermore, we collect only a fraction $\alpha_{\rm col}$ of the emitted light (collection efficiency, which depends on the far-field radiation pattern and on the collection optics). We obtain an expression for the collected photon rate for each NV centre, 
\begin{eqnarray}
I_{\rm col}\approx\alpha_{\rm col}Qk_L
\end{eqnarray}
where $Q=\frac{k_r}{k_r+k_{nr}}$ is the quantum efficiency of the emitter. Thus, the PL signal measured in the experiments is proportional to the local laser intensity (via $k_L$), to the quantum efficiency $Q$, and to the collection efficiency $\alpha_{\rm col}$. Below we discuss how these quantities can vary in the presence of the metallic wires.

We first examine the spatial modulation of the laser intensity (vacuum wavelength $\lambda=532$~nm) due to reflections at the diamond/metal interface. To analyse this interference effect, we solve the wave optics problem in the plane wave approximation. The light initially travels in the $z$ direction in a transparent medium of refractive index $n_1$ (the diamond) and hits a flat interface with an absorbing medium of complex refractive index $n_2$ (Cr or Ti) at normal incidence (Fig.~\ref{FigInterferences}a). For an incident plane wave with electric field amplitude $E_i=E_0e^{i(n_1k_0z-\omega t)}$ where $k_0=\frac{2\pi}{\lambda}$, the transmitted and reflected waves are $E_t=tE_0e^{i(n_2k_0z-\omega t)}$ and $E_r=rE_0e^{i(-n_1k_0z-\omega t)}$, respectively, where $t=\frac{2n_1}{n_1+n_2}$ and $r=\frac{n_1-n_2}{n_1+n_2}$ are the Fresnel coefficients~\cite{Jackson}. The total amplitude in the transparent medium is $E_{\rm tot}=E_i+E_r$ which gives a partial standing wave of intensity
\begin{eqnarray} \label{eq:interf}
|E_{\rm tot}|^2=|E_0|^2\left[1+|r|^2+2|r|\cos\left(4\pi n_1\frac{z}{\lambda}-\theta\right)\right]
\end{eqnarray}
where $|r|$ and $\theta$ are the magnitude and phase angle of $r$, respectively, i.e. $r=|r|e^{i\theta}$. 

This interference pattern is shown in Fig.~\ref{FigInterferences}b for the diamond/metal interface, and for comparison for the diamond/air interface. The refractive indices are taken as follows: $n_1=2.425$ for diamond~\cite{Phillip1964}, $n_2=3.03+3.33i$ for Cr and $n_2=2.48+3.35i$ for Ti~\cite{Johnson1974}. On the metal side, the intensity is rapidly attenuated, which justifies why we can neglect the wave reflected at the Cr/Au or Ti/Au interface: for instance, the reflection at the Cr/Au interface is 17\% with a transmission after a round trip through the 10-nm-thick Cr layer of 21\%, giving less than 4\% left from this reflected wave at the diamond/Cr interface. On the diamond side, there is little difference between Ti and Cr, which both induce a large standing wave with a visibility of about 90\%, i.e. the laser intensity is about 10 times larger at the anti-nodes than at the nodes. The intensity right at the interface is about 30\% of the maximum (i.e., of the intensity at an anti-node), and it reaches nearly 50\% of the maximum at a distance $z=-8$~nm (corresponding to the shallowest NVs, as in sample \#2) and more than 90\% at $z=-28$~nm (the deepest NVs, as in sample \#1). Moreover, the intensity with the metal is even larger than with air for $-18<z<-74$~nm. Thus, this laser interference effect alone cannot explain the strong reduction in PL observed in all samples under the metal compared to the bare diamond surface (by a factor 2-4 typically).

Instead, we attribute the strong PL quenching under the wire to a reduction in the quantum efficiency $Q$. Indeed, optical emitters near a metal may couple to evanescent field components providing an additional non-radiative decay channel increasing $k_{nr}$~\cite{Novotny,Buchler2005}. In particular, using the refractive indices given above, we predict (by solving Helmholtz equation) that a surface plasmon polariton (SPP) mode exists at the diamond-Cr interface with a spatial extent into the diamond of $\approx100$~nm ($1/e$ decay constant for the electric field amplitude), making it a prime candidate to explain the PL quenching of the NVs in our samples. The generated SPPs propagate along the metal-diamond interface but are rapidly dissipated due to ohmic losses. Another avenue for non radiative decay is via coupling to electronic excitations~\cite{Tisler2013}, which is likely the dominant effect in the case of graphene on diamond as in sample \#5, and also eventually dissipates as heat. We note that the presence of the interface modifies the local density of optical states (LDOS)~\cite{Novotny,Buchler2005}, hence the radiative decay rate $k_r$, due to a similar interference effect as for the laser light. However, this effect is small in comparison to the change in $k_{nr}$ for the NV-metal distances considered here ($h_{\rm NV}\sim8-28$~nm). Finally, the collection efficiency $\alpha_{\rm col}$ is also affected by the presence of the interface, which modifies the angular emission pattern, although the presence of the metal is expected to increase $\alpha_{\rm col}$ rather than decrease it (by redirecting more light to the collection side). 

On the other hand, for NV-metal distances of the order of the wavelength, the laser interference effect is the dominant effect governing the measured PL intensity $I_{\rm col}$, and this is what gives rise to the fringe pattern seen in sample \#5b with the Al$_2$O$_3$ ramp. To model this situation, we use the plane wave approximation at normal incidence as before and include  both the diamond/oxide and the oxide/Cr interfaces (Fig.~\ref{FigInterferences}c), with a refractive index $n_2=1.684$ for Al$_2$O$_3$~\cite{Boidin2016}. The standing wave pattern is plotted in Fig.~\ref{FigInterferences}d in the case where the oxide has a thickness $h_{\rm ox}=200$~nm. Because of the small reflection at the diamond/oxide interface, there is a small change in the amplitude of the standing wave in the diamond compared to in the oxide, which depends on the oxide thickness due to multiple reflection effects, although the main effect governing the laser intensity in the diamond remains the reflection at the oxide/metal interface. In Fig.~\ref{FigInterferences}e (green line), we plot the intensity at $z=-h_{\rm ox}-h_{\rm NV}$ with $h_{\rm NV}=12$~nm (as in sample \#5b), as a function of the oxide thickness of $h_{\rm ox}$. We find that the minimum laser intensity in the NV plane (obtained for $h_{\rm ox}=130$, 288, 446, 604, 762~nm etc.) is less than 10\% of the maximum (at $h_{\rm ox}=45$, 203, 361, 519, 677~nm etc.). Such a contrast is larger than observed in the experiment (Fig.~\ref{FigRamp}c), which can be explained by the spread in $z$ of the NV layer. Figure~\ref{FigInterferences}e also shows the intensity in the metal (red line), revealing small oscillations caused by Fabry-P{\'e}rot resonances through the oxide, and the intensity in the NV plane in the case where no metal is present (blue dashed line), resulting in a reduced contrast of the fringe pattern as seen experimentally. The position of the nodes and anti-nodes listed above were used in Fig.~\ref{FigRamp}g to estimate the NV-metal distance along the Al$_2$O$_3$ ramp. 

We note that, according to Fig.~\ref{FigInterferences}e, the maxima in the laser intensity at the NVs coincide with minima in the laser intensity in the metal. Therefore, a decrease in $I_d$ corresponds to a decrease in the laser intensity in the metal, while we saw in Sec.~\ref{sec:LaserDep} that an increase in the total laser power (i.e. in both the metal and the NV layer) resulted in a decrease in $I_d$. This is why we interpreted the correlation between PL intensity and current leakage $I_d$ observed in Fig.~\ref{FigRamp} as evidence that the leakage is dictated by the laser intensity in the NV plane rather than in the metal. Nevertheless, because the NV centres are close to the diamond surface ($h_{\rm NV}\sim12$~nm in sample \#5b), this experiment alone does not allow us to discriminate a diamond surface effect from an effect involving the implanted defects.      


\section{Application of Biot-Savart and Amp{\`e}re's laws} \label{sec:Ampere}

\begin{figure}[b!]
	\begin{center}
		\includegraphics[width=0.48\textwidth]{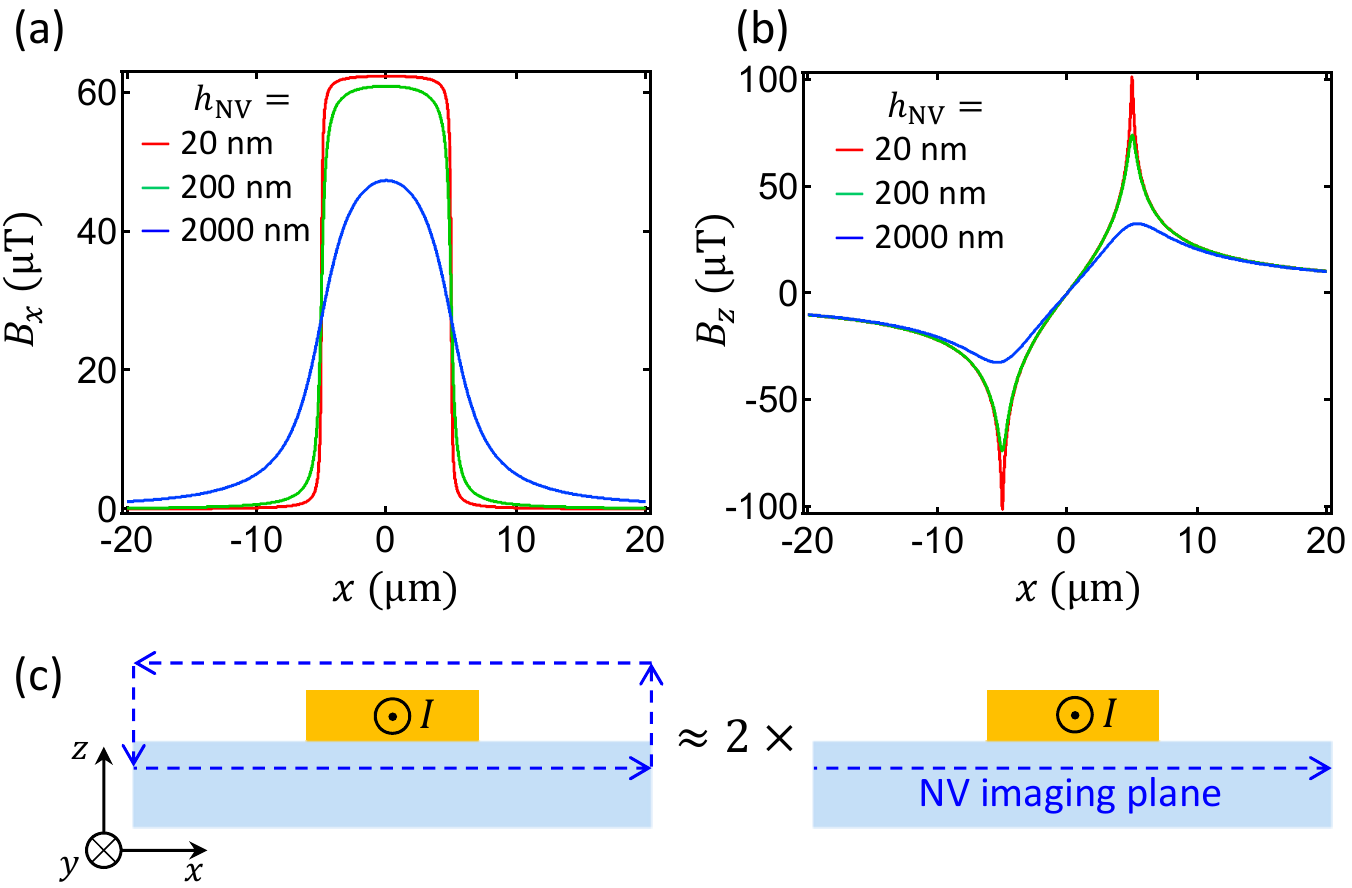}
		\caption{(a,b) Calculated $B_x$ (a) and $B_z$ (b) profiles as a function of $x$ for different probing distances $h_{\rm NV}$, using the following parameters: $I=1$~mA, $w=10~\mu$m, $t=100$~nm. (c) Schematic showing the equivalence used to evaluate the integral in Amp{\`e}re's law.}
		\label{FigLaws}
	\end{center}
\end{figure} 

In Sec.~\ref{sec:anomaly}, we compared the measured magnetic field to theory by using the Biot-Savart law, Eq.~(\ref{eq:BS}). 
For the geometry of Fig. \ref{FigIntro}b with a uniform current density inside the wire, ${\bf J}=-\frac{I}{wt}\hat{\bf e}_y$, the only non-vanishing components of the magnetic field are $B_x$ and $B_z$. These are plotted in Figs.~\ref{FigLaws}a and \ref{FigLaws}b, respectively, as a function of the lateral position $x$ for various probe distances $h_{\rm NV}$ obtained by numerical integration of Eq.~(\ref{eq:BS}) with $I=1$~mA, $w=10~\mu$m and $t=100$~nm. The value of $h_{\rm NV}$ affects the field only near the edges of the wire (within a distance of the order of $h_{\rm NV}$ from the edge), a consequence of the convolution with the resolution function~\cite{Casola2018}. This can be seen in the analytical expressions obtained in the thin-wire limit ($t\ll w$),
\begin{eqnarray}
B_x &= &\frac{\mu_0I}{2\pi w}\left[\tan^{-1}\left(\frac{w-2x}{2h_{\rm NV}}\right)+\tan^{-1}\left(\frac{w+2x}{2h_{\rm NV}}\right)\right] \nonumber \\
B_z &= &\frac{\mu_0I}{4\pi w}\log\left[\frac{(w+2x)^2+4h_{\rm NV}^2}{(w-2x)^2+4h_{\rm NV}^2}\right].  \label{eq:BSanalZ}
\end{eqnarray}  
It comes that for $h_{\rm NV}\ll w$, $B_x$ is constant under the wire with a value $B_x=\frac{\mu_0 I}{2w}$ independent of $h_{\rm NV}$, whereas the peak value of $B_z$ at the edges $x=\pm w/2$ scales as $\log(1+\frac{w^2}{d^2})$. In practice, the sharp peaks in $B_z$ are not resolved in the measurements because of the finite spatial resolution. In Fig.~\ref{FigAnomaly}f,g,o,p, we accounted for this effect by applying a convolution to the calculated $B_z$ profile with a Gaussian function with a full width at half maximum of 1~$\mu$m~\cite{Tetienne2018b}. Moreover, this makes the calculation rather insensitive to the exact value of $h_{\rm NV}$ used (for $h_{\rm NV}<50$~nm), which also justifies why the spread in $h_{\rm NV}$ due to the implantation process (typically $\pm h_{\rm NV}/2$) can be safely neglected.    

Another way to compare theory and experiment is to use Amp{\`e}re's circuital law in its integral form,
\begin{eqnarray} \label{eq:Ampere}
\oint_C{\bf B}_I\cdot d{\bf l} = \mu_0\int_S {\bf J}({\bf r}) dS = \mu_0 I_{\rm enc}~,
\end{eqnarray}  
where $C$ is a closed curve, $S$ is a surface enclosed by $C$, and $I_{\rm enc}$ is the total enclosed current. Assuming that ${\bf J}({\bf r})$ is symmetric with respect to some plane parallel to $xy$ (which is automatically verified in the thin-wire limit), one can choose $C$ to be also symmetric with respect to this plane while coinciding with the NV plane on one side, as shown in Fig.~\ref{FigLaws}c. We can then approximate the left-hand-side integral in Eq.~(\ref{eq:Ampere}) as $2\int_{-x_b}^{+x_b}B_x(x)dx$ where $B_x(x)$ is the field profile measured at a given distance $h_{\rm NV}$ from the wire, $x=\pm x_b$ are the bounds of the measurements such that $x_b\gg w$, and one has $I_{\rm enc}=I$. Here and in the Biot-Savart law above, we neglected the weak diamagnetic response of diamond (magnetic susceptibility of $-2.1\times10^{-5}$), i.e. the diamond is assumed to be magnetically transparent.  With these approximations, we then define the deviation from Amp{\`e}re's law as
\begin{eqnarray} \label{eq:chi}
\chi \doteq 1-\frac{2\int_{-x_b}^{+x_b}B_x(x)dx}{\mu_0 I}~,
\end{eqnarray}  
which was used as a metric to quantify the discrepancy between experiment and theory in Sec.~\ref{sec:anomaly}.

\section{Relationships between magnetic field components} \label{sec:relationships}

In Ref.~\cite{Lima2009}, Lima and Weiss derived relationships between the magnetic field components in a given plane $z$, starting from the differential form of Amp{\`e}re's law in a source-free region, $\nabla\cross{\bf B}=0$. Here we seek to derive these relationships using the Biot-Savart law instead. The interest is two-fold. First, it will make explicit where the assumption regarding the relative location of the sources (described by a current density ${\bf J}$) comes from. Second, it is more general as the Biot-Savart law is valid also for time-dependent sources in the limit of infinite speed of light $c\rightarrow\infty$~\cite{French1963,Terry1982,Charitat2003,Hill2011,Rosser2013,Wolsky2015}, while $\nabla\cross{\bf B}=0$ assumes that ${\bf B}$ does not vary with time (otherwise the vacuum displacement current $\epsilon_0\frac{\partial{\bf E}}{\partial t}$ would have to be included). 

Let us consider the general problem of a current distribution contained in an unbounded slab delimited by the planes $z=z_1$ and $z=z_2>z_1$. Expressing the current density by its Cartesian components ${\bf J}=(J_x,J_y,J_z)$, the Biot-Savart law gives 
\begin{widetext}
\begin{eqnarray} \label{eq:BScart}
B_x(x,y,z) &=& \frac{\mu_0}{4\pi}\int_{-\infty}^{+\infty}dx'\int_{-\infty}^{+\infty}dy'\int_{z_1}^{z_2}dz'\frac{(z-z')J_y(x',y',z')-(y-y')J_z(x',y',z')}{[(x-x')^2+(y-y')^2+(z-z')^2]^{3/2}} \\
B_y(x,y,z) &=& \frac{\mu_0}{4\pi}\int_{-\infty}^{+\infty}dx'\int_{-\infty}^{+\infty}dy'\int_{z_1}^{z_2}dz'\frac{(x-x')J_z(x',y',z')-(z-z')J_x(x',y',z')}{[(x-x')^2+(y-y')^2+(z-z')^2]^{3/2}} \\
B_z(x,y,z) &=& \frac{\mu_0}{4\pi}\int_{-\infty}^{+\infty}dx'\int_{-\infty}^{+\infty}dy'\int_{z_1}^{z_2}dz'\frac{(y-y')J_x(x',y',z')-(x-x')J_y(x',y',z')}{[(x-x')^2+(y-y')^2+(z-z')^2]^{3/2}}~. \label{eq:BScart2}
\end{eqnarray}
Defining the two-dimensional Fourier transform of a generic function $F(x,y,z)$ in the $xy$ plane as
\begin{eqnarray} 
f(k_x,k_y,z) = \int_{-\infty}^{+\infty}\int_{-\infty}^{+\infty}dxdyF(x,y,z)e^{i(k_xx+k_yy)}~ 
\end{eqnarray}
where ${\bf k}=(k_x,k_y)$ is the spatial frequency vector, we can rewrite the Biot-Savart law in the Fourier space as
\begin{eqnarray} \label{eq:BSfourier}
b_x(k_x,k_y,z) &=& \frac{\mu_0}{2}\int_{z_1}^{z_2}dz'e^{-k|z-z'|}\left[{\rm sgn}(z-z')j_y(k_x,k_y,z')-i\frac{k_y}{k}j_z(k_x,k_y,z')\right]   \\ 
b_y(k_x,k_y,z) &=& \frac{\mu_0}{2}\int_{z_1}^{z_2}dz'e^{-k|z-z'|}\left[i\frac{k_x}{k}j_z(k_x,k_y,z')-{\rm sgn}(z-z')j_x(k_x,k_y,z')\right]   \label{eq:BSfourier2} \\
b_z(k_x,k_y,z) &=& \frac{\mu_0}{2}\int_{z_1}^{z_2}dz'e^{-k|z-z'|}\left[i\frac{k_y}{k}j_x(k_x,k_y,z')-i\frac{k_x}{k}j_y(k_x,k_y,z')\right] \label{eq:BSfourier3}
\end{eqnarray}
where $k=\sqrt{k_x^2+k_y^2}$ and sgn stands for the signum function defined such that sgn$(z)=z/|z|$ if $z\neq0$ and sgn$(0)=0$. These equations are valid as long as $k\neq0$ (i.e. except for the spatial DC component), and are typically found without the sgn$(z-z')$ factor, e.g. in Ref.~\cite{Roth1989}, because one generally probes only one side of the sources hence sgn$(z-z')$ has a constant value. If we decompose the total magnetic field into its contributions from the sources above (${\bf B}^+$, such that $z_1>z$) and below (${\bf B}^-$, such that $z_2<z$) the $z$ plane, i.e. ${\bf B}={\bf B}^++{\bf B}^-$, we see from Eqs.~(\ref{eq:BSfourier}-\ref{eq:BSfourier3}) that we can write the following relationship between the magnetic field components in the Fourier plane,
\begin{eqnarray} \label{eq:relation}
i\frac{k_x}{k}b_x^\pm(k_x,k_y,z)+i\frac{k_y}{k}b_y^\pm(k_x,k_y,z)=\pm b_z^\pm(k_x,k_y,z)~.
\end{eqnarray}
This equality applies independently to the components of ${\bf B}^+$ and to the components of ${\bf B}^-$, with a sign difference ($\pm$) in the right-hand-side term between these two cases. We note that this relation can be directly derived from Gauss's law for magnetism, $\nabla\cdot{\bf B}=0$, where the $\pm$ sign then comes from the choice of upward/downward continuation when evaluating the $\frac{\partial b_z}{\partial z}$ term~\cite{Lima2009}. Further inspection of Eqs.~(\ref{eq:BSfourier}-\ref{eq:BSfourier3}) shows that there is no other relationship between the field components if ${\bf J}$ is not specified, hence those are {\it not} completely inter-related in general. Moreover, in a situation where the sources are distributed both above and below the $z$ plane, then there is no relationship at all between the components of the {\it total} field.

To obtain Eqs.~(\ref{eq:bx},\ref{eq:by}) from the Biot-Savart law, one must use the continuity condition for the current, $\nabla\cdot{\bf J}=0$, which is valid only in the magnetostatic approximation (precisely, when $\frac{\partial\rho}{\partial t}=0$ where $\rho$ is the electric charge density). In the Fourier space, $\nabla\cdot{\bf J}=0$ becomes $-ik_xj_x-ik_yj_y+\frac{\partial j_z}{\partial z}=0$. Injecting this into Eq.~(\ref{eq:BSfourier3}) to eliminate $j_y$, we obtain
\begin{eqnarray} \label{eq:BSfourier4}
b_z(z) &=& \frac{\mu_0}{2}\int_{z_1}^{z_2}dz'e^{-k|z-z'|}\left[i\frac{k_y}{k}j_x(z')-i\frac{k_x}{k}\left(-\frac{k_x}{k_y}j_x(z')-\frac{i}{k_y}\frac{\partial j_z}{\partial z'}\right)\right] 
\end{eqnarray}
where we dropped the $(k_x,k_y)$ indices for clarity. Using an integration by parts and the fact that $j_z(z_1)=j_z(z_2)=0$ by assumption that the sources are confined to the slab, we get that
\begin{eqnarray} \label{eq:BSfourier5}
\int_{z_1}^{z_2}dz'e^{-k|z-z'|}\frac{\partial j_z}{\partial z'} =  -\int_{z_1}^{z_2}dz'{\rm sgn}(z-z')e^{-k|z-z'|}kj_z(z')~.  
\end{eqnarray}
We can then simplify Eq.~(\ref{eq:BSfourier4}),
\begin{eqnarray} 
b_z(z) &=& \frac{\mu_0}{2}\int_{z_1}^{z_2}dz'e^{-k|z-z'|}\left[i\frac{k}{k_y}j_x(z')+{\rm sgn}(z-z')\frac{k_x}{k_y}j_z(z')\right]  \\
\frac{ik_y}{k}b_z(z) &=& \frac{\mu_0}{2}\int_{z_1}^{z_2}dz'e^{-k|z-z'|}{\rm sgn}(z-z')\left[-{\rm sgn}(z-z')j_x(z')+i\frac{k_x}{k}j_z(z')\right]~, 
\end{eqnarray}
\end{widetext}
which by identification with Eq.~(\ref{eq:BSfourier2}) gives Eq.~(\ref{eq:by}). Likewise, eliminating $j_x$ in Eq.~(\ref{eq:BSfourier3}) leads to Eq.~(\ref{eq:bx}). 

In summary, the components of the magnetic field are completely inter-related {\it only} (i) in a magnetostatic situation {\it and} (ii) if the sources are located on a single side of the measurement plane, with a difference in sign in these relationships depending on which side the sources are on. The latter is a consequence of the symmetry properties of the Biot-Savart law, see Eqs.~(\ref{eq:BScart}-\ref{eq:BScart2}), where the different terms are either even or odd functions of $(z-z')$. In the case of a current in a wire, it simply means that measuring the magnetic field above or below the wire changes the sign of the planar components without changing the out-of-plane component. In the general case where there are sources on both sides of the measurement plane, there is no relationship between the total out-of-plane component ($B_z$) and the in-plane components, however the in-plane components are still related via $k_yb_x=k_xb_y$ in the magnetostatic approximation. 

In our experiments, we measure a time-averaged magnetic field because the measurement is repeated a large number of times ($\sim10^6$ times, defining one measurement as one $\pi$-flip on the NV spins) and therefore we are only sensitive to the time average of the current density. The quantities ${\bf B}$ and ${\bf J}$ throughout the paper thus refer to the time-averaged magnetic field and current density, respectively. We note that there is also a fluctuating component in the current density due to the thermal motion of the charge carriers leading to a fluctuating magnetic field~\cite{Kolkowitz2015}. These fluctuations average to zero and are present even in the absence of DC current ($I=0$). As such, they are not expected to affect the measurement of the current-induced time-averaged magnetic field as determined by frequency shifts in the ODMR spectrum (see examples ODMR spectra with the current on/off in Fig.~\ref{FigODMR}c).

\section{Violation of Gauss's law for magnetism} \label{sec:gauss}

\begin{figure}[t!]
	\begin{center}
		\includegraphics[width=0.45\textwidth]{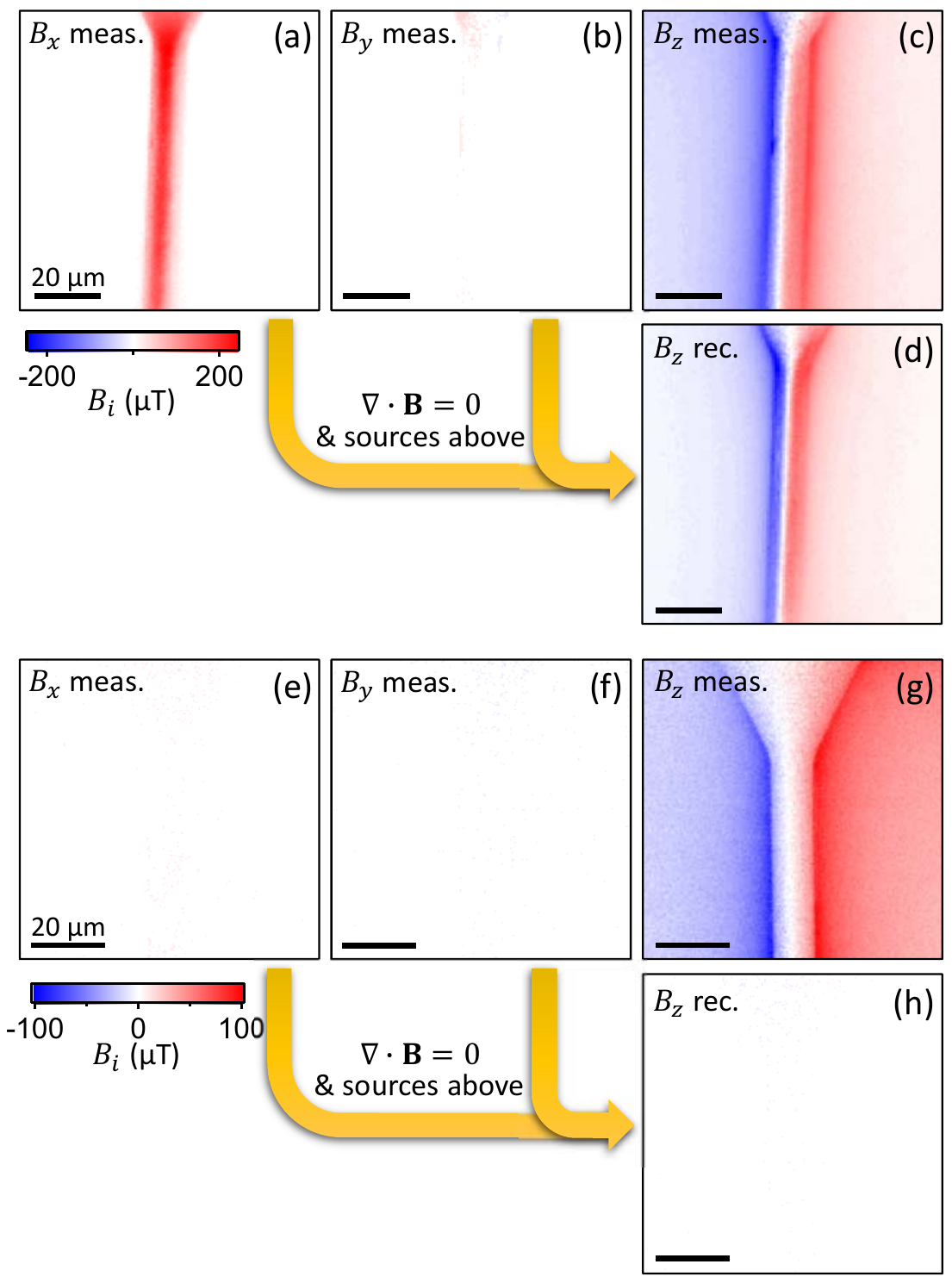}
		\caption{\h (a-c) Maps of the magnetic field components $B_x$ (a), $B_y$ (b) and $B_z$ (c) reproduced from Fig.~\ref{FigAnomaly}c-e, corresponding to sample \#1. (d) Map of the reconstructed $B_z$ component based on the measured $B_x$ and $B_y$ and Eq.~(\ref{eq:Gauss}). (e-h) Same as (a-d) but for sample \#2, with (a-c) reproduced from Fig.~\ref{FigAnomaly}l-n. }
		\label{FigGauss}
	\end{center}
\end{figure} 

{\h 
Gauss's law for magnetism, $\nabla\cdot{\bf B}=0$, is given in the Fourier space by Eq.~(\ref{eq:relation}). In the case where all the sources are located above the measurement plane, it reads \begin{eqnarray} \label{eq:Gauss}
ik_xb_x+ik_yb_y=kb_z~.
\end{eqnarray}
This equality, which reflects the fact that the magnetic field has no divergence, or that magnetic monopoles do not exist, is unconditionally true apart from the assumption on the location of the sources. We can use it to reconstruct the $B_z$ component from the measured $B_x$ and $B_y$ and compare to the measured $B_z$. This is shown in Fig.~\ref{FigGauss} applied to the data of samples \# 1 and \#2. Clearly, Gauss's law is not satisfied if we assume that all the magnetic field sources are above the NV plane. 
}

\section{Relationship between magnetic field and current density} \label{sec:relationships2}

Eqs.~(\ref{eq:BSfourier}-\ref{eq:BSfourier3}) can be inverted in a few special cases~\cite{Roth1989}. For instance, if the current density ${\bf J}(x,y,z)$ does not depend on the vertical position $z$ (which implies that $J_z$ is identically null), we can integrate $z$ out to obtain, separating the sources that are above (${\bf J}^+$) and below (${\bf J}^-$) the NV plane as previously,
\begin{eqnarray} \label{eq:link}
b_x^\pm(k_x,k_y) &=& \mp\frac{\mu_0}{2}g(k)j_y^\pm(k_x,k_y)   \\ 
b_y^\pm(k_x,k_y) &=& \pm\frac{\mu_0}{2}g(k)j_x^\pm(k_x,k_y) \\
b_z^\pm(k_x,k_y) &=& \frac{\mu_0}{2}g(k)\left[i\frac{k_y}{k}j_x^\pm(k_x,k_y)\right. \nonumber  \\
& & \left. -i\frac{k_x}{k}j_y^\pm(k_x,k_y)\right] \label{eq:link2}
\end{eqnarray}
where $g(k)=\frac{e^{-kh_{\rm min}}\left(1-e^{-kt}\right)}{k}$ is a geometric factor, $h_{\rm min}$ the minimum distance between the slab and the NV plane (i.e. the minimum of $|z-z'|$) and $t$ is the thickness of the slab (i.e. $t=|z_2-z_1|$). Alternatively, if we allow the current density to vary with $z$ but assume that the exponent in the propagation factor $e^{-k|z-z'|}$ is such that $k|z-z'|\ll 1$, then we obtain
\begin{eqnarray} \label{eq:link3}
b_x^\pm(k_x,k_y) &=& \mp\frac{\mu_0}{2}\tilde{j}_y^\pm(k_x,k_y)   \\ 
b_y^\pm(k_x,k_y) &=& \pm\frac{\mu_0}{2}\tilde{j}_x^\pm(k_x,k_y) \\
b_z^\pm(k_x,k_y) &=& \frac{\mu_0}{2}\left[i\frac{k_y}{k}\tilde{j}_x^\pm(k_x,k_y)\right. \nonumber  \\
& & \left. -i\frac{k_x}{k}\tilde{j}_y^\pm(k_x,k_y)\right] \label{eq:link4}
\end{eqnarray}
where we introduced the projected current density
\begin{eqnarray}
\tilde{j}_p^\pm(k_x,k_y)=\int_{z_1}^{z_2}dz~j_p^\pm(k_x,k_y,z)
\end{eqnarray}
and we automatically have $\tilde{j}_z^\pm=0$  because the current is confined in the slab. The condition $k|z-z'|\ll 1$ requires that $h_{\rm max}\ll\Delta x_{\rm min}$ where $h_{\rm max}$ is the maximum distance between the slab and the NV plane (i.e. the maximum of $|z-z'|$) and $\Delta x_{\rm min}$ is the lateral spatial resolution of the measurement, which sets the maximum $k$-value accessible. In our experiments, $\Delta x_{\rm min}\approx500$~nm roughly matched to the pixel size, which implies that $k|z-z'|\ll 1$ is a very good approximation as long as $h_{\rm max}\lesssim100$~nm. From Eqs.~(\ref{eq:link3}-\ref{eq:link4}), we can then link the total magnetic field ${\bf b}={\bf b}^++{\bf b}^-$ to the current densities ${\bf j}^+$ and ${\bf j}^-$, 
\begin{eqnarray} \label{eq:link5}
b_x = b_x^++b_x^- &=& -\frac{\mu_0}{2}\left(\tilde{j}_y^+-\tilde{j}_y^-\right) \\ 
b_y = b_y^++b_y^- &=& \frac{\mu_0}{2}\left(\tilde{j}_x^+-\tilde{j}_x^-\right)  \label{eq:link8} \\
b_z = b_z^++b_z^- &=& -\frac{\mu_0}{2}\frac{ik}{k_x}\left(\tilde{j}_y^++\tilde{j}_y^-\right)  \label{eq:link7} \\ 
b_z = b_z^++b_z^- &=& \frac{\mu_0}{2}\frac{ik}{k_y}\left(\tilde{j}_x^++\tilde{j}_x^-\right) \label{eq:link6}
\end{eqnarray}
where we used the continuity of the current density in Eq.~(\ref{eq:link4}). Using Eqs.~(\ref{eq:link5}-\ref{eq:link6}), the projected current densities $\tilde{\bf J}^+$ and $\tilde{\bf J}^-$ can be fully determined from the measurement of the total magnetic field, with no experimental parameter otherwise. In practice, we first use Eqs.~(\ref{eq:link5},\ref{eq:link8}) to determine the difference $\tilde{\bf J}^w\doteq\tilde{\bf J}^+-\tilde{\bf J}^-$ from the measured in-plane field components $B_x$ and $B_y$, and Eqs.~(\ref{eq:link7},\ref{eq:link6}) to determine the sum $\tilde{\bf J}\doteq\tilde{\bf J}^++\tilde{\bf J}^-$ from the out-of-plane component $B_z$. We can then deduce $\tilde{\bf J}^d\doteq\tilde{\bf J}-\tilde{\bf J}^w$, $\tilde{\bf J}^-=\tilde{\bf J}^d/2$ and $\tilde{\bf J}^+=\tilde{\bf J}-\tilde{\bf J}^-$.

\section{Truncation artefacts in the reconstructed current density} \label{sec:truncation}

\begin{figure*}[t!]
	\begin{center}
		\includegraphics[width=0.99\textwidth]{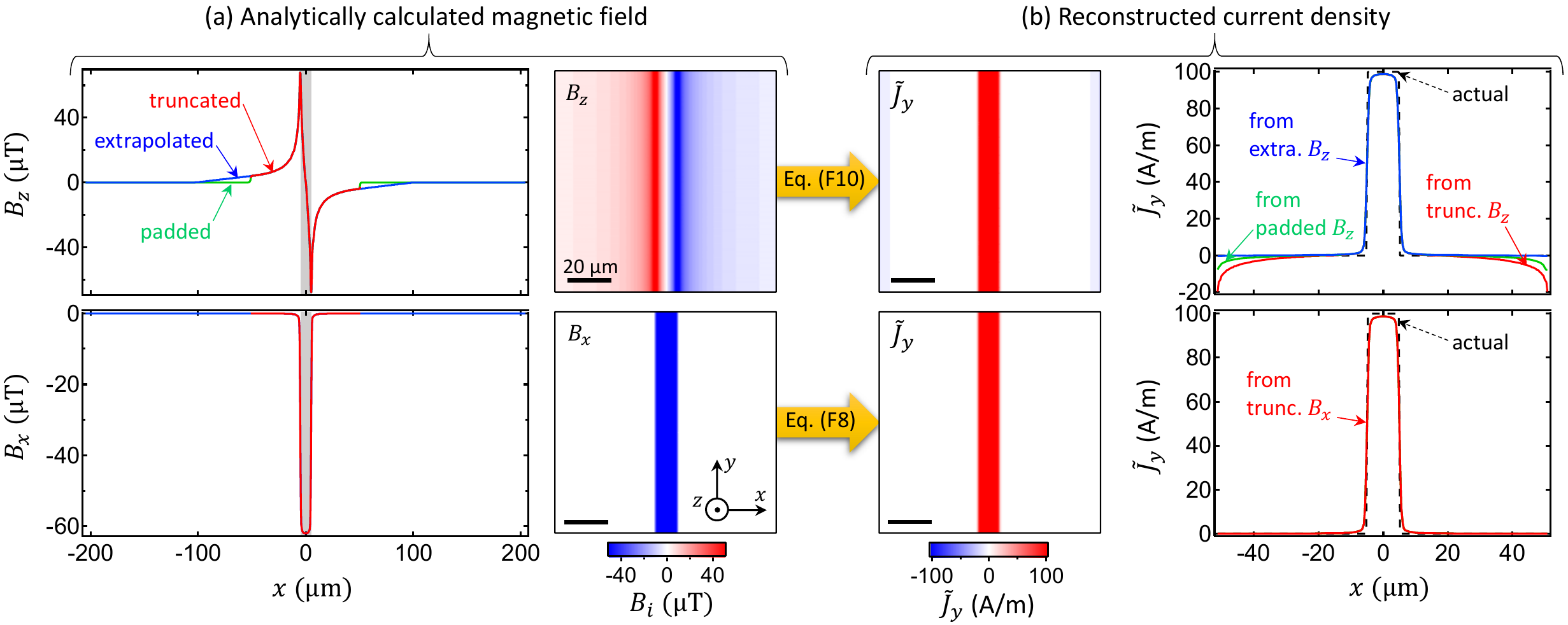}
		\caption{(a) Magnetic field from an infinitely long thin wire calculated using Eqs.~(\ref{eq:BSanalZ}). The width of the wire is $w=10~\mu$m, the probe distance is $h_{\rm NV}=1$~nm and the total current is $I=1$~mA. The images show the $B_z$ (top) and $B_x$ (bottom) field components within a $250\times 250$ pixels image with a pixel size of $400\times 400$~nm$^2$. The graphs show the profiles of $B_z$ (top) and $B_z$ (bottom) vs $x$. The red lines show the profile limited to the size of the images (`truncated'); the green lines extend the truncated profile by padding with zeros (`padded'); the blue lines extend the truncated profile by linearly extrapolating the end until the field reaches zero (`extrapolated'). (b) Current density $\tilde{J}_y$ reconstructed from the magnetic field plotted in (a). The images show $\tilde{J}_y$ reconstructed from the truncated $B_z$ (top) and from the truncated $B_x$ (bottom). The graphs show the profile of $\tilde{J}_y$ vs $x$ as reconstructed using $B_z$ (top) or $B_x$ (bottom). Red, green and blue lines correspond to the current density reconstructed from the truncated field, padded field and extrapolated field, respectively. The dashed line is the actual current density assumed in the calculation of the magnetic field. For $\tilde{J}_y$ calculated from $B_z$, an offset was added to ensure the different curves coincide at the centre of the wire.}
		\label{FigTruncation}
	\end{center}
\end{figure*} 

Here we analyse the artefacts in the reconstructed current density arising from truncation in the measured magnetic field. To do so, we consider a current $I$ flowing in an infinitely long straight wire of width $w=10~\mu$m and thickness $t\ll w$. In this limit, the magnetic field at a distance $h_{\rm NV}$ is given by Eqs.~(\ref{eq:BSanalZ}). When reconstructing the current density using Eqs.~(\ref{eq:link5}-\ref{eq:link6}), artefacts arise because the magnetic field is measured in a finite region of space near the wire. Specifically, in our experiments we typically record images with about $250\times 250$ pixels and a pixel size of $400\times 400$~nm$^2$, hence a $100\times 100~\mu$m$^2$ field of view. Profiles and images of the calculated magnetic field in this scenario are shown in Fig.~\ref{FigTruncation}a. As clearly seen in the graphs, the problem is particularly pronounced for the $B_z$ component, which decays more slowly than $B_x$ ($\sim1/x$ against $\sim1/x^3$) and still evaluates to about 5\% of its maximum value at the edges of the image (i.e. $50~\mu$m away for the centre of the wire), against $10^{-4}$ for $B_x$. 

In Fig.~\ref{FigTruncation}b, we compare the current density $\tilde{J}_y$ reconstructed through different strategies. In the images and the red lines in the graphs, we use the truncated field as directly measured. While the $\tilde{J}_y$ obtained from $B_x$ is faithful to the actual current density (dashed line in the graphs) except for the smoothing that comes from the finite pixel size (400 nm), that from the truncated $B_z$ exhibits a strong deviation (with a negative sign) near the edges. This would lead to a significant error in the estimation of the total integrated current, here by 21\% using $B_z$ (only 0.1\% using $B_x$). Another issue is that since the inversion of Eq.~(\ref{eq:link6}) is not valid for $k=0$, the DC offset in $\tilde{J}_y$ (when calculated from $B_z$) must be adjusted based on physical considerations, typically by requiring that the current density be null far from the wire, which is complicated by the presence of these edge artefacts. Note that in Fig.~\ref{FigTruncation}b we chose the DC offset such that $\tilde{J}_y$ is identical at the centre of the wire across the different reconstruction strategies.   

There are several solutions to this problem. One solution commonly employed is zero padding, which involves extending the magnetic field beyond the measured region by padding with zeros (green line in the graphs in Fig.~\ref{FigTruncation}a). As shown in the graphs in Fig.~\ref{FigTruncation}b (green line), zero padding (applied in the $x$ direction only) reduces the artefacts but does not suppress them completely (still 7\% error in the integrated current). Another approach is to extrapolate the magnetic field. Here we use a simple linear extrapolation based on the slope of the field at the edge of the images, and extend the field until it reaches zero beyond which we pad with zeros (blue lines in the graphs in Fig.~\ref{FigTruncation}a). As shown in the graphs in Fig.~\ref{FigTruncation}b (blue line), this extrapolation (applied in the $x$ direction only) efficiently suppresses the artefacts, with a remaining error of only 1.5\% in the integrated current. 

In our experiments, the reconstructed current density is not confined inside the metallic wire but instead leaks into the diamond and spreads laterally over several micrometres. This means that the magnetic field $B_z$ decays even slower than in the normal wire case, accentuating the truncation problem and making the use of extrapolation indispensable. We note that more complicated extrapolation schemes could in principle be applied, however they would not increase the accuracy in the absence of a model describing how the current density (hence the magnetic field) should decay. In practice, we applied the linear extrapolation by evaluating the slope using an average of the last 10 pixels of the image in order to average out the noise. In the $y$ direction, we did not apply any padding or extrapolation in the case of a straight wire because the truncated data is equivalent to a periodic condition, hence does not produce any artefact in the ideal case (see images in Fig.~\ref{FigTruncation}b). In sample \#5 where the current flows both along $x$ and $y$, we applied a linear extrapolation in both directions (i.e. at all four boundaries of the image). Finally, we note that other sources of artefacts may come from long-range contributions to the measured magnetic field due to the current flowing in remote wires (in particular, in the leads connected to the imaged wire).

\section{Uncertainties}  \label{sec:uncertainties}

Here we estimate the uncertainties associated with the different quantities determined from experiment. The total magnetic field ${\bf B}_{\rm tot}={\bf B}_0+{\bf B}_I$ as determined by fitting of the ODMR data is susceptible to systematic errors of the order of tens of $\mu$T for two main reasons: (i) the presence of strain or of residual electric field due to surface band bending~\cite{Broadway2018}, and (ii) an asymmetry in the line shape of the ODMR lines, for instance due to partial polarisation of the $^{15}$N nuclear spin of the NV centres. This may explain the small artefacts seen near the edges of the wire in the ${\bf B}_0$ maps in Fig.~\ref{FigB0}. However, these systematic errors are unchanged (to first order) when measuring ${\bf B}_{\rm tot}$ or ${\bf B}_0$, so that they produce a negligible correction ($<1~\mu$T) to the current-induced field ${\bf B}_I={\bf B}_{\rm tot}-{\bf B}_0$. In ${\bf B}_I$, the main source of systematic error arises from temperature drifts causing the bias field ${\bf B}_0$ to change between the two measurements~\cite{Broadway2018b}, which manifests as an overall offset in the magnetic field maps of up to $2~\mu$T typically. As for random (statistical) errors in ${\bf B}_I$, they are dominated by the photon count noise in the ODMR data and result in a typical uncertainty of $1~\mu$T for a single pixel, determined by evaluating the standard deviation of the magnetic field in a small uniform region of the sample (a measure of the pixel-to-pixel noise). The systematic error can be readily converted into an uncertainty for the value of $\chi$ via Eq.~(\ref{eq:chi}), for example we obtain an absolute uncertainty of 0.06 for a $2~\mu$T uncertainty in $B_x$, a current of $I=5$~mA and a $100~\mu$m wide image.    
     
When reconstructing the current density, errors from the measured magnetic field are propagated and additional errors are introduced. According to Eqs.~(\ref{eq:link5}-\ref{eq:link6}), an erroneous offset of $\sim2~\mu$T in $B_x$ ($B_y$) translates into an offset of $\sim3$~A/m in the difference $\tilde{J}_y^w\doteq\tilde{J}_y^+-\tilde{J}_y^-$ ($\tilde{J}_x^w$), whereas an overall offset in $B_z$ has no consequence since the $k=0$ component is not determined. For $\tilde{\bf J}^w$, this is the main source of error because the truncation artefacts are negligible in $B_x$ and $B_y$ as shown in Sec.~\ref{sec:truncation}. As a result, the systematic error in the integrated current $I_w$ for a $100~\mu$m wide image can be up to 0.3 mA, independent of the absolute value of $I_w$. 

The $B_z$ field component is used to infer the total current density, for instance Eq.~(\ref{eq:link7}) gives the $y$-component, $\tilde{J}_y=\tilde{J}_y^++\tilde{J}_y^-$. Here the main sources of error are: (i) truncation artefacts, and (ii) error in the estimation of the DC offset ($k=0$). Truncation artefacts affect the current density calculated near the edges of the images, and can be mitigated by appropriate extrapolation as shown in Sec.~\ref{sec:truncation}. The DC offset is more problematic. Indeed, it is normally set by physical considerations, by requiring that the current density be null far from the current-carrying wire, but this may be incorrect in the present case where the current leaks over large distances, especially in sample \#2 where it seems that $\tilde{J}_y$ has not completely decayed at the boundaries of the image. Furthermore, small errors near the edges due to truncation artefacts may translate into an error in the offset, since we use the edges to determine it. In particular, in most samples we find that the reconstructed $\tilde{J}_y$ differ slightly between the left and right boundaries by up to $\sim6$~A/m typically, which we attribute to residual truncation artefacts (even with the extrapolation). Choosing the offset to cancel the mean value of $\tilde{J}_y$ at the left and right boundaries, we thus have a possible error in the offset of $\pm3$~A/m. As a result, the systematic error in the integrated current $I_{\rm tot}$ for a $100~\mu$m wide image can be up to 0.3 mA, relatively independent of the absolute value of $I_{\rm tot}$.  

Thus, the uncertainty in $I_{\rm tot}$ (determined from $B_z$) and in $I_w$ (determined from $B_x$) is about 0.3~mA each, which gives an uncertainty of 0.4~mA in the difference $I_d=I_{\rm tot}-I_w$. These are the values quoted in the main text.

\bibliographystyle{naturemag}
\bibliography{bib}	

\begin{thebibliography}{10}
\expandafter\ifx\csname url\endcsname\relax
  \def\url#1{\texttt{#1}}\fi
\expandafter\ifx\csname urlprefix\endcsname\relax\def\urlprefix{URL }\fi
\providecommand{\bibinfo}[2]{#2}
\providecommand{\eprint}[2][]{\url{#2}}

\bibitem{Doherty2013}
\bibinfo{author}{Doherty, M.~W.} \emph{et~al.}
\newblock \bibinfo{title}{{The nitrogen-vacancy colour centre in diamond}}.
\newblock \emph{\bibinfo{journal}{Physics Reports}}
  \textbf{\bibinfo{volume}{528}}, \bibinfo{pages}{1--45}
  (\bibinfo{year}{2013}).

\bibitem{Rondin2014}
\bibinfo{author}{Rondin, L.} \emph{et~al.}
\newblock \bibinfo{title}{{Magnetometry with nitrogen-vacancy defects in
  diamond}}.
\newblock \emph{\bibinfo{journal}{Rep. Prog. Phys.}}
  \textbf{\bibinfo{volume}{77}}, \bibinfo{pages}{56503} (\bibinfo{year}{2014}).

\bibitem{Casola2018}
\bibinfo{author}{Casola, F.}, \bibinfo{author}{{Van Der Sar}, T.} \&
  \bibinfo{author}{Yacoby, A.}
\newblock \bibinfo{title}{{Probing condensed matter physics with magnetometry
  based on nitrogen-vacancy centres in diamond}}.
\newblock \emph{\bibinfo{journal}{Nature Reviews Materials}}
  \textbf{\bibinfo{volume}{3}}, \bibinfo{pages}{17088} (\bibinfo{year}{2018}).

\bibitem{Acosta2010}
\bibinfo{author}{Acosta, V.~M.} \emph{et~al.}
\newblock \bibinfo{title}{Temperature dependence of the nitrogen-vacancy
  magnetic resonance in diamond}.
\newblock \emph{\bibinfo{journal}{Phys. Rev. Lett.}}
  \textbf{\bibinfo{volume}{104}}, \bibinfo{pages}{070801}
  (\bibinfo{year}{2010}).

\bibitem{Toyli2012}
\bibinfo{author}{Toyli, D.~M.} \emph{et~al.}
\newblock \bibinfo{title}{Measurement and control of single nitrogen-vacancy
  center spins above 600 k}.
\newblock \emph{\bibinfo{journal}{Phys. Rev. X}} \textbf{\bibinfo{volume}{2}},
  \bibinfo{pages}{031001} (\bibinfo{year}{2012}).

\bibitem{Rondin2013}
\bibinfo{author}{Rondin, L.} \emph{et~al.}
\newblock \bibinfo{title}{{Stray-field imaging of magnetic vortices with a
  single diamond spin}}.
\newblock \emph{\bibinfo{journal}{Nature Communications}}
  \textbf{\bibinfo{volume}{4}}, \bibinfo{pages}{2279} (\bibinfo{year}{2013}).
\newblock \eprint{1302.7307}.

\bibitem{Tetienne2014}
\bibinfo{author}{Tetienne, J.-P.} \emph{et~al.}
\newblock \bibinfo{title}{{Nanoscale imaging and control of domain-wall hopping
  with a nitrogen-vacancy center microscope.}}
\newblock \emph{\bibinfo{journal}{Science}} \textbf{\bibinfo{volume}{344}},
  \bibinfo{pages}{1366--9} (\bibinfo{year}{2014}).

\bibitem{Tetienne2015}
\bibinfo{author}{Tetienne, J.-P.} \emph{et~al.}
\newblock \bibinfo{title}{{The nature of domain walls in ultrathin ferromagnets
  revealed by scanning nanomagnetometry}}.
\newblock \emph{\bibinfo{journal}{Nature Communications}}
  \textbf{\bibinfo{volume}{6}}, \bibinfo{pages}{6733} (\bibinfo{year}{2015}).

\bibitem{Dussaux2016}
\bibinfo{author}{Dussaux, A.} \emph{et~al.}
\newblock \bibinfo{title}{{Local dynamics of topological magnetic defects in
  the itinerant helimagnet FeGe}}.
\newblock \emph{\bibinfo{journal}{Nature Communications}}
  \textbf{\bibinfo{volume}{7}}, \bibinfo{pages}{12430} (\bibinfo{year}{2016}).

\bibitem{Gross2016}
\bibinfo{author}{Gross, I.} \emph{et~al.}
\newblock \bibinfo{title}{{Direct measurement of interfacial
  Dzyaloshinskii-Moriya interaction in X|CoFeB|MgO heterostructures with a
  scanning NV magnetometer (X=Ta, TaN, and W)}}.
\newblock \emph{\bibinfo{journal}{Physical Review B}}
  \textbf{\bibinfo{volume}{94}}, \bibinfo{pages}{064413}
  (\bibinfo{year}{2016}).

\bibitem{Dovzhenko2018}
\bibinfo{author}{Dovzhenko, Y.} \emph{et~al.}
\newblock \bibinfo{title}{{Magnetostatic twists in room-temperature skyrmions
  explored by nitrogen-vacancy center spin texture reconstruction}}.
\newblock \emph{\bibinfo{journal}{Nature Communications}}
  \textbf{\bibinfo{volume}{9}}, \bibinfo{pages}{2712} (\bibinfo{year}{2018}).

\bibitem{Gross2017}
\bibinfo{author}{Gross, I.} \emph{et~al.}
\newblock \bibinfo{title}{{Real-space imaging of non-collinear
  antiferromagnetic order with a single-spin magnetometer}}.
\newblock \emph{\bibinfo{journal}{Nature}} \textbf{\bibinfo{volume}{549}},
  \bibinfo{pages}{252--256} (\bibinfo{year}{2017}).

\bibitem{Waxman2014}
\bibinfo{author}{Waxman, A.} \emph{et~al.}
\newblock \bibinfo{title}{Diamond magnetometry of superconducting thin films}.
\newblock \emph{\bibinfo{journal}{Phys. Rev. B}} \textbf{\bibinfo{volume}{89}},
  \bibinfo{pages}{054509} (\bibinfo{year}{2014}).

\bibitem{Thiel2016}
\bibinfo{author}{Thiel, L.} \emph{et~al.}
\newblock \bibinfo{title}{{Quantitative nanoscale vortex-imaging using a
  cryogenic quantum magnetometer}}.
\newblock \emph{\bibinfo{journal}{Nat. Nanotechnol.}}
  \textbf{\bibinfo{volume}{11}}, \bibinfo{pages}{677--681}
  (\bibinfo{year}{2016}).

\bibitem{Pelliccione2016}
\bibinfo{author}{Pelliccione, M.} \emph{et~al.}
\newblock \bibinfo{title}{{Scanned probe imaging of nanoscale magnetism at
  cryogenic temperatures with a single-spin quantum sensor}}.
\newblock \emph{\bibinfo{journal}{Nat. Nanotechnol.}}
  \textbf{\bibinfo{volume}{11}}, \bibinfo{pages}{700--705}
  (\bibinfo{year}{2016}).

\bibitem{Schlussel2018}
\bibinfo{author}{{Schlussel}, Y.} \emph{et~al.}
\newblock \bibinfo{title}{{Widefield imaging of superconductor vortices with
  electron spins in diamond}}.
\newblock \emph{\bibinfo{journal}{ArXiv e-prints}}  (\bibinfo{year}{2018}).
\newblock \eprint{1803.01957}.

\bibitem{VanderSar2015}
\bibinfo{author}{van~der Sar, T.}, \bibinfo{author}{Casola, F.},
  \bibinfo{author}{Walsworth, R.} \& \bibinfo{author}{Yacoby, A.}
\newblock \bibinfo{title}{{Nanometre-scale probing of spin waves using
  single-electron spins}}.
\newblock \emph{\bibinfo{journal}{Nature Communications}}
  \textbf{\bibinfo{volume}{6}}, \bibinfo{pages}{7886} (\bibinfo{year}{2015}).

\bibitem{Du2017}
\bibinfo{author}{Du, C.} \emph{et~al.}
\newblock \bibinfo{title}{{Control and local measurement of the spin chemical
  potential in a magnetic insulator}}.
\newblock \emph{\bibinfo{journal}{Science}} \textbf{\bibinfo{volume}{357}},
  \bibinfo{pages}{195--198} (\bibinfo{year}{2017}).

\bibitem{Page2018}
\bibinfo{author}{{Page}, M.~R.} \emph{et~al.}
\newblock \bibinfo{title}{{Optically Detected Ferromagnetic Resonance in
  Metallic Ferromagnets via Nitrogen Vacancy Centers in Diamond}}.
\newblock \emph{\bibinfo{journal}{ArXiv e-prints}}  (\bibinfo{year}{2016}).
\newblock \eprint{1607.07485}.

\bibitem{Kolkowitz2015}
\bibinfo{author}{Kolkowitz, S.} \emph{et~al.}
\newblock \bibinfo{title}{Probing johnson noise and ballistic transport in
  normal metals with a single-spin qubit}.
\newblock \emph{\bibinfo{journal}{Science}} \textbf{\bibinfo{volume}{347}},
  \bibinfo{pages}{1129--1132} (\bibinfo{year}{2015}).

\bibitem{Agarwal2017}
\bibinfo{author}{Agarwal, K.} \emph{et~al.}
\newblock \bibinfo{title}{Magnetic noise spectroscopy as a probe of local
  electronic correlations in two-dimensional systems}.
\newblock \emph{\bibinfo{journal}{Phys. Rev. B}} \textbf{\bibinfo{volume}{95}},
  \bibinfo{pages}{155107} (\bibinfo{year}{2017}).

\bibitem{Ariyaratne2018}
\bibinfo{author}{{Ariyaratne}, A.}, \bibinfo{author}{{Bluvstein}, D.},
  \bibinfo{author}{{Myers}, B.~A.} \& \bibinfo{author}{{Bleszynski Jayich},
  A.~C.}
\newblock \bibinfo{title}{{Nanoscale electrical conductivity imaging using a
  nitrogen-vacancy center in diamond}}.
\newblock \emph{\bibinfo{journal}{Nat. Commun.}} \textbf{\bibinfo{volume}{9}},
  \bibinfo{pages}{2406} (\bibinfo{year}{2018}).

\bibitem{Nowodzinski2015}
\bibinfo{author}{Nowodzinski, A.} \emph{et~al.}
\newblock \bibinfo{title}{Nitrogen-vacancy centers in diamond for current
  imaging at the redistributive layer level of integrated circuits}.
\newblock \emph{\bibinfo{journal}{Microelectronics Reliability}}
  \textbf{\bibinfo{volume}{55}}, \bibinfo{pages}{1549 -- 1553}
  (\bibinfo{year}{2015}).
\newblock \bibinfo{note}{Proceedings of the 26th European Symposium on
  Reliability of Electron Devices, Failure Physics and Analysis}.

\bibitem{Chang2017}
\bibinfo{author}{Chang, K.}, \bibinfo{author}{Eichler, A.},
  \bibinfo{author}{Rhensius, J.}, \bibinfo{author}{Lorenzelli, L.} \&
  \bibinfo{author}{Degen, C.~L.}
\newblock \bibinfo{title}{Nanoscale imaging of current density with a
  single-spin magnetometer}.
\newblock \emph{\bibinfo{journal}{Nano Letters}} \textbf{\bibinfo{volume}{17}},
  \bibinfo{pages}{2367--2373} (\bibinfo{year}{2017}).

\bibitem{Tetienne2017}
\bibinfo{author}{Tetienne, J.-P.} \emph{et~al.}
\newblock \bibinfo{title}{Quantum imaging of current flow in graphene}.
\newblock \emph{\bibinfo{journal}{Science Advances}}
  \textbf{\bibinfo{volume}{3}}, \bibinfo{pages}{e1602429}
  (\bibinfo{year}{2017}).

\bibitem{Roth1989}
\bibinfo{author}{Roth, B.~J.}, \bibinfo{author}{Sepulveda, N.~G.} \&
  \bibinfo{author}{Wikswo, J.~P.}
\newblock \bibinfo{title}{{Using a magnetometer to image a two-dimensional
  current distribution}}.
\newblock \emph{\bibinfo{journal}{Journal of Applied Physics}}
  \textbf{\bibinfo{volume}{65}}, \bibinfo{pages}{361--372}
  (\bibinfo{year}{1989}).

\bibitem{Meltzer2018}
\bibinfo{author}{{Meltzer}, A.~Y.}, \bibinfo{author}{{Levin}, E.} \&
  \bibinfo{author}{{Zeldov}, E.}
\newblock \bibinfo{title}{{Direct Reconstruction of Two-Dimensional Currents in
  Thin Films from Magnetic-Field Measurements}}.
\newblock \emph{\bibinfo{journal}{Physical Review Applied}}
  \textbf{\bibinfo{volume}{8}}, \bibinfo{pages}{064030} (\bibinfo{year}{2017}).

\bibitem{Chen2016}
\bibinfo{author}{Chen, S.} \emph{et~al.}
\newblock \bibinfo{title}{Electron optics with p-n junctions in ballistic
  graphene}.
\newblock \emph{\bibinfo{journal}{Science}} \textbf{\bibinfo{volume}{353}},
  \bibinfo{pages}{1522--1525} (\bibinfo{year}{2016}).

\bibitem{Bandurin2016}
\bibinfo{author}{Bandurin, D.~A.} \emph{et~al.}
\newblock \bibinfo{title}{{Negative local resistance caused by viscous electron
  backflow in graphene}}.
\newblock \emph{\bibinfo{journal}{Science}} \textbf{\bibinfo{volume}{351}},
  \bibinfo{pages}{1055--8} (\bibinfo{year}{2016}).

\bibitem{Steinert2010}
\bibinfo{author}{Steinert, S.} \emph{et~al.}
\newblock \bibinfo{title}{{High sensitivity magnetic imaging using an array of
  spins in diamond}}.
\newblock \emph{\bibinfo{journal}{Rev. Sci. Instrum.}}
  \textbf{\bibinfo{volume}{81}}, \bibinfo{pages}{043705}
  (\bibinfo{year}{2010}).

\bibitem{Maertz2010}
\bibinfo{author}{Maertz, B.~J.}, \bibinfo{author}{Wijnheijmer, a.~P.},
  \bibinfo{author}{Fuchs, G.~D.}, \bibinfo{author}{Nowakowski, M.~E.} \&
  \bibinfo{author}{Awschalom, D.~D.}
\newblock \bibinfo{title}{{Vector magnetic field microscopy using nitrogen
  vacancy centers in diamond}}.
\newblock \emph{\bibinfo{journal}{Appl. Phys. Lett.}}
  \textbf{\bibinfo{volume}{96}}, \bibinfo{pages}{30--32}
  (\bibinfo{year}{2010}).

\bibitem{Pham2011}
\bibinfo{author}{Pham, L.~M.} \emph{et~al.}
\newblock \bibinfo{title}{{Magnetic field imaging with nitrogen-vacancy
  ensembles}}.
\newblock \emph{\bibinfo{journal}{New J. Phys.}} \textbf{\bibinfo{volume}{13}},
  \bibinfo{pages}{045021} (\bibinfo{year}{2011}).

\bibitem{Simpson2016}
\bibinfo{author}{Simpson, D.~A.} \emph{et~al.}
\newblock \bibinfo{title}{{Magneto-optical imaging of thin magnetic films using
  spins in diamond}}.
\newblock \emph{\bibinfo{journal}{Sci. Rep.}} \textbf{\bibinfo{volume}{6}},
  \bibinfo{pages}{22797} (\bibinfo{year}{2016}).

\bibitem{Chipaux2015}
\bibinfo{author}{Chipaux, M.} \emph{et~al.}
\newblock \bibinfo{title}{{Magnetic imaging with an ensemble of NV centers in
  diamond}}.
\newblock \emph{\bibinfo{journal}{Eur. Phys. J. D}}
  \textbf{\bibinfo{volume}{69}}, \bibinfo{pages}{166} (\bibinfo{year}{2015}).

\bibitem{Glenn2017}
\bibinfo{author}{Glenn, D.~R.} \emph{et~al.}
\newblock \bibinfo{title}{Micrometer-scale magnetic imaging of geological
  samples using a quantum diamond microscope}.
\newblock \emph{\bibinfo{journal}{Geochemistry, Geophysics, Geosystems}}
  \textbf{\bibinfo{volume}{18}}, \bibinfo{pages}{3254--3267}
  (\bibinfo{year}{2017}).

\bibitem{Tetienne2018b}
\bibinfo{author}{Tetienne, J.-P.} \emph{et~al.}
\newblock \bibinfo{title}{{Proximity-induced artefacts in magnetic imaging with
  nitrogen-vacancy ensembles in diamond}}.
\newblock \emph{\bibinfo{journal}{Sensors}} \textbf{\bibinfo{volume}{18}},
  \bibinfo{pages}{1290} (\bibinfo{year}{2018}).

\bibitem{Lima2009}
\bibinfo{author}{Lima, E.~A.} \& \bibinfo{author}{Weiss, B.~P.}
\newblock \bibinfo{title}{{Obtaining vector magnetic field maps from
  single-component measurements of geological samples}}.
\newblock \emph{\bibinfo{journal}{Journal of Geophysical Research: Solid
  Earth}} \textbf{\bibinfo{volume}{114}}, \bibinfo{pages}{B06102}
  (\bibinfo{year}{2009}).

\bibitem{data}
 \urlprefix\url{https://cloudstor.aarnet.edu.au/plus/s/6hsVkzF54oH9MOL}.

\bibitem{Lehtinen2016}
\bibinfo{author}{Lehtinen, O.} \emph{et~al.}
\newblock \bibinfo{title}{{Molecular dynamics simulations of shallow nitrogen
  and silicon implantation into diamond}}.
\newblock \emph{\bibinfo{journal}{Phys. Rev. B}} \textbf{\bibinfo{volume}{93}},
  \bibinfo{pages}{35202} (\bibinfo{year}{2016}).

\bibitem{Ma2010}
\bibinfo{author}{Ma, W.}, \bibinfo{author}{Miao, T.} \& \bibinfo{author}{Zhang,
  X.}
\newblock \bibinfo{title}{Thermal and electrical transport characteristics of
  polycrystalline gold nanofilms}.
\newblock \emph{\bibinfo{journal}{2010 14th International Heat Transfer
  Conference}} \textbf{\bibinfo{volume}{6}}, \bibinfo{pages}{337--343}
  (\bibinfo{year}{2010}).

\bibitem{Barjon2009}
\bibinfo{author}{Barjon, J.} \emph{et~al.}
\newblock \bibinfo{title}{Resistivity of boron doped diamond}.
\newblock \emph{\bibinfo{journal}{physica status solidi (RRL)}}
  \textbf{\bibinfo{volume}{3}}, \bibinfo{pages}{202--204}
  (\bibinfo{year}{2009}).

\bibitem{Isberg2002}
\bibinfo{author}{Isberg, J.}, \bibinfo{author}{Hammersberg, J.},
  \bibinfo{author}{Johansson, E.}, \bibinfo{author}{Twitchen, D.~J.} \&
  \bibinfo{author}{Whitehead, A.~J.}
\newblock \bibinfo{title}{{High Carrier Mobility in Single-Crystal
  Plasma-Deposited Diamond.pdf}}.
\newblock \emph{\bibinfo{journal}{Science}} \textbf{\bibinfo{volume}{297}},
  \bibinfo{pages}{1670--1673} (\bibinfo{year}{2002}).

\bibitem{Nesladek2008}
\bibinfo{author}{Nesladek, M.}, \bibinfo{author}{Bogdan, A.},
  \bibinfo{author}{Deferme, W.}, \bibinfo{author}{Tranchant, N.} \&
  \bibinfo{author}{Bergonzo, P.}
\newblock \bibinfo{title}{{Charge transport in high mobility single crystal
  diamond}}.
\newblock \emph{\bibinfo{journal}{Diamond and Related Materials}}
  \textbf{\bibinfo{volume}{17}}, \bibinfo{pages}{1235--1240}
  (\bibinfo{year}{2008}).

\bibitem{Tetienne2018}
\bibinfo{author}{Tetienne, J.-P.} \emph{et~al.}
\newblock \bibinfo{title}{{Spin properties of dense near-surface ensembles of
  nitrogen-vacancy centers in diamond}}.
\newblock \emph{\bibinfo{journal}{Physical Review B}}
  \textbf{\bibinfo{volume}{97}}, \bibinfo{pages}{085402}
  (\bibinfo{year}{2018}).

\bibitem{Pakes2014}
\bibinfo{author}{Pakes, C.~I.}, \bibinfo{author}{Garrido, J.~A.} \&
  \bibinfo{author}{Kawarada, H.}
\newblock \bibinfo{title}{{Diamond surface conductivity: Properties, devices,
  and sensors}}.
\newblock \emph{\bibinfo{journal}{MRS Bulletin}} \textbf{\bibinfo{volume}{39}},
  \bibinfo{pages}{542--548} (\bibinfo{year}{2014}).

\bibitem{Broadway2018}
\bibinfo{author}{Broadway, D.~A.} \emph{et~al.}
\newblock \bibinfo{title}{{Spatial mapping of band bending in semiconductor
  devices using in situ quantum sensors}}.
\newblock \emph{\bibinfo{journal}{Nature Electronics}}
  \textbf{\bibinfo{volume}{1}}, \bibinfo{pages}{502--507}
  (\bibinfo{year}{2018}).

\bibitem{Dreau2011}
\bibinfo{author}{Dr{\'{e}}au, A.} \emph{et~al.}
\newblock \bibinfo{title}{{Avoiding power broadening in optically detected
  magnetic resonance of single NV defects for enhanced dc magnetic field
  sensitivity}}.
\newblock \emph{\bibinfo{journal}{Phys. Rev. B}} \textbf{\bibinfo{volume}{84}},
  \bibinfo{pages}{195204} (\bibinfo{year}{2011}).

\bibitem{Teraji2015}
\bibinfo{author}{Teraji, T.}
\newblock \bibinfo{title}{High-quality and high-purity homoepitaxial diamond
  (100) film growth under high oxygen concentration condition}.
\newblock \emph{\bibinfo{journal}{Journal of Applied Physics}}
  \textbf{\bibinfo{volume}{118}}, \bibinfo{pages}{115304}
  (\bibinfo{year}{2015}).

\bibitem{Lillie2018}
\bibinfo{author}{{Lillie}, S.~E.} \emph{et~al.}
\newblock \bibinfo{title}{{Magnetic noise from ultra-thin abrasively deposited
  materials on diamond}}.
\newblock \emph{\bibinfo{journal}{ArXiv e-prints}}  (\bibinfo{year}{2018}).
\newblock \eprint{1808.04085}.

\bibitem{Doherty2012}
\bibinfo{author}{Doherty, M.~W.} \emph{et~al.}
\newblock \bibinfo{title}{{Theory of the ground-state spin of the NV$^-$ center
  in diamond}}.
\newblock \emph{\bibinfo{journal}{Physical Review B}}
  \textbf{\bibinfo{volume}{85}}, \bibinfo{pages}{205203}
  (\bibinfo{year}{2012}).

\bibitem{Broadway2018c}
\bibinfo{author}{{Broadway}, D.~A.} \emph{et~al.}
\newblock \bibinfo{title}{{Microscopic imaging of elastic deformation in
  diamond via in-situ stress tensor sensors}}.
\newblock \emph{\bibinfo{journal}{ArXiv e-prints}}
  \bibinfo{pages}{arXiv:1812.01152} (\bibinfo{year}{2018}).
\newblock \eprint{1812.01152}.

\bibitem{Jackson}
\bibinfo{author}{{Jackson}, J.~D.}
\newblock \emph{\bibinfo{title}{{Classical Electrodynamics, 3rd Edition}}}
  (\bibinfo{publisher}{Wiley}, \bibinfo{year}{1998}).

\bibitem{Phillip1964}
\bibinfo{author}{Phillip, H.~R.} \& \bibinfo{author}{Taft, E.~A.}
\newblock \bibinfo{title}{Kramers-kronig analysis of reflectance data for
  diamond}.
\newblock \emph{\bibinfo{journal}{Phys. Rev.}} \textbf{\bibinfo{volume}{136}},
  \bibinfo{pages}{A1445--A1448} (\bibinfo{year}{1964}).

\bibitem{Johnson1974}
\bibinfo{author}{Johnson, P.~B.} \& \bibinfo{author}{Christy, R.~W.}
\newblock \bibinfo{title}{Optical constants of transition metals: Ti, v, cr,
  mn, fe, co, ni, and pd}.
\newblock \emph{\bibinfo{journal}{Phys. Rev. B}} \textbf{\bibinfo{volume}{9}},
  \bibinfo{pages}{5056--5070} (\bibinfo{year}{1974}).

\bibitem{Novotny}
\bibinfo{author}{Novotny, L.} \& \bibinfo{author}{Hecht, B.}
\newblock \emph{\bibinfo{title}{Principles of Nano-Optics}}
  (\bibinfo{publisher}{Cambridge University Press}, \bibinfo{year}{2006}).

\bibitem{Buchler2005}
\bibinfo{author}{Buchler, B.~C.}, \bibinfo{author}{Kalkbrenner, T.},
  \bibinfo{author}{Hettich, C.} \& \bibinfo{author}{Sandoghdar, V.}
\newblock \bibinfo{title}{{Measuring the quantum efficiency of the optical
  emission of single radiating dipoles using a scanning mirror}}.
\newblock \emph{\bibinfo{journal}{Physical Review Letters}}
  \textbf{\bibinfo{volume}{95}}, \bibinfo{pages}{063003}
  (\bibinfo{year}{2005}).

\bibitem{Tisler2013}
\bibinfo{author}{Tisler, J.} \emph{et~al.}
\newblock \bibinfo{title}{{Single defect center scanning near-field optical
  microscopy on graphene}}.
\newblock \emph{\bibinfo{journal}{Nano Lett.}} \textbf{\bibinfo{volume}{13}},
  \bibinfo{pages}{3152--3156} (\bibinfo{year}{2013}).

\bibitem{Boidin2016}
\bibinfo{author}{Boidin, R.}, \bibinfo{author}{Halenkovič, T.},
  \bibinfo{author}{Nazabal, V.}, \bibinfo{author}{Beneš, L.} \&
  \bibinfo{author}{Němec, P.}
\newblock \bibinfo{title}{Pulsed laser deposited alumina thin films}.
\newblock \emph{\bibinfo{journal}{Ceramics International}}
  \textbf{\bibinfo{volume}{42}}, \bibinfo{pages}{1177 -- 1182}
  (\bibinfo{year}{2016}).

\bibitem{French1963}
\bibinfo{author}{French, A.~P.} \& \bibinfo{author}{Tessman, J.~R.}
\newblock \bibinfo{title}{Displacement currents and magnetic fields}.
\newblock \emph{\bibinfo{journal}{American Journal of Physics}}
  \textbf{\bibinfo{volume}{31}}, \bibinfo{pages}{201--204}
  (\bibinfo{year}{1963}).

\bibitem{Terry1982}
\bibinfo{author}{Terry, W.~K.}
\newblock \bibinfo{title}{The connection between the charged-particle current
  and the displacement current}.
\newblock \emph{\bibinfo{journal}{American Journal of Physics}}
  \textbf{\bibinfo{volume}{50}}, \bibinfo{pages}{742--745}
  (\bibinfo{year}{1982}).

\bibitem{Charitat2003}
\bibinfo{author}{Charitat, T.} \& \bibinfo{author}{Graner, F.}
\newblock \bibinfo{title}{About the magnetic field of a finite wire}.
\newblock \emph{\bibinfo{journal}{European Journal of Physics}}
  \textbf{\bibinfo{volume}{24}}, \bibinfo{pages}{267} (\bibinfo{year}{2003}).

\bibitem{Hill2011}
\bibinfo{author}{Hill, S.~E.}
\newblock \bibinfo{title}{Reanalyzing the ampère-maxwell law}.
\newblock \emph{\bibinfo{journal}{The Physics Teacher}}
  \textbf{\bibinfo{volume}{49}}, \bibinfo{pages}{343--345}
  (\bibinfo{year}{2011}).

\bibitem{Rosser2013}
\bibinfo{author}{Rosser, G.}
\newblock \emph{\bibinfo{title}{Interpretation of classical electromagnetism}},
  vol.~\bibinfo{volume}{78} (\bibinfo{publisher}{Springer Science \& Business
  Media}, \bibinfo{year}{2013}).

\bibitem{Wolsky2015}
\bibinfo{author}{Wolsky, A.~M.}
\newblock \bibinfo{title}{On a charge conserving alternative to maxwell’s
  displacement current}.
\newblock \emph{\bibinfo{journal}{European Journal of Physics}}
  \textbf{\bibinfo{volume}{36}}, \bibinfo{pages}{035019}
  (\bibinfo{year}{2015}).

\bibitem{Broadway2018b}
\bibinfo{author}{Broadway, D.~A.} \emph{et~al.}
\newblock \bibinfo{title}{High precision single qubit tuning via
  thermo-magnetic field control}.
\newblock \emph{\bibinfo{journal}{Applied Physics Letters}}
  \textbf{\bibinfo{volume}{112}}, \bibinfo{pages}{103103}
  (\bibinfo{year}{2018}).

\end{thebibliography}
	   
\end{document}